\documentclass{aa}
\usepackage{latexsym, graphicx, natbib}
\usepackage{txfonts}

\begin{document}
   \title{The Simulated \hbox{\rm H\,{\sc i}} Sky at low redshift.}

   \subtitle{}

   \author{A. Popping
          \inst{1, 2}
          \and
          R. Dav$\mathrm{\acute{e}}$ \inst{3}
	  \and
	  R. Braun\inst{2}
	  \and
	  B. D. Oppenheimer\inst{3}
          }

   \offprints{A. Popping}

   \institute{Kapteyn Astronomical Institute, P.O. Box 800, 9700 AV Groningen, the Netherlands\\
              \email{popping@astro.rug.nl}
         \and
             Australia Telescope National Facility, CSIRO, P.O. Box 76, Epping, NSW 1710, Australia\\
         \and
	     Astronomy Department, University of Arizona, Tucson, AZ 85721
             }

   \date{}

 
  \abstract
{Observations of intergalactic neutral hydrogen can provide a wealth
  of information about structure and galaxy formation, potentially
  tracing accretion and feedback processes on Mpc scales.  Below
  a column density of $N_{\rm{HI}} \sim 10^{19}$ cm$^{-2}$, the
  ``edge'' or typical observational limit for \hbox{\rm H\,{\sc
      i}} emission from galaxies, simulations predict a cosmic web of
  extended emission and filamentary structures.  Current observations
  of this regime are limited by telescope sensitivity, which will soon
  advance substantially.}
{We study the distribution of neutral hydrogen and its 21cm emission
  properties in a cosmological hydrodynamic simulation, to gain more
  insights into the distribution of \hbox{\rm H\,{\sc i}} below
  $N_{\rm{HI}} \sim 10^{19}$ cm$^{-2}$.  Such Lyman Limit systems are
  expected to trace out the cosmic web, and are relatively
  unexplored.}
{Beginning with a 32 $h^{-1}$ Mpc simulation, we extract the neutral
  hydrogen component by determining the neutral fraction, including a
  post-processed correction for self-shielding based on the thermal
    pressure. We take into account molecular hydrogen, assuming an
    average density ratio $\Omega_{H_2}/\Omega_{HI} = 0.3$ at z~=~0. The
  statistical properties of the \hbox{\rm H\,{\sc i}} emission are
  compared with observations, to assess the reliability of the
  simulation.  We then make predictions for upcoming surveys.}
{ The simulated \hbox{\rm H\,{\sc i}} distribution
  robustly describes the full column density range between
  $N_{\rm{HI}} \sim 10^{14}$ and $N_{\rm{HI}} \sim 10^{21}$ cm$^{-2}$
  and agrees very well with available measurements from
    observations. Furthermore there is good correspondence in the
  statistics when looking at the two-point correlation function and
  the \hbox{\rm H\,{\sc i}} mass function. The reconstructed maps
    are used to simulate observations of existing and future
    telescopes by adding noise and taking account of the sensitivity
    of the telescopes.}
{The general agreement in statistical properties of \hbox{\rm H\,{\sc
      i}} suggests that neutral hydrogen as modeled in this
  hydrodynamic simulation is a fair representation of that in the
  Universe. Our method can be applied to other simulations, to compare
  different models of accretion and feedback.  Future \hbox{\rm
    H\,{\sc i}} observations will be able to probe the regions where
  galaxies connect to the cosmic web.}


   \maketitle
%

\section{Introduction}

Current cosmological models ascribe only about 4\% of the density in the
Universe to baryons \citep{2007ApJS..170..377S}.  The majority of
these baryons reside outside of galaxies; stars and cold galactic gas
may account for about about one third \citep{1998ApJ...503..518F}.
Intergalactic baryons have historically been traced in absorption,
such as the Ly$\alpha$ forest arising from diffuse photoionized
gas that may account for up to 30\% of baryons.  The remaining
baryons are predicted to exist in a warm-hot intergalactic medium
(WHIM) \citep[e.g.][]{1999ApJ...511..521D,2001ApJ...552..473D,
1999ApJ...514....1C}, which is shock-heated during the collapse of
density perturbations that give rise to the cosmic web.  Still, absorption
probes yield only one-dimensional redshift-space information, or in rare
cases several probes through common structures.  Mapping intergalactic
baryons in emission can in principle provide morphological and kinematic
information on accreting (and perhaps outflowing) gas within the
cosmic web.

Unfortunately, emission from intergalactic baryons is difficult to
observe, because current telescope sensitivities result in a detection
limit of column densities $N_{\rm{HI}} \ga 10^{19}$ cm$^{-2}$, which
are the realm of Damped Ly$\alpha$ (DLA) systems and sub-DLAs.  Below
column densities of $\sim N_{\rm{HI}} \sim 10^{19.5}$ cm$^{-2}$, the
neutral fraction of hydrogen decreases rapidly due to the transition
from optically-thick to optically-thin gas ionized by the metagalactic
ultraviolet flux. At lower densities the gas is no longer affected by
self shielding and the atoms are mostly ionized.  This sharp decline
in neutral fraction from almost unity to less than a percent happens
within a few kpc \citep{1994ApJ...423..196D}.  Below $N_{\rm{HI}} \sim
10^{17.5}$ cm$^{-2}$ the gas is optically thin and the decline in
neutral fraction with total column is much more gradual. A consequence
of this rapid decline in neutral fraction is a plateau in the
\hbox{\rm H\,{\sc i}} column density distribution function between
$N_{\rm{HI}} \sim 10^{17.5}$ and $N_{\rm{HI}} \sim 10^{19.5}$
cm$^{-2}$, where the relative surface area at these columns shows only
modest growth. This behaviour is confirmed in QSO absorption studies
tabulated by \cite{2002ApJ...567..712C} and in \hbox{\rm H\,{\sc i}}
emission by \cite{2004A&A...417..421B}. Below $N_{\rm{HI}} \sim
10^{17.5}$ cm$^{-2}$ the relative surface area increases rapidly,
reaching about a factor of 30 larger at $N_{\rm{HI}} \sim 10^{17}$
compared with $N_{\rm{HI}} \sim 10^{19}$ cm$^{-2}$.

This plateau in the distribution function is a critical issue for
  observers of neutral hydrogen in emission. Although telescope
  sensitivities have increased substantially over the past decades,
  the detected surface area of galaxies observed in the 21cm line has
  only increased modestly \citep[eg.][]{1994ApJ...423..196D}. Clearly
  there is a flattening in the distribution function near $N_{\rm{HI}}
  \sim 10^{19.5}$ cm$^{-2}$ which has limited the ability of even deep
  observations to detect hydrogen emission from a larger area. By
  establishing that a steeper distribution function is again expected below
  about $N_{\rm{HI}} \sim 10^{17.5}$, it provides a clear technical
  target for what the next generation of radio telescopes needs to
  achieve to effectively probe diffuse gas.

Exploration of the $N_{\rm{HI}} < 10^{17.5}$ cm$^{-2}$ regime is
essential for gaining a deeper understanding of the repository of
baryons that drive galaxy formation and evolution.  This gas, residing
in filamentary structures, is the reservoir that fuels future star
formation, and could provide a direct signature of smooth cold-mode
accretion predicted to dominate gas acquisition in star-forming
galaxies today \citep[]{2005MNRAS.363....2K, 2008arXiv0808.0553D,
2008arXiv0809.1430K}.  Furthermore, the trace neutral fraction in this
phase may provide a long-lived fossil record of tidal interactions and
feedback processes such as galactic winds and AGN-driven cavities.

Several new large facilities to study 21cm emission are under
development today.  In view of the observational difficulties in probing
the low \hbox{\rm H\,{\sc i}} column regime, it is particularly
important to have reliable numerical simulations to aid in planning
new observational campaigns, and eventually to help interpret such
observations within a structure formation context.  While simulations
of galaxy formation are challenging, historically they have had much
success predicting the more diffuse baryons residing in the cosmic
web~\citep[e.g.][]{1999ApJ...511..521D}.  If such simulations display
statistical agreement with key existing \hbox{\rm H\,{\sc i}} emission
data, then they can be used to make plausible predictions for the types of
structures that may be detected, along with suggesting optimum observing
strategies.

In this paper, we employ a state-of-the art cosmological hydrodynamic
simulation to study \hbox{\rm H\,{\sc i}} emission from filamentary
large-scale structure and the galaxies within them.  The simulation
used here include a well-constrained prescription for galactic
outflows that has been shown to reproduce the observed metal and
\hbox{\rm H\,{\sc i}} absorption line properties from $z\sim
6\rightarrow 0$ \citep{2006MNRAS.373.1265O,
2008MNRAS.387..577O,2009MNRAS.395.1875O}.  We develop a method to
produce \hbox{\rm H\,{\sc i}} maps from these simulations, and compare
statistical properties of the reconstructed \hbox{\rm H\,{\sc i}} data
with the statistics of real \hbox{\rm H\,{\sc i}} observations, to
assess the reliability of the simulation. For this purpose we will
primarily use the \hbox{\rm H\,{\sc i}} Parkes All Sky Survey (HIPASS)
\cite{2001MNRAS.322..486B} since this is the largest available
\hbox{\rm H\,{\sc i}} survey.  This work is intended to provide an
initial step towards a more thorough exploration of model constraints
that will be enabled by comparisons with present and future \hbox{\rm
H\,{\sc i}} data.

Note that current spatial resolution of simulations having
cosmologically-representative volumes cannot reproduce a galaxy as
would be seen with \hbox{\rm H\,{\sc i}} observations having sub-kpc
resolution. Therefore we do not consider the internal kinematics or
detailed shapes of the objects associated with simulated galaxies.  We can
only assess the statistical properties of the diffuse \hbox{\rm H\,{\sc
i}} phase and predict how the gas is distributed on multi-kpc scales.
We particularly focus on lower column density material that may be probed
with future \hbox{\rm H\,{\sc i}} surveys, which primarily reside in
cosmic filaments within which galaxies are embedded.

Our paper is organized as follows. In section two we briefly describe
the particular simulation that has been analyzed.  In section three we
describe our method to extract the neutral hydrogen from the
simulations. The neutral fraction is determined from both a general
ionization balance as well as a local self-shielding correction. We
also model the transition from atomic to molecular hydrogen We
present our results, showing the statistical properties of the
recovered \hbox{\rm H\,{\sc i}} in section four, where they are
compared with similar statistics obtained from observations.  The
distribution of neutral hydrogen is compared with the distribution of
dark matter and stars in this section as well. In the fifth section we
discuss the results and outline the implications.  Finally, section
six reiterates our main conclusions.

\section{Simulation Code}

A modified version of the N-body+hydrodynamic code Gadget-2 is
employed, which uses a tree-particle-mesh algorithm to compute
gavitational forces on a set of particles, and an entropy-conserving
formulation of Smoothed Particle Hydrodynamics (SPH:
\cite{2002MNRAS.333..649S}) to simulate pressure forces and shocks
in the baryonic gaseous particles. This Lagrangian code is fully
adaptive in space and time, allowing simulations with a large dynamic
range necessary to study both high-density regions harboring galaxies
and the lower-density IGM. It includes a prescription for star
formation following \cite{2003MNRAS.339..312S} and galactic
outflows as described below.  The code has been described in detail
in \cite{2006MNRAS.373.1265O} and \cite{2008MNRAS.387..577O}; we
will only summarise the properties here.

The novel feature of our simulation is that it includes a
well-constrained model for galactic outflows.  The implementation
follows \cite{2003MNRAS.339..289S}, but employs scalings of outflow
speed and mass loading factor with galaxy mass as expected for
momentum-driven winds \citep{2005ApJ...618..569M}.  Our simulations
using these scalings have been shown to successfully reproduce a wide
range of IGM and galaxy data, including IGM enrichment as traced by
$z\sim2-6$ \ion{C}{iv} absorbers~\citep{2006MNRAS.373.1265O}, the
galaxy mass-metallicity relation~\citep{2008MNRAS.385.2181F}, the
early galaxy luminosity function and its
evolution~\citep{2006MNRAS.370..273D}, \ion{O}{vi} absorption at
low-$z$~\citep{2009MNRAS.395.1875O}, and enrichment and entropy
levels in galaxy groups~\cite{2008MNRAS.391..110D}.  Such outflows
are expected to impact the distribution of gas in the large-scale
structure around galaxies out to typically $\sim 100$kpc
\citep{2008MNRAS.387..577O, 2009MNRAS.395.1875O}, so are
important for studying the regions expected to yield detectable
\hbox{\rm H\,{\sc i}} emission.

The simulation used here is run with cosmological parameters
consistent with the 3-year WMAP results \citep{2007ApJS..170..377S}.
The parameters are $\Omega_0 = 0.25$, $\Omega_{\Lambda} = 0.75$, $\Omega_b
= 0.044$, $H_0 = 71$ km s$^{-1}$ Mpc$^{-1}$, $\sigma_8 = 0.83$, and $n
= 0.95$.  The periodic cubic volume has a box length of 32 $h^{-1}$ Mpc
(comoving), and the gravitational softening length is set to 2.5 $h^{-1}$
kpc (comoving).  Dark matter and gas are represented using $256^3$
particles each,  yielding a mass per dark matter and gas particle of
$1.57\times 10^8 M_\odot$ and $3.35\times 10^7 M_\odot$, respectively.
The simulation was started in the linear regime at $z=129$, with initial
conditions established using a random realization of the power spectrum
computed following \cite{1999ApJ...511....5E}, and evolved to $z=0$.

\section{Method for Making \hbox{\rm H\,{\sc i}} Maps}

We now describe the algorithm used to extract the neutral hydrogen
component from this simulation. The method developed is general, and
can be applied to any simulation that has a similar set of output
parameters. In our analysis, we set the total hydrogen number density
to $n_{\rm{H}} = 0.74\rho_g/m_{\rm{H}}$ where $m_{\rm{H}}$ is the mass
of the hydrogen atom. The factor 0.74 assumes a helium abundance of $Y
= 0.26$ by mass, and that all the helium is in the form of \hbox{\rm
He\,{\sc i}} and \hbox{\rm He\,{\sc ii}} with a similar neutral
fraction as hydrogen. Apart from this factor the presence of helium is
not taken into account in our calculations.  We will describe how we
determine the neutral fraction of the gas particles, including
applying a correction for self shielding in high density regions and
taking into account molecular hydrogen formation where relevant. This
allows reconstruction of the neutral hydrogen distribution, by mapping
the particles onto a three dimensional grid.

\subsection{Neutral Fraction of Hydrogen}

We begin by calculating the neutral fraction from the density and
temperature of gas in the simulations, together with the \hbox{\rm
H\,{\sc i}} photo-ionization rate provided by the cosmic UV background.
\cite{1994ApJ...423..196D} found that the radial structure of the column
density of \hbox{\rm H\,{\sc i}} is more sensitive to the extragalactic
radiation field than to the distribution of mass in the host galaxy.
When calculating the neutral fraction, we assume that all photoionization
is due to radiation external to the disk and that internal stellar sources
are not significant. In this case the nebular model as described in
\cite{1989agna.book.....O} is a very good approximation, since the typical
number density in the outer parts of galaxies, approximately $10^{-2}$
cm$^{-3}$, is so low that collisional ionization is negligible. When
going further out, the densities become even lower. Inside galaxies
the volume densities are so high, that the neutral fraction is of order
unity owing to self-shielding.

In the IGM, hydrogen becomes ionized when the extreme ultraviolet (UV)
radiation ionizes and heats the surrounding gas. On the other hand,
the recombination of electrons leads to neutralization. The degree of
ionization is determined by the balance between photo-ionization and
radiative recombination. Only photons more energetic than $h\nu > 13.6$
eV can ionize hydrogen. The ionization equilibrium equation is given by e.g. \cite{1989agna.book.....O} as:
\begin{equation}
n_{\rm{H}} \int^{\infty}_{\nu_0} \frac{4\pi J_{\nu}}{h\nu}\alpha(\nu)
d\nu = n_e n_p \beta(T),
\end{equation}
where $J_{\nu}$ is the mean intensity of ionizing photons,
$\alpha(\nu)$ is the ionization cross section and $\beta(T)$ is the
recombination rate coefficient to all levels of atomic hydrogen at
temperature $T$.  $\nu_0$ is the ionization threshold frequency. Only
radiation with frequency $\nu > \nu_0$ is effective in photoionization
of hydrogen from the ground level. Summarising, the integral
represents the number of photoionizations  per unit time and
the right hand side of the equation gives the number of
recombinations per unit time. The neutral hydrogen,
$n_{\rm{H}}$, electron, $n_e$, and proton, $n_p$, densities are related
through the charge conservation and hydrogen abundance equations,
\begin{equation}
n_e = n_p = (1-\xi)n
\end{equation}
where $\xi$ is the neutral fraction, so
\begin{equation}
n_{\rm{H}} = \xi n
\end{equation}
with $n$ the total density.

We can write the ionization balance for neutral Hydrogen as 
\begin{equation}
\xi n \Gamma_{\rm{HI}} = (1-\xi)^2n^2 \beta(T)
\end{equation}
where $\Gamma_{\rm{HI}}$ is the ionization rate for neutral hydrogen.\\
With this equation it is easy to determine the neutral fraction, which
is given by
\begin{equation}
\xi = \frac{2C + 1 - \sqrt{(2C+1)^2-4C^2}}{2C}
\end{equation}
using
\begin{equation}
C = \frac{n \beta(T)}{\Gamma_{\rm{HI}}}
\end{equation}

Obviously the neutral fractions that we calculate are closely related
to the values we use for the photoionization and recombination rate.
The photoionization rate at low redshift is not well constrained
observationally; by combining Ly$\alpha$ forest data and simulations,
\cite{2001ApJ...553..528D} obtain a photoionization rate of
$\Gamma_{\rm{HI}} \sim 10^{-13.3 \pm 0.7}$ s$^{-1}$ for redshift
$z \sim 0.17$.  We will use the photoionization rate given by the
CUBA model of \cite{2001cghr.confE..64H}, which is $\Gamma_{\rm{HI}}
\sim 10^{-13}$ s$^{-1}$ for $z \sim 0$.

The recombination rate coefficients are dependent on temperature.
We make use of an analytic function described by
\cite{1996ApJS..103..467V}, that fits the coefficients in the
temperature range form 3 K to $10^{10}$ K:
\begin{equation}
\beta(T) = a \Big[
\sqrt{T/T_0}\big(1+\sqrt{T/T_0}\big)^{1-b}\big(1+\sqrt{T/T_1}\big)^{1+b}
\Big] ^{-1}
\end{equation}
where $a$, $b$, $T_0$ and $T_1$ are the fitting parameters. For the
\hbox{\rm H\,{\sc i}} ion the fitting parameters are: $a=7.982 \times
10^{-11}$ cm$^3$ s$^{-1}$, $b=0.7480$, $T_0=3.148$ K and $T_1 = 7.036
\times 10^{5}$ K.\\

\begin{figure}[t!]
  \includegraphics[width=0.5\textwidth]{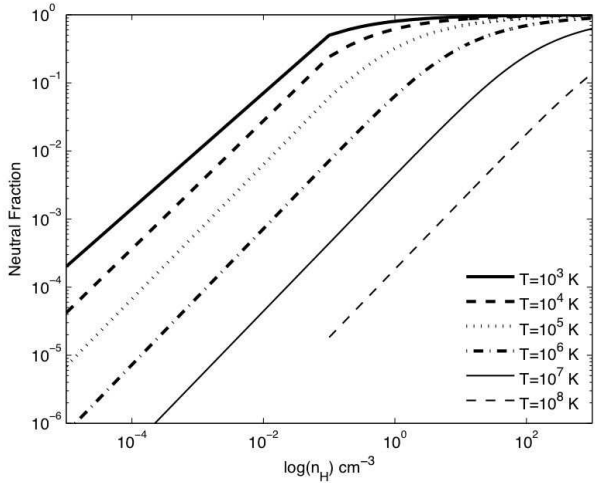}
  \caption{Neutral fraction as a function of density for different
    temperatures between $10^4$ and $10^9$ K.}
  \label{neutralbal}
\end{figure}

The neutral fraction is plotted for different temperatures as function
of density of \hbox{\rm H} atoms in Fig.~\ref{neutralbal}.  For temperatures
below $10^4$ K, the neutral fraction is still significant (a few
percent) at reasonably low densities of 0.01 cm$^{-3}$ but at higher
temperatures most of the gas is ionized, and the neutral fraction
drops very quickly.

\subsection{Molecular Hydrogen}
When gas has cooled sufficiently, it coexists in the
  molecular (H$_2$) and atomic (\hbox{\rm H\,{\sc i}}) phases. The
  H$_2$ regions are found in dense molecular clouds where star
  formation occurs. Unfortunately there is large uncertainty in the
  average amount of H$_2$ in galaxies, as estimates have to rely on indirect
  tracers and conversion factors, for which the dependancies are not
  well-understood. As a result, there is substantial variance in
  estimates of the average density ratio at z~=~0, $\eta_{universe}
  = \Omega_{H_2} / \Omega_{HI}$ (e.g. 0.42 and 0.26 stated
  respectively by \cite{2003ApJ...582..659K} and
  \cite{2009MNRAS.tmp..289O}). It is beyond the scope of this paper to
  revisit these determinations, therefore we will adopt a value of
  $\Omega_{H_2} / \Omega_{HI} = 0.3$ that falls within the error bars
  of current estimates. Given the observed local value of the atomic
  mass density, of $\rho_{HI}=6.1\times 10^7$ $h$ M$_{\odot}$
  Mpc$^{-3}$ \citep{2003AJ....125.2842Z}, this implies a molecular
  mass density of 
  $\rho_{H_2}=1.8\times 10^7$ $h$ M$_{\odot}$ Mpc$^{-3}$.

To define the regions of molecular hydrogen, we use a threshold
based on the thermal pressure $(P/k = nT)$. \cite{2002ApJ...569..157W},
\cite{2004ApJ...612L..29B} and more recently
\cite{2006ApJ...650..933B} have made the case that the amount of molecular
hydrogen that is formed in galaxies is determined by 
only one parameter, the interstellar gas pressure. In hydrostatic
pressure equilibrium, the hydrostatic pressure is balanced by the sum
of all contributions to the gas pressure: magnetic pressure, cosmic
ray pressure and kinetic pressure terms (of which the thermal pressure
is relatively small) (e.g. \cite{1996ASPC..106....1W} and references
therein). However, thermal pressure is directly coupled to energy
dissipation via radiation, and therefore thermal pressure can track
the total pressure due to various equipartition mechanisms. An
evaluation of the various contributions to the total hydrostatic
pressure is given by \cite{1990ApJ...365..544B}.

\begin{figure}[t!]
  \includegraphics[width=0.5\textwidth]{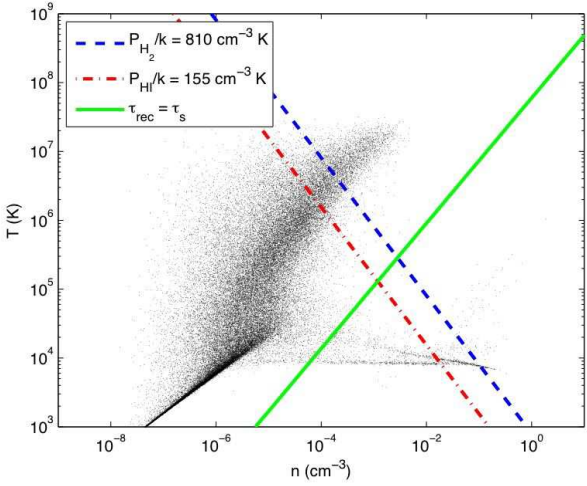}
  \caption{Temperatures are plotted against densities for every
      200$^{\textrm{th.}}$ particle in the simulation. The dashed
      (blue) and dash-dotted (red) lines correspond to constant
      thermal pressures of $P/k = 155$ and 810 cm$^{-3}$ K, that were
      found empirically to reproduce the observed mass densities of
      atomic and molecular gas at z~=~0. The solid (green) line shows
      where the recombination time is equal to the sound-crossing time
      at a physical scale of one kpc. Particles above/left of the
      green line are unlikely to be neutral or molecular.}
  \label{temp_dens}
\end{figure}

Two lines of constant thermal pressure are shown in
Fig.~\ref{temp_dens} where temperatures are plotted against density
for individual particles in the simulation. When following these
lines, they cross two regions, one with high densities and low
temperatures and one with moderate densities, but very high
temperatures. These two regions are distinguished by the solid green
line in Fig.~\ref{temp_dens}, where the radiative recombination time
is equivalent to the sound-crossing time on a kiloparsec scale. The
radiative recombination time is given in \cite{2005pcim.book.....T}
by:

\begin{equation}
\tau_{\textrm{rec}} \simeq 2 \times 10^2 \frac{T^{0.6}}{n_e} \textrm{ years}
\end{equation}
where $n_e$ is the electron density which is comparable to the total
density for low neutral fractions. The sound crossing time is given by
$\tau_{\textrm{s}} = R/C_{\textrm{s}}$, where $R$ is the relevant
scale (assumed here to be one kpc) and $C_{\textrm{s}} \simeq 1.4
\times 10^4 T^{1/2}$ cm s$^{-1}$ is the sound velocity. All particles
right of the green line have a recombination time that is shorter than
the sound crossing time. Particles with a recombination time that
is larger than the sound crossing time are unlikely to be neutral or
molecular.

\begin{figure}[t!]
  \includegraphics[width=0.5\textwidth]{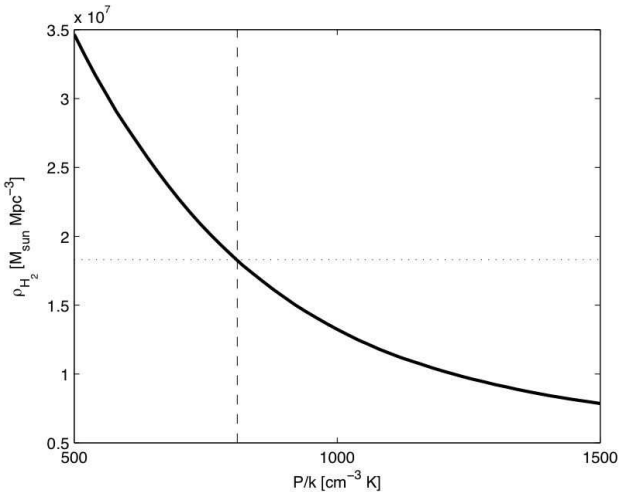}
  \caption{The average molecular density at z~=~0 is plotted
      against the threshold thermal pressure, where molecular hydrogen
      is assumed to form from atomic, while also satisfying the
      condition that $\tau_{\textrm{rec}}<\tau_{\textrm{s}}$. The
      dashed line indicates a pressure of $P/k = 810$ cm$^{-3}$ K,
      where $\rho_{H_2}=1.8\times 10^7$ $h$ M$_{\odot}$ Mpc$^{-3}$
      which is shown by the dotted line.}
  \label{H2pres}
\end{figure}

For each particle the thermal pressure can be calculated and particles
with a pressure exceeding the threshold value and satisfying
$\tau_{\textrm{rec}}<\tau_{\textrm{s}}$ are considered molecular. By
exploring different pressure values as shown in Fig.~\ref{H2pres}, the
threshold can be tuned to yield the required molecular mass density,
$\rho_{H_2}=1.8\times 10^7$ $h$ M$_{\odot}$ Mpc$^{-3}$. The
threshold thermal pressure value we empirically determine is $P/k =
810$ cm$^{-3}$ K. We must stress, that this value is very likely not
a real physical value, as the resolution in our simulation is not
sufficient to resolve the scales of molecular clouds. Molecular clouds
have smaller scales with higher densities, which will likely have
significantly enhanced pressures.

\begin{figure}[t!]
  \includegraphics[width=0.5\textwidth]{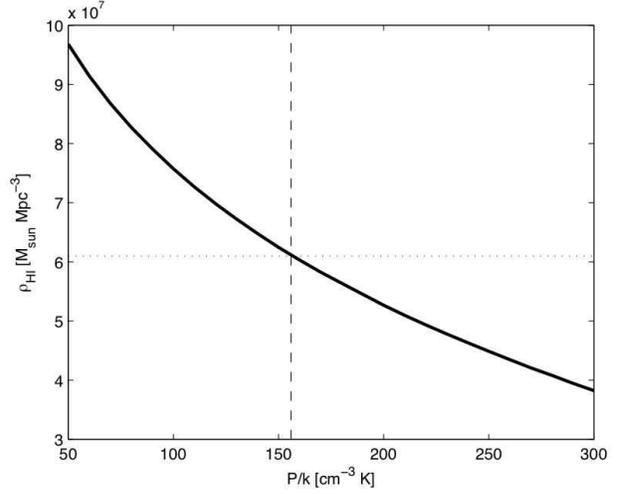}
  \caption{The average \hbox{\rm H\,{\sc i}} mass density at
      z~=~0 is plotted against the threshold thermal pressure, where
      atomic hydrogen is assumed to recombine from ionized, while also
      satisfying the condition that
      $\tau_{\textrm{rec}}<\tau_{\textrm{s}}$ (and accounting for a
      molecular density of $\rho_{H_2}=1.8\times 10^7$ $h$ M$_{\odot}$
      Mpc$^{-3}$). At a pressure of $P/k = 155$ cm$^{-3}$ K (dashed
      line), $\rho_{HI}=6.1\times 10^7$ $h$ M$_{\odot}$ Mpc$^{-3}$
      (dotted line), which is consistent with the \hbox{\rm H\,{\sc
      i}} density from \cite{2003AJ....125.2842Z}.}
  \label{HIpres}
\end{figure}

\subsection{Correction for Self-Shielding}

Although the ionization state and kinetic temperature are
  determined self-consistently within the simulation, it has been
  necessary to assume that each gas particle is subjected to the same
  all-pervasive radiation field. At both extremely low and high
  particle densities this approximation is sufficient, since local
  conditions will dominate. However, at intermediate densities, the
  ``self-shielding'' of particles by their neighbours may play a
  critical role in permitting local recombination, when the same
  particle would be substantially ionized in isolation.  Present
cosmological simulations are not capable of solving the full radiative
transfer equations, although it is now becoming possible to
post-process radiative transfer on individual galaxies
\citep[e.g.][]{2008MNRAS.390.1349P}.  Because we want to study
emission from the IGM as well as galaxies, we must instead adopt a
simple correction based on density and temperature to
approximate self-shielding. We adopt a similar approach to the
  one that was used to model the atomic to molecular transition above,
  using the thermal pressure as a proxy for the hydrostatic
  pressure. Only gas at a sufficiently high thermal pressure and for
  which the recombination time is shorter than the sound crossing time
  on kpc scales is assumed to recombine. Particles that satisfy the
  pressure and timescale condition are considered to be fully
  self-shielded, and their neutral fraction is set to unity.

We will assume that the highest pressure regions which satisfy
$\tau_{\textrm{rec}}<\tau_{\textrm{s}}$ have already provided
$\rho_{H_2}=1.8\times 10^7$ $h$ M$_{\odot}$ Mpc$^{-3}$ as discussed
above. We subsequently calculate the atomic density as
function of the thermal pressure
threshold, as shown in Fig.~\ref{HIpres}. It is empirically found,
that a threshold value of $P/k = 155$ cm$^{-3}$ K results in an
\hbox{\rm H\,{\sc i}} density of $\rho=6.1\times 10^7$ $h$
M$_{\odot}$ Mpc$^{-3}$, that is similar to the derived value in
\cite{2003AJ....125.2842Z}. We will adopt this threshold value for our
further analysis.

The typical densities and temperatures where self-shielding becomes
  important are not accurately defined. When looking at Fig.~\ref{temp_dens},
  the typical temperatures and densities which satisfy our empirical thermal
  pressure criterion for local recombination are temperatures of $\sim10^4$
  K and densities of $\sim 0.01$ cm$^{-3}$. These values agree well with
  various estimates from literature
  (e.g. \cite{1997ApJ...490..564W},\cite{2003ApJ...587..278W}, and
  \cite{2005PhDT........17P}.)

\subsection{Gridding Method}
To reconstruct the density fields, we have employed a grid-based method,
in which the value of the density field is calculated at a set of
locations defined on a regular grid. The mass of each particle is
spread over this grid in accordance with a particular weighting
function $W$, to yield
\begin{equation}
\widehat{\rho}\Big(\frac{\mathbf{n}}{M}\Big)
= \frac{M^3}{N} \sum_{i=1}^N m_i W 
\Big(\mathbf{x}_i - \frac{\mathbf{n}}{M} \Big)
\end{equation}
where $\mathbf{n} = (n_x,n_y,n_z)$ denotes the grid-cell, $M$ is the
number of cells of the grid in each dimension, $N$ is the number of
particles and $m_i$ is the mass of particle $i$.\\

We adopt the weighting function directly from what is used for SPH in
Gadget-2, namely a spline kernel defined by \cite{1985A&A...149..135M}:
\begin{equation}
W(r,h) = \frac{8}{\pi h^3} \left\{ \begin{array}{lr}
1-6\Big( \frac{r}{h}\Big)^2+6\Big(\frac{r}{h}\Big)^3\textrm{,}
& 0\leq\frac{r}{h}\leq\frac{1}{2}\\
 & \\
2\Big(1-\frac{r}{h}\Big)^3\textrm{,}
& \frac{1}{2} < \frac{r}{h} \leq 1\\
 & \\
0\textrm{,}
& \frac{r}{h} >1\\
\end{array} \right.
\end{equation}
where $r$ is the distance from the position of a particle and $h$ is
the smoothing length for each particle (which in Gadget-2 is the
radius that encloses 32 gas particle masses).  Furthermore we set a
limit to the size of the smoothing length: the smoothing length of a
particle has to be at least 1.5 times the resolution of the
grid-cells, which means that a particle is distributed over at least
three grid-cells in each dimension.  This adversely affects the
highest density regions in the reconstructed field (when insufficient
gridding resolution is employed), but gives a more realistic
representation of resolved objects and transitions without shot noise
or step functions.  We note that our procedure explicitly conserves
total mass.

\section{Results}

Reconstructed density fields are gridded in three dimensions, for
the total hydrogen component (ionized plus neutral) the neutral
component and the molecular component.  This makes it possible to
compare the distribution of the total and neutral hydrogen budget and
permits determination of neutral fractions for the volume and column
densities. Initially the full 32 $h^{-1}$ Mpc cubes is gridded with a
cell size of 80 kpc. This allows visualization of the distribution on
large scales and determination of the average density of neutral
hydrogen in the simulation volume $\bar{\rho}_{\rm{HI}}$. The degree
of clustering can be determined by looking at the two-point
correlation function. However, this low resolution grid is not
suitable to resolve the high density regions and small structures, as
we will describe later. High density regions of the simulation volume
were selected and gridded with a cell-size of 2 kpc. The \hbox{\rm
H\,{\sc i}} column density distribution function and the \hbox{\rm
H\,{\sc i}} mass function can be determined from these regions. The
properties of the simulated \hbox{\rm H\,{\sc i}} gas will be
described and the statistics will be compared with the statistical
properties of observational data, mostly from the \hbox{\rm H\,{\sc
i}} Parkes All Sky Survey (HIPASS) (\cite{2001MNRAS.322..486B}).

Apart from gas or SPH particles, the simulations contain star and dark
matter particles as well. We will adopt a relatively simple gridding
scheme to reconstruct the distribution of stars and dark matter. This
can be very useful to verify whether the stars, but especially the gas
(or reconstructed \hbox{\rm H\,{\sc i}}) trace the distribution of
Dark Matter.

\subsection{Mean  \hbox{\rm H\,{\sc i}} Density}

The average \hbox{\rm H\,{\sc i}} density is an important
property, as this single number gives the amount of neutral hydrogen
that is reconstructed without any further analysis. The \hbox{\rm
H\,{\sc i}} density is very well determined from the 1000 brightest
HIPASS galaxies in \cite{2003AJ....125.2842Z} They deduce an \hbox{\rm
H\,{\sc i}} density due to galaxies in the local universe of
$\bar{\rho}_{\rm{HI}} = (6.9 \pm 1.1)\cdot 10^7$ $h$ M$_\odot$ Mpc$^{-3}$
or $\bar{\rho}_{\rm{HI}} = (6.1 \pm 1.0)\cdot 10^7$ $h$ M$_\odot$
Mpc$^{-3}$ when taking into account biases like selection bias,
Eddington effect, \hbox{\rm H\,{\sc i}} self absorption and cosmic
variance. From \cite{2002ApJ...567..247R} a value of
$\bar{\rho}_{\rm{HI}} = 7.1\cdot 10^7$ $h$ M$_\odot$ Mpc$^{-3}$ can be
derived for the average \hbox{\rm H\,{\sc i}} density in the
universe. We will adopt the value of $\bar{\rho}_{\rm{HI}} = (6.1 \pm
1.0)\cdot 10^7$ $h$ M$_\odot$ Mpc$^{-3}$. 
The pressure thresholds for molecular and atomic hydrogen are tuned to
reproduce this density as is described earlier in this paper.

\subsection{ \hbox{\rm H\,{\sc i}} Distribution Function}

As mentioned above, the low and intermediate column densities
$(N_{\rm{HI}} < 10^{19}$ cm$^{-2})$ do not have a very significant
contribution to the total mass budget of \hbox{\rm H\,{\sc i}}. 

For comparison with our simulation, the \hbox{\rm H\,{\sc i}}
distribution function derived from QSO absorption line data will be
used as tabulated in \cite{2002ApJ...567..712C}. For the QSO data the
column density distribution function $f(N_{\rm{HI}})$ is defined such
that $f(N_{\rm{HI}})dN_{\rm{HI}}dX$ is the number of absorbers with
column density between $N_{\rm{HI}}$ and $N_{\rm{HI}} + dN_{\rm{HI}}$
over an absorption distance interval $dX$. We derive $f(N_{\rm{HI}})$
from the statistics of our reconstructed \hbox{\rm H\,{\sc i}}
emission. The column density distribution function in a reconstructed
cube can be calculated from,

\begin{equation}
 f(N_{\rm{HI}}) = \frac{c}{H_0 dz} \frac{A(N_{\rm{HI}})}{dN_{\rm{HI}}} \textrm{ cm}^2,
\end{equation}
where $dX = dz H_0/c$ and $A(N_{\rm{HI}})$ is the surface area subtended
by \hbox{\rm H\,{\sc i}} in the column density interval $dN_{\rm{HI}}$
centered on $N_{\rm{HI}}$. 

\begin{figure*}[!t]
  \includegraphics[width=0.5\textwidth]{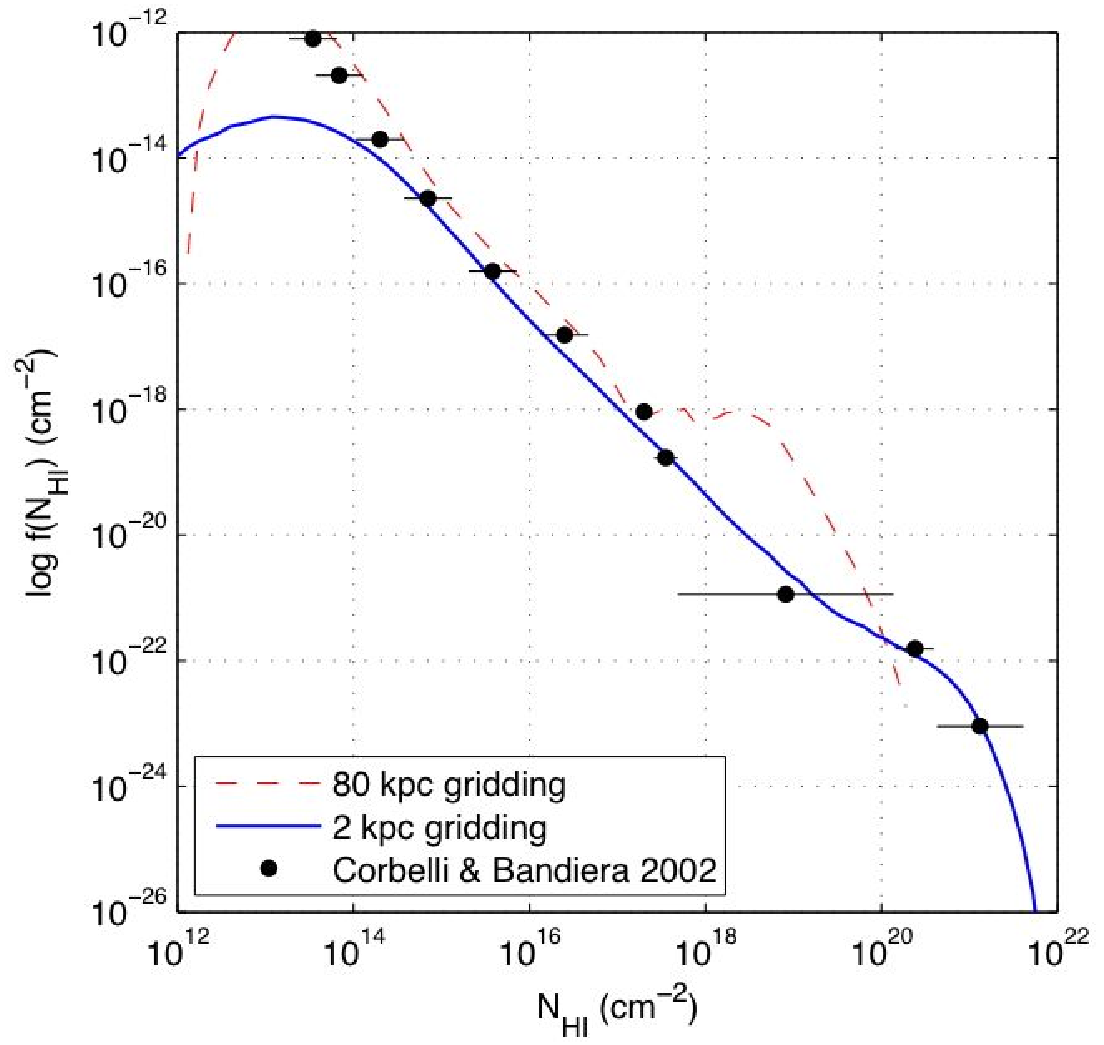}
  \includegraphics[width=0.5\textwidth]{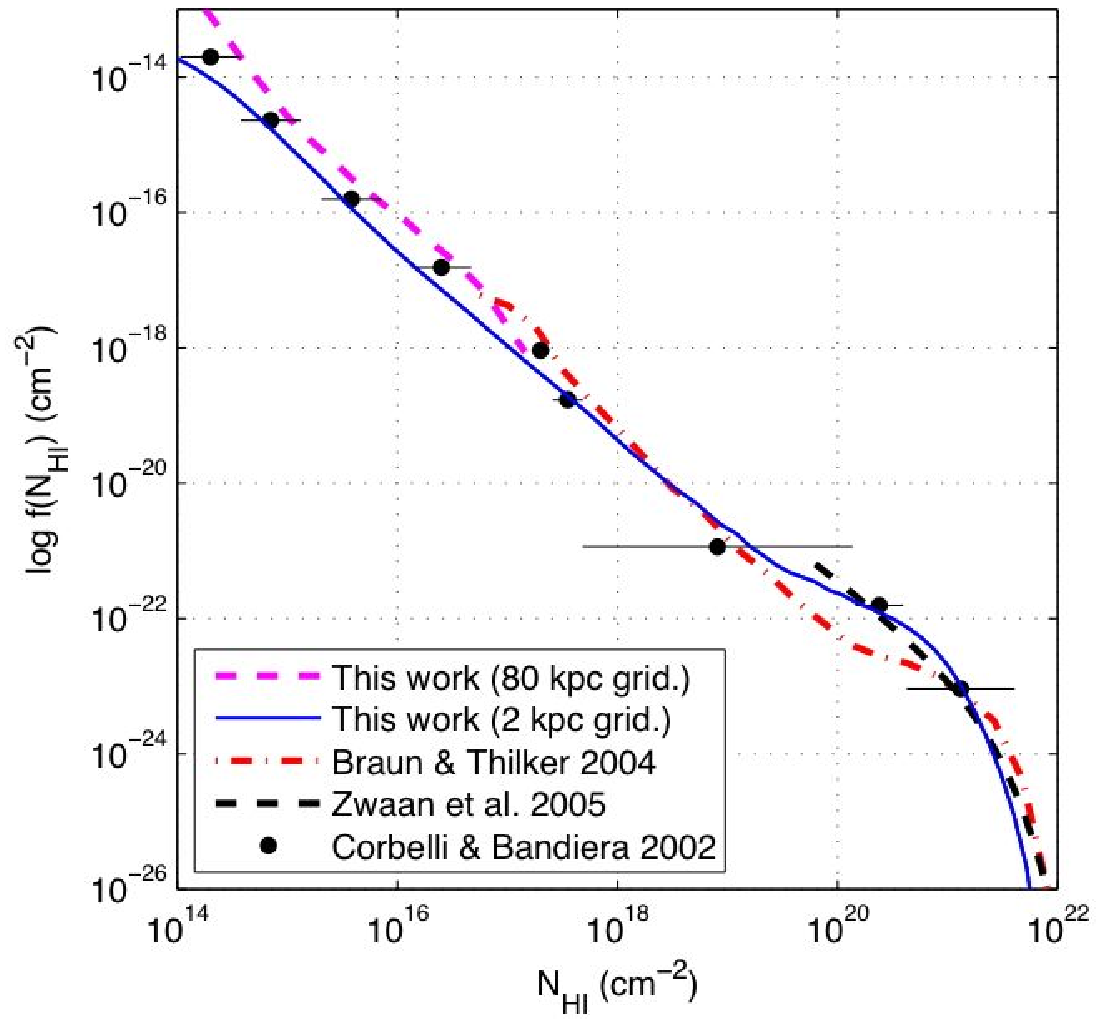}
    \caption{Left panel: \hbox{\rm H\,{\sc i}} distribution function
      after gridding to 80 kpc (dashed (red) line). The solid (blue)
      line corresponds to data gridded to a 2 kpc cell size. Filled
      dots correspond to the QSO absorption line data
      \citep{2002ApJ...567..712C}. Right panel: Combined \hbox{\rm
        H\,{\sc i}} distribution functions of the simulation, gridded
      to a resolution of 2 kpc (solid (blue) line) and 80 kpc (dashed
      (purple) line). Overlaid are distribution function from
      observational data of M31 \citep{2004A&A...417..421B}, WHISP
      (\cite{2002A&A...390..829S}, \cite{2005MNRAS.364.1467Z}) and QSO
      absorption lines \citep{2002ApJ...567..712C} respectively. The
      reconstructed \hbox{\rm H\,{\sc i}} distribution function
      corresponds very well to all observed distribution functions }
  \label{distr}
\end{figure*}

As the simulations contain \hbox{\rm H\,{\sc i}} column densities over
the full range between $N_{\rm{HI}} = 10^{14}$ and $10^{21}$
cm$^{-2}$, we can plot the \hbox{\rm H\,{\sc i}} column density
distribution function $f(N_{\rm{HI}})$ over this entire range with
excellent statistics, in contrast to what has been achieved
observationally. In the left panel of Fig.~\ref{distr} we overlay the
\hbox{\rm H\,{\sc i}} distribution functions we derive from the
simulations with the data values obtained from QSO absorption lines as
tabulated by \cite{2002ApJ...567..712C} (black dots). The horizontal
lines on the QSO data points correspond to the bin-size over which
each data point has been derived. Vertical error bars are not shown,
as these have the same size as the dot.  Around $N_{\rm{HI}} =
10^{19}$ cm$^{-2}$ there is only one data bin covering two orders of
magnitude in column density, illustrating the difficulty of sampling
this region with observations. This corresponds to the transition
between optically thick and thin gas, where only a small increase in
surface covering is associated with a large decrease in the column
density.

The dashed (red) line corresponds to data gridded to a 80 kpc cell size.
At low column densities the simulated distribution function agrees
very well with the QSO absorption line data. The transition from
optically thick to optically thin gas happens within just a few kpc of
radius in a galaxy disk \citep{1994ApJ...423..196D}. Clearly a
reconstructed cube with a 80~kpc cell size does not have enough
resolution to resolve such transitions. Some form of plateau can be
recognised in the coarsely gridded data above $N_{\rm{HI}} = 10^{16}$
cm$^{-2}$, however it is not a smooth transition. Furthermore because
of the large cell size, no high column density regions can be
reconstructed at all. The cores of galaxies have high column
densities, but these are severely diluted within the 80 kpc voxels.

To circumvent these limitations, structures with an \hbox{\rm H\,{\sc
i}} mass exceeding $5\times10^8$ $M_\odot$ in an 80 kpc voxel have
been identified for individual high resolution gridding. This mass
limit is chosen to match the mass-resolution of the simulation. The
mass of a typical gas particle is $\sim 2.5 \cdot 10^7$ $M_\odot$,
when taking into account the abundance of hydrogen with respect to
helium, we need at least 20 gas particles to form a $5\times10^8$
$M_\odot$ structure. As the neutral fraction is much less than one for
most of the particles, the number of particles in one object is much
larger. We find 719 structures above the mass limit and grid a
300~kpc box around each object with a cell size of 2~kpc.

We emphasize that gridding to a higher resolution does not mean
  that the physics is computed at a higher resolution. We are still
  limited by the simplified physics and finite mass resolution of the
  particles. A method of accounting for structure or clumping below
  the resolution of the simulation is described in
  e.g. \cite{2006MNRAS.372..679M}. To derive the clumping factor, they
  have used another simulation, with the same number of particles, but
  a much smaller computational volume, and thus higher resolution. In
  our analysis, we accept that we cannot resolve the smallest
  structures, since we are primarily interested in the diffuse outer
  portions of galactic disks. We have chosen a 2~kpc voxel size, as
  this number represents the nominal spatial resolution of the
  simulation. The simulation has a gravitational softening length of
  2.5 kpc $h^{-1}$, but note that the smoothing lengths can go as low
  as 10\% of the gravitational softening length.

Distribution functions are plotted for simulated \hbox{\rm H\,{\sc i}}
using the two different voxel sizes of 80 and 2 kpc in the left panel
of Fig.~\ref{distr} . When using a 80 kpc voxel size, the
reconstructed maps are unable to resolve structures with high
densities, causing erratic behaviour at column densities above
$N_{\rm{HI}} \sim 10^{17}$ cm$^{-2}$. When using the smaller
voxel size of 2 kpc, there is an excellent fit to the observed data
between about $N_{\rm{HI}} = 10^{15} \textrm{ and } 10^{20.5}$ cm$^{-2}$. The
lower column densities are not reproduced within the sub-cubes (although
they are in the coarsely-gridded full simulation cube), while the
finite mass and spatial resolution of the simulation do not allow a
meaningful distribution function to be determined above about
$N_{\rm{HI}} = 10^{21}$ cm$^{-2}$.

Below $N_{\rm{HI}} = 10^{20}$ cm$^{-2}$ a transition can be seen with
the distribution function becoming flatter. The effect of
self-shielding is decreasing, which limits the amount of neutral
hydrogen at these column densities. Around $N_{\rm{HI}} = 10^{17}$
cm$^{-2}$ the optical depth to photons at the hydrogen ionisation edge
is equal to 1 \citep{2002ApJ...568L..71Z}. Self-shielding no longer
has any effect below this column density and a second transition can
be seen. Now the neutral fraction is only determined by the balance
between photo-ionisation and radiative recombination. The distribution
function is increasing again as a power law toward the very low column
densities of the Lyman-alpha forest. The slope in this regime agrees
very well with the QSO data. Note that the 2~kpc gridded data are
slightly offset to lower occurrences compared to the 80 kpc gridded
data. This is because we only considered the vicinity of the largest
mass concentrations in the simulation for high resolution
sampling. For the same reason the function is not representative below
$N_{\rm{HI}} \sim 3 \times 10^{14}$ cm$^{-2}$, while for the full,
80~kpc gridded cube it can be traced to $N_{\rm{HI}} \sim 5 \times
10^{13}$ cm$^{-2}$.  Of course, lower column density systems can be
produced in these simulations when artificial spectra are
constructed~\citep[e.g.][]{2001ApJ...553..528D, 2009MNRAS.395.1875O},
but our focus here is on the high column density systems that are
well-described by our gridding approach.

\begin{figure}[t!]
  \includegraphics[width=0.5\textwidth]{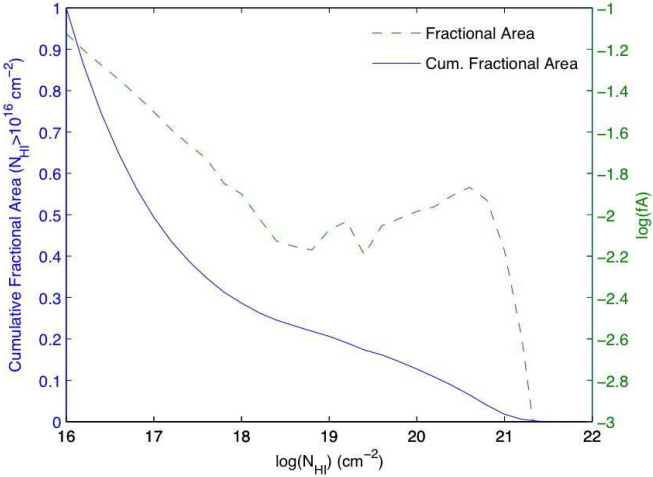}
  \caption{Fractional area of reconstructed \hbox{\rm H\,{\sc i}}
  (dashed line, right-hand axis) and cumulated surface area (solid
  line, left-hand axis) plotted against column density on a
  logarithmic scale with a bin size of $d\log(N_{HI})=0.2$. The
  probability of detecting emission with a column density near
  $N_{\rm{HI}} \sim 10^{17}$ cm$^{-2}$ is significantly larger than
  around $N_{\rm{HI}} \sim 10^{19.5}$ cm$^{-2}$. The cumulated surface
  area is normalized to that at a column density of $N_{HI}=10^{16}$
  cm$^{-2}$. At column densities of $N_{\rm{HI}} \sim 10^{17}$
  cm$^{-2}$, the area subtended by \hbox{\rm H\,{\sc i}} emission is
  much larger than at a limit of $N_{\rm{HI}} \sim 10^{19.5}$
  cm$^{-2}$, which is the sensitivity limit of most current
  observations of nearby galaxies.}
  \label{area}
\end{figure}

The distribution functions after gridding to 2~kpc (solid
line), and the low column density end of the 80~kpc gridding (dotted
line) are plotted again in the right panel of Fig.~\ref{distr}, but
now with several observed distributions overlaid. The high column
density regime is covered by the WHISP data
\citep{2002A&A...390..829S, 2005A&A...442..137N} in \hbox{\rm H\,{\sc
i}} emission; a Schechter function fit to this data by
\cite{2005MNRAS.364.1467Z} is shown by the dashed line. The
dash-dotted line shows \hbox{\rm H\,{\sc i}} emission data from the
extended M31 environment after combining data from a range of
different telescopes \citep{2004A&A...417..421B}. Since this curve is
based on only a single, highly inclined system, it may not be as
representative as the curves based on larger statistical samples. Our
simulated data agrees very well with the various observed data
sets. The distribution function indicates that there is less \hbox{\rm
H\,{\sc i}} surface area with a column density of $N_{\rm{HI}} \sim
10^{19}$ cm$^{-2}$ than at higher column densities of a few times
$10^{20}$ cm$^{-2}$. This is indeed the case, which can be seen if the
relative occurrence of different column densities is plotted. In
Fig.~\ref{area} the fractional area is plotted (dashed line) as
function of column density on logarithmic scale, which is given by:
\begin{equation}
fA = \frac{A(N_{HI})}{d\log(N_{HI})}.
\end{equation}
The surface area first increases from the highest column densities
(which are poorly resolved in any case above $10^{21}$ cm$^{-2}$) down
to a column density of a few times $10^{20}$ cm$^{-2}$, but then
remains relatively constant (per logarithmic bin). Only below column
densities of a few times $10^{18}$ cm$^{-2}$ does the surface area per
bin start to increase again, indicating that the probability of
detecting emission with a column density near $N_{\rm{HI}} \sim
10^{17}$ cm$^{-2}$ is significantly larger compared
to detecting emission with a column density of $N_{\rm{HI}} \sim
10^{19}$ cm$^{-2}$. Also of interest are plots of the cumulative
\hbox{\rm H\,{\sc i}} mass and surface area.

The solid line in Fig.~\ref{area} shows the total surface area
subtended by \hbox{\rm H\,{\sc i}} exceeding the indicated column
density. The plot is normalised to unity at a column density of
$N_{\rm{HI}} = 10^{16}$ cm$^{-2}$. At high column densities the
cumulative fractional area increases only moderately. Below a column
density of $N_{\rm{HI}} \sim 10^{18}$ cm$^{-2}$ there is a clear bend
and the function starts to increase more rapidly. At column densities of
$N_{\rm{HI}} \sim 10^{17}$ cm$^{-2}$, the area subtended by \hbox{\rm
H\,{\sc i}} emission is much larger than at a limit of $N_{\rm{HI}} \sim
10^{19.5}$ cm$^{-2}$, which corresponds to the sensitivity limit of most
current observations of nearby galaxies.

\subsection{\hbox{\rm H\,{\sc i}} column density}

\begin{figure*}[t!]
  \includegraphics[width=0.50\textwidth]{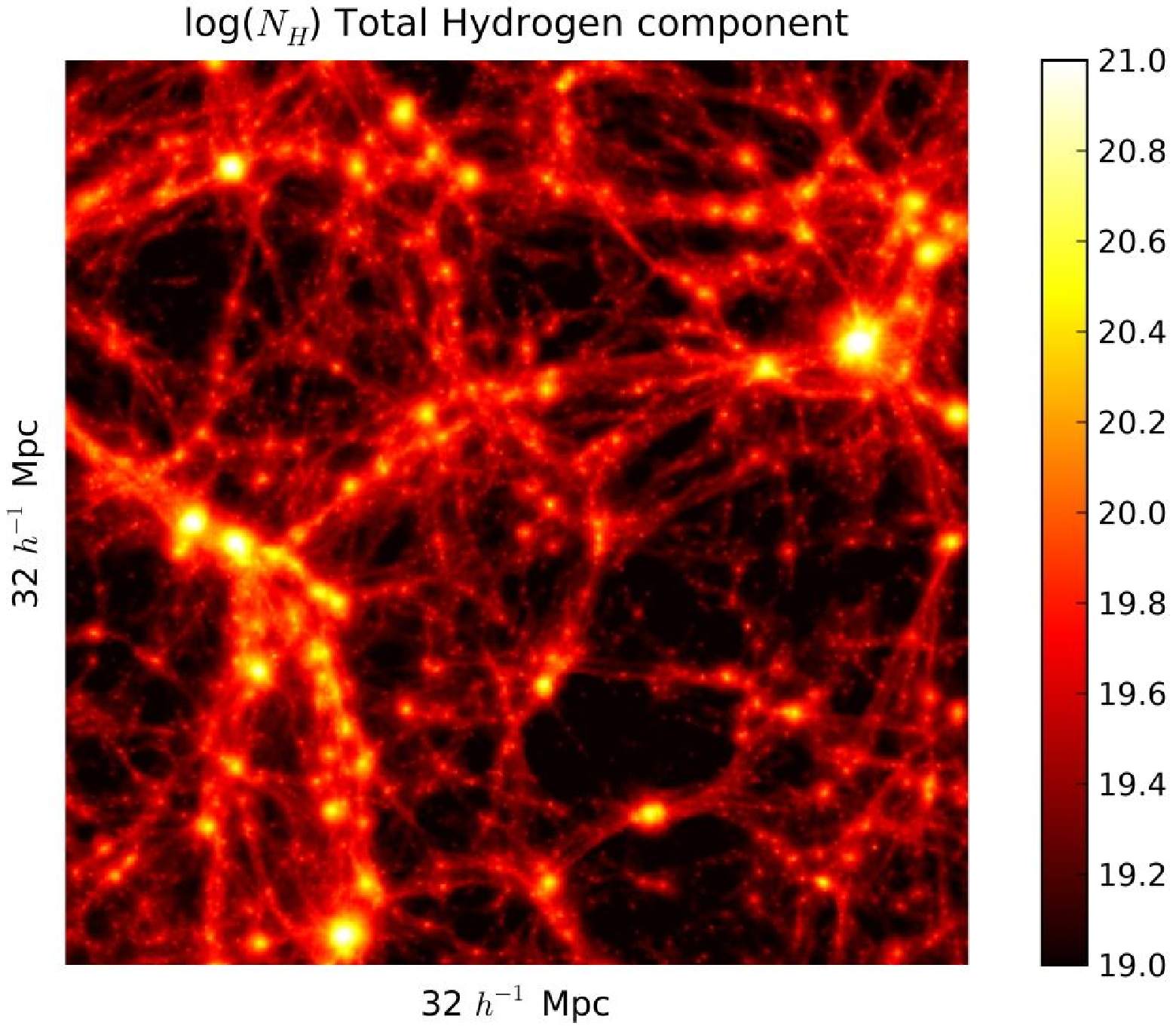}  
\includegraphics[width=0.5\textwidth]{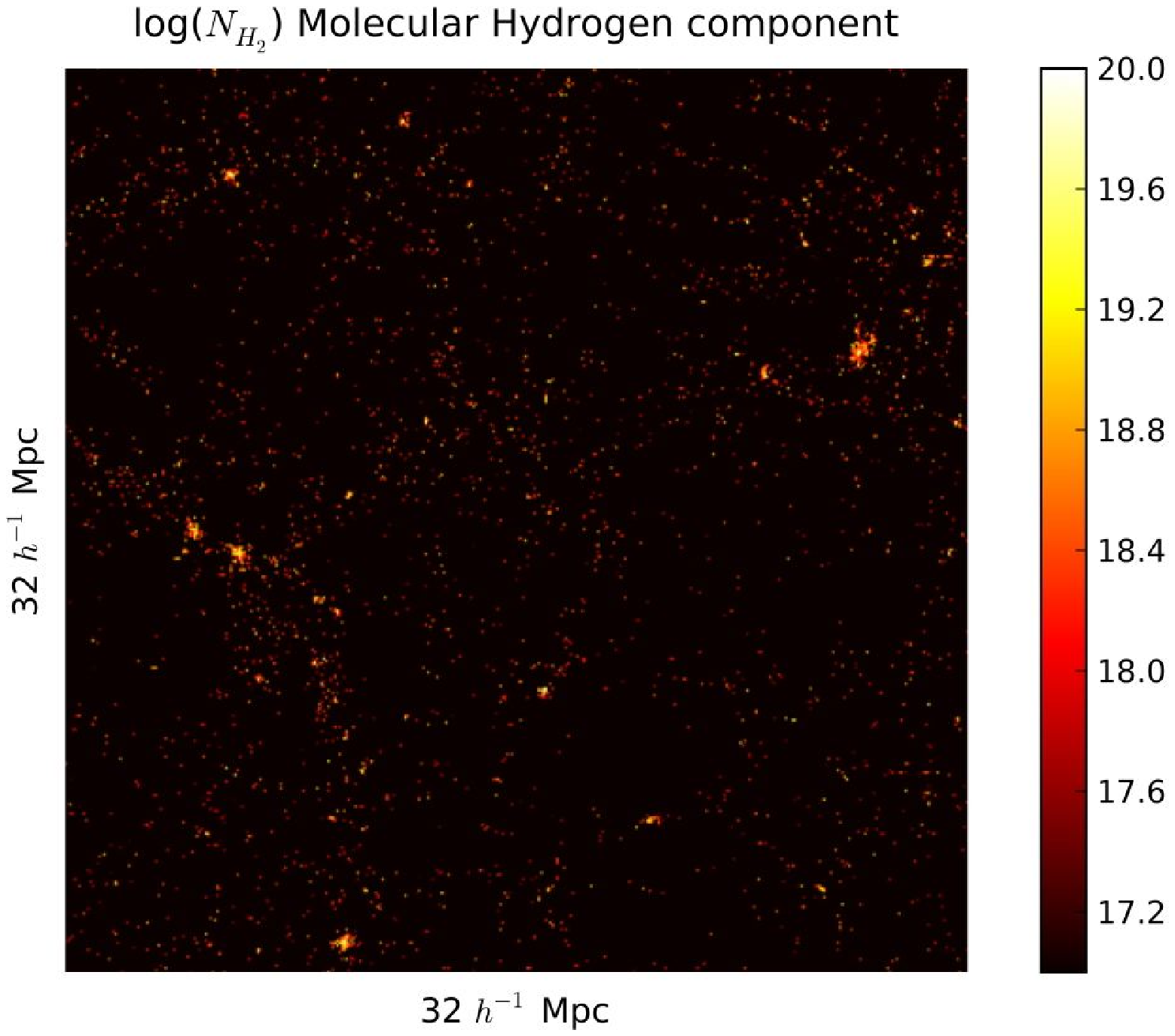}
 \begin{center}
  \includegraphics[width=0.5\textwidth]{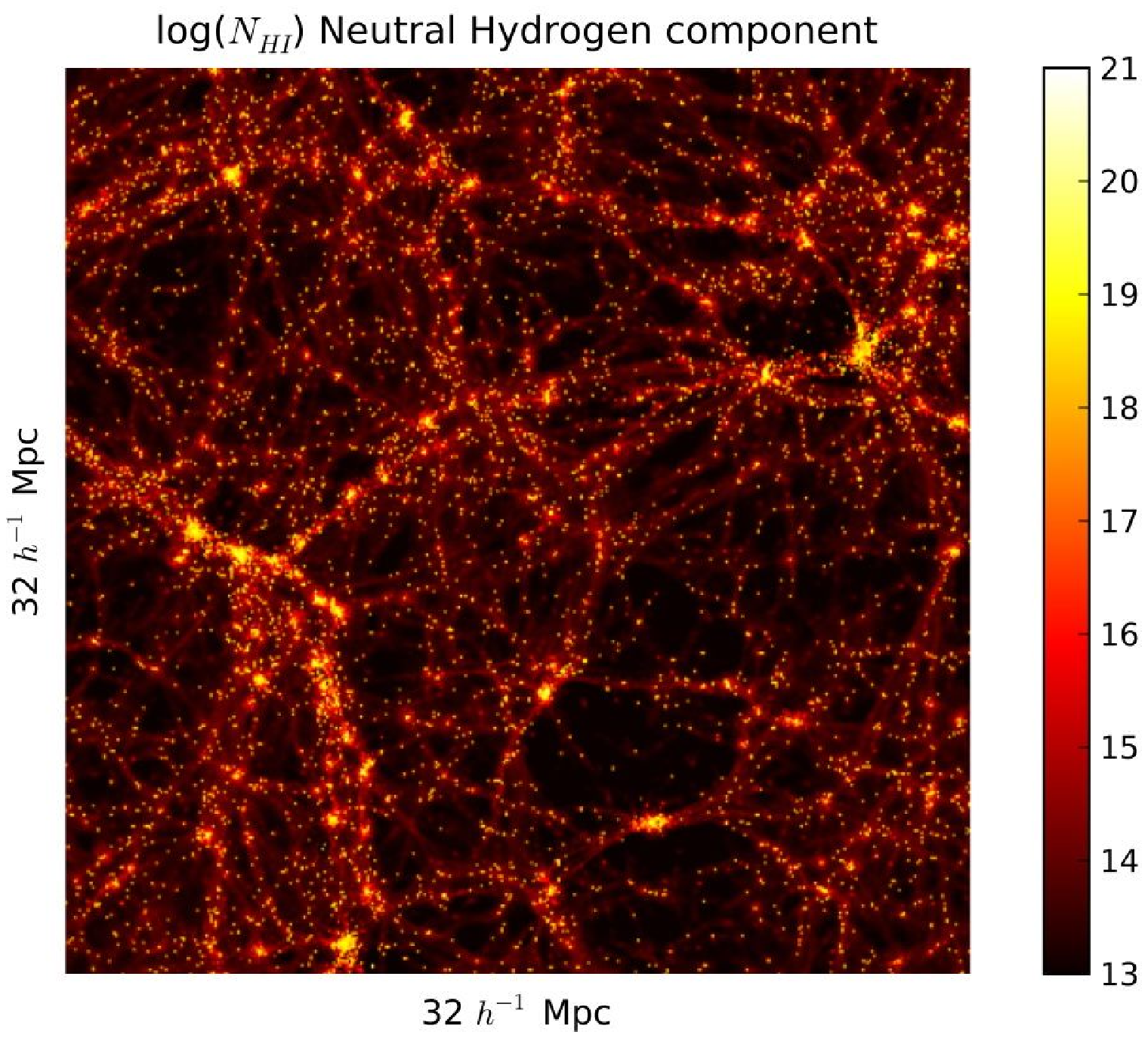}
 \end{center} 
\caption{Top left panel: Column density of total \hbox{\rm H} gas
    integrated over a depth of 32 $h^{-1}$ Mpc on a logarithmic scale,
    gridded to a resolution of 80 kpc. Top right panel: Molecular
    hydrogen component. Only very dense regions in the total hydrogen
    component contain molecular hydrogen. Bottom panel: Neutral atomic
    hydrogen component of the same region. In the neutral hydrogen
    distribution the highest densities are comparable to the densities
    in the total hydrogen distribution, but there is a very sharp
    transition to low neutral column densities as the gas becomes
    optically thin. Note the very different scales, the total
    hydrogen spanning only 2 orders of magnitude and the neutral
    hydrogen, eight.}
  \label{mom80kpc}
\end{figure*}

In Fig.~\ref{mom80kpc} column density maps are shown of the total and
the neutral hydrogen distribution. The maps are integrated over the
full 32 $h^{-1}$ Mpc depth of the cube, with the colorbar showing
logarithmic column density in units of cm$^{-2}$. The total hydrogen
map reaches maximum values of $N_{\rm{H}} \sim 10^{21}$ cm$^{-2}$,
while the connecting filaments have column densities of approximately
an order of magnitude less. In the intergalactic medium, the column
densities are still quite high, $N_{\rm{H}} \sim 10^{19}$ cm$^{-2}$,
yielding a very large mass fraction when the large surface area of the
intergalactic medium is taken into account.

In the column density map of neutral hydrogen it can be seen that it
is primarily the peaks which remain. At the locations
of the peaks of the total hydrogen map, we can see peaks in the
\hbox{\rm H\,{\sc i}} map with comparable column densities, that
correspond to
the massive galaxies and groups. The filaments connecting the galaxies
can still be recognized, but with neutral column densities of the
order of $N_{\rm{HI}} \sim 10^{16}$ cm$^{-2}$. Here the gas is still
relatively dense, but not dominated by self-shielding, resulting
in a lower neutral fraction. In the intergalactic regime, the neutral
fraction drops dramatically. The gas is highly ionized with neutral
columns of only $N_{\rm{HI}} \sim 10^{14}$ cm$^{-2}$, yielding only a
very small neutral mass contribution.\\

\begin{figure*}[t!]

  \includegraphics[width=0.35\textwidth]{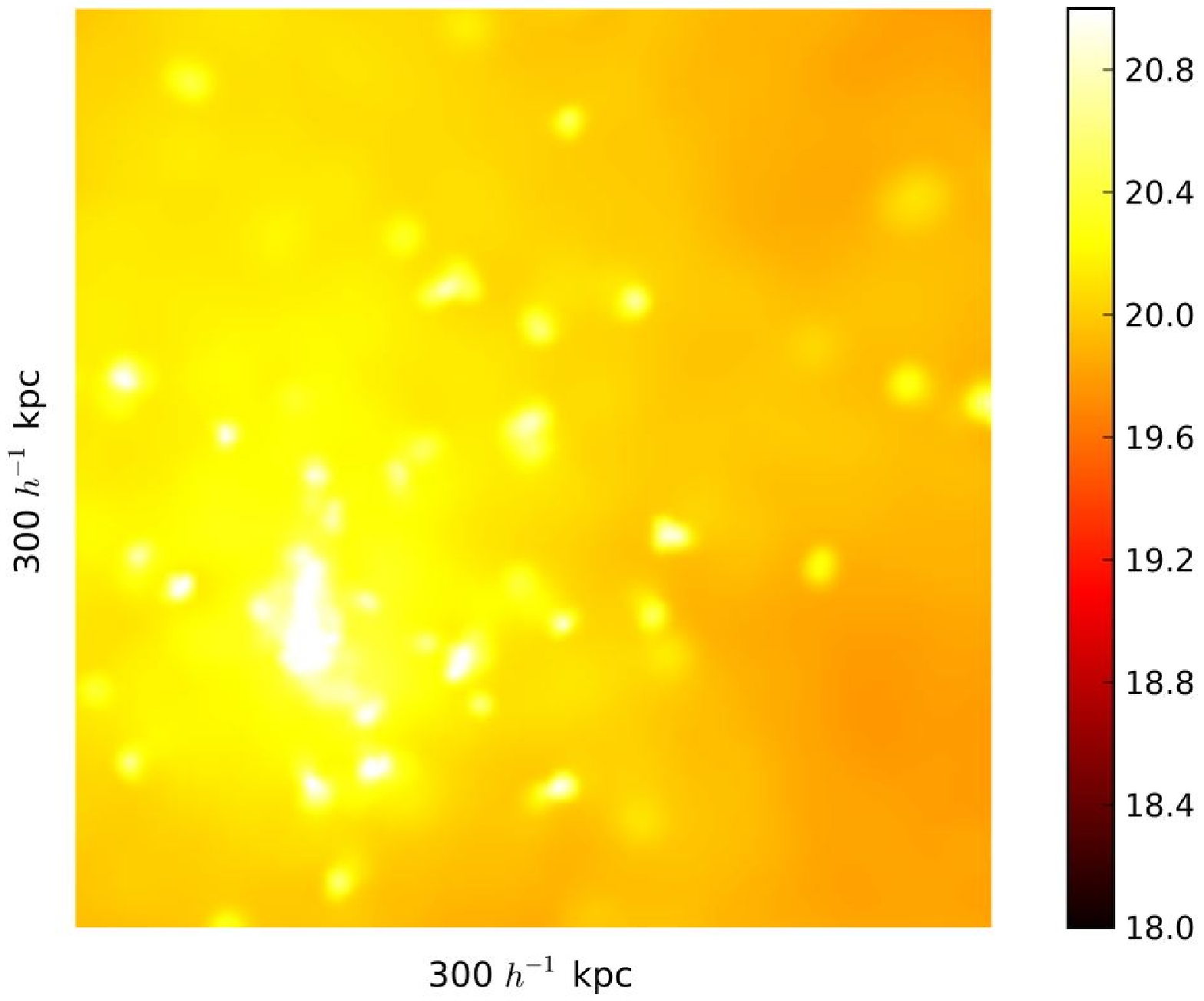}
  \includegraphics[width=0.35\textwidth]{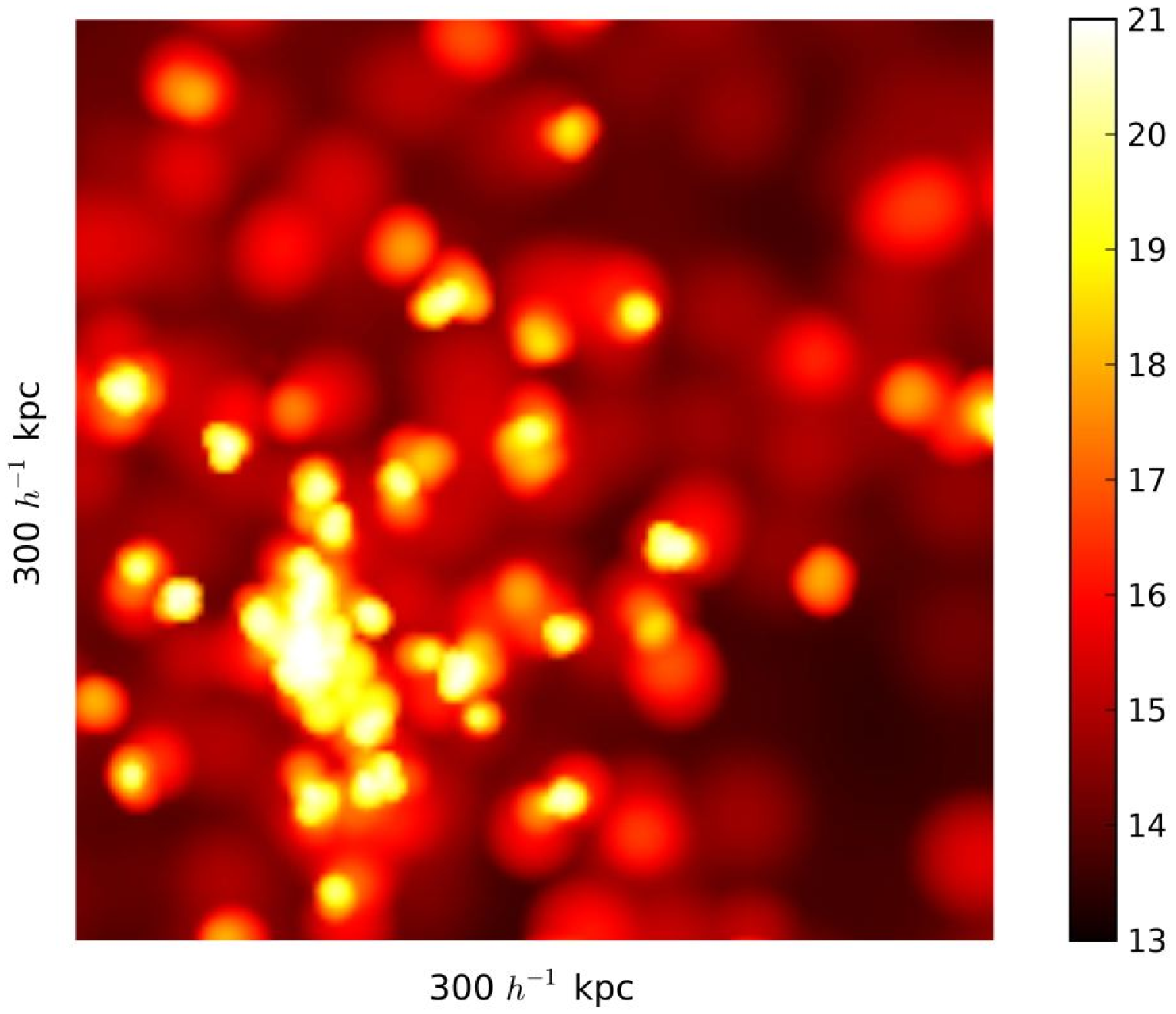}
  \includegraphics[width=0.29\textwidth]{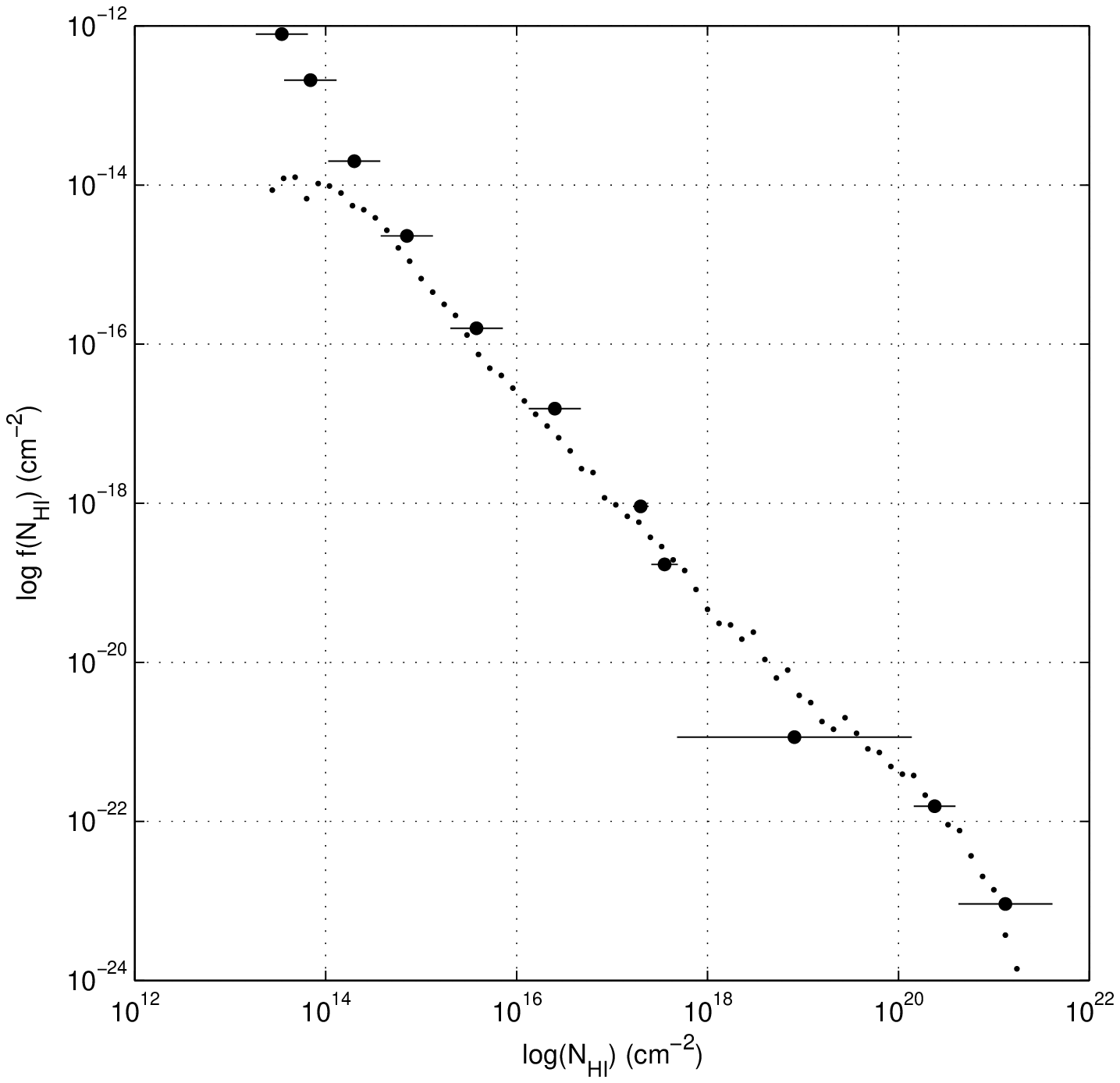}

  \includegraphics[width=0.35\textwidth]{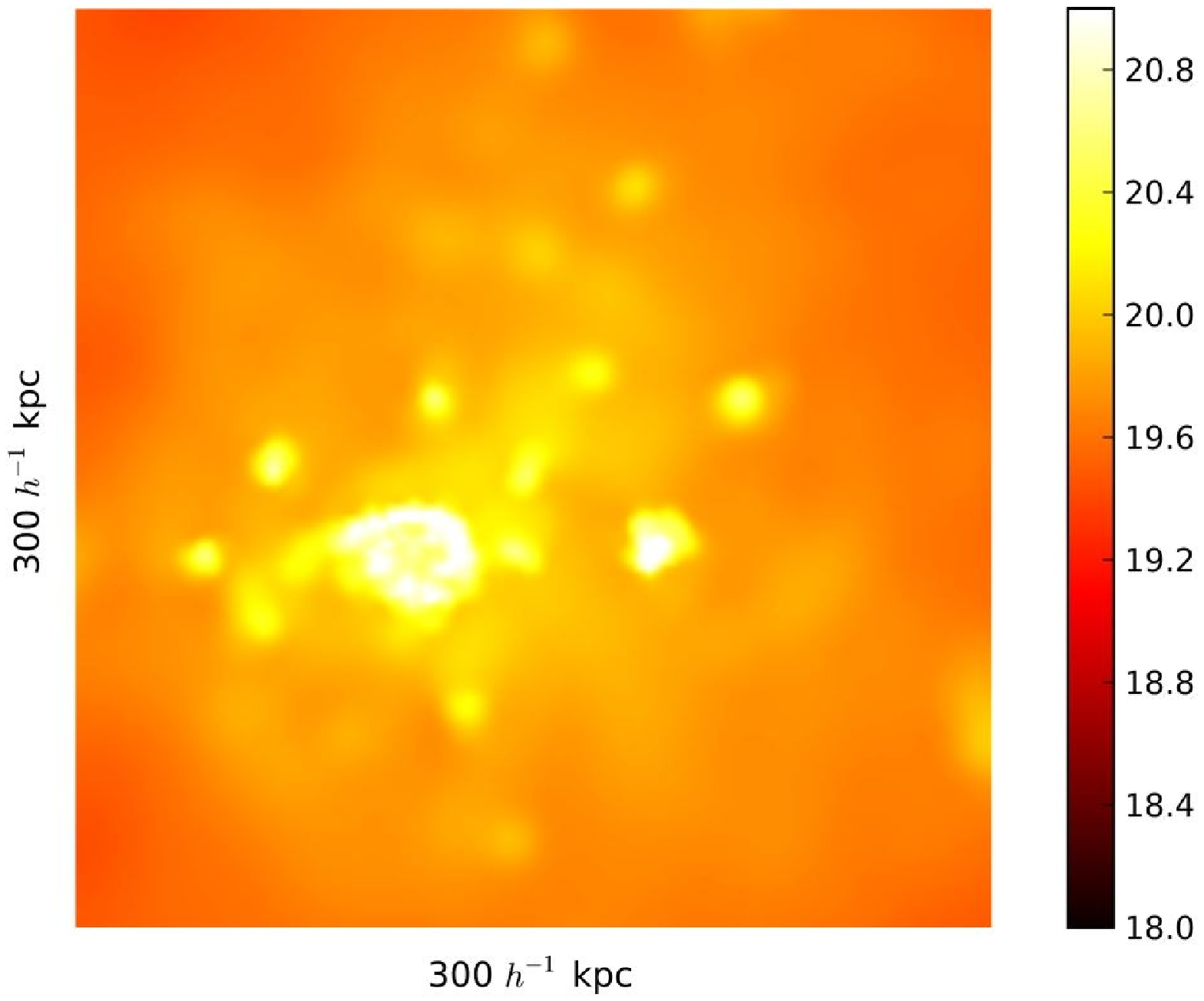}
  \includegraphics[width=0.35\textwidth]{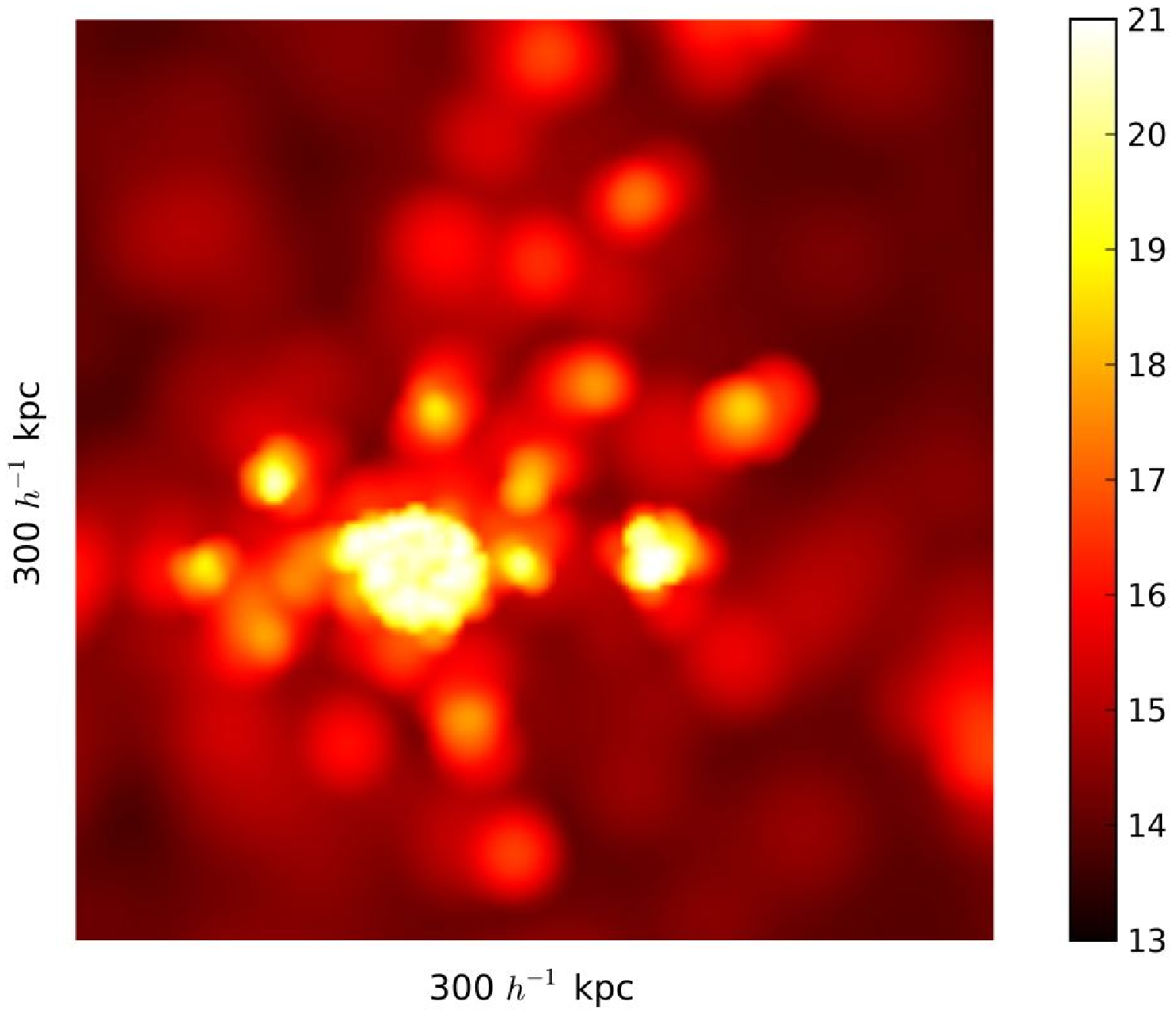}
  \includegraphics[width=0.29\textwidth]{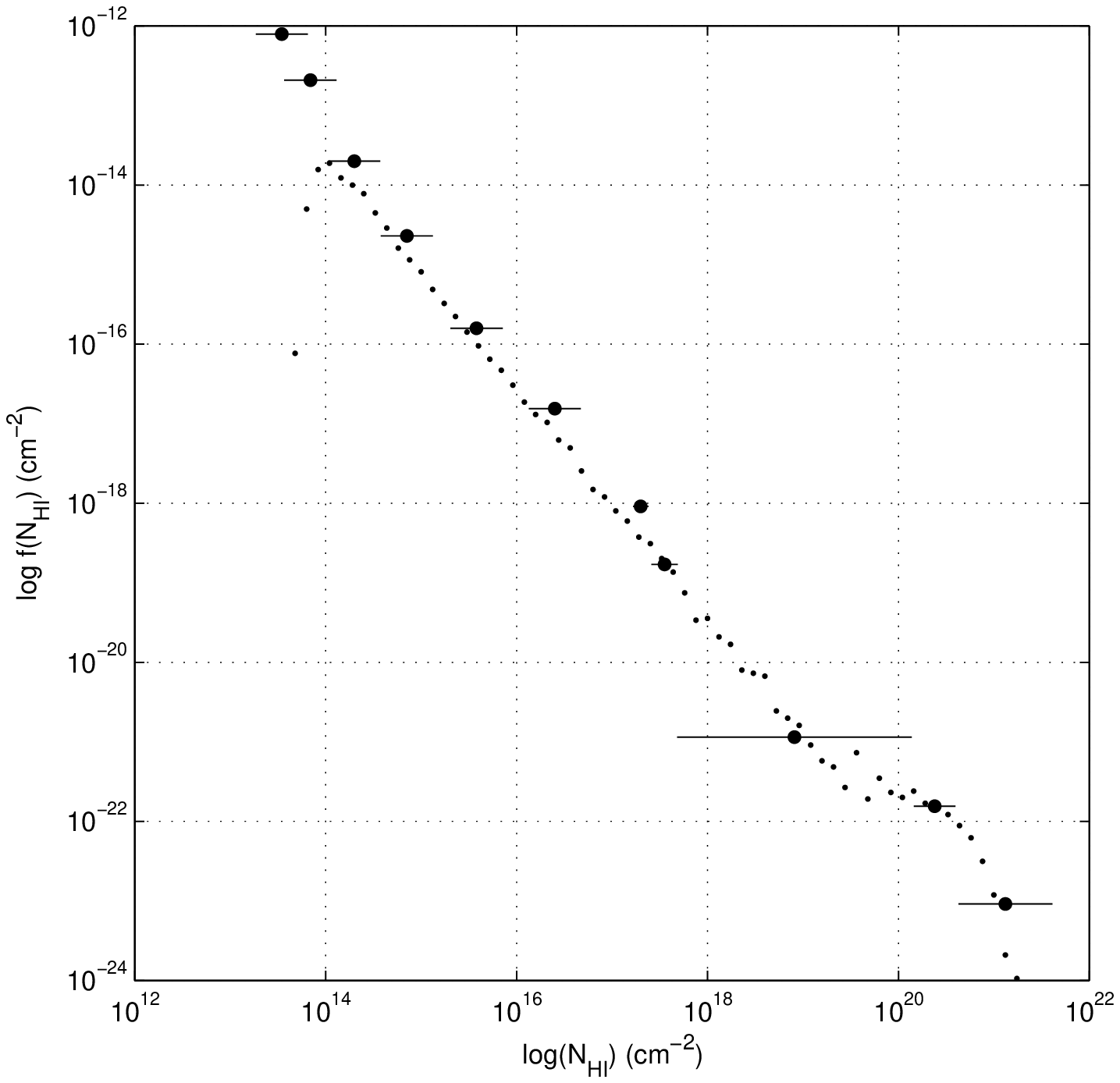}

  \includegraphics[width=0.35\textwidth]{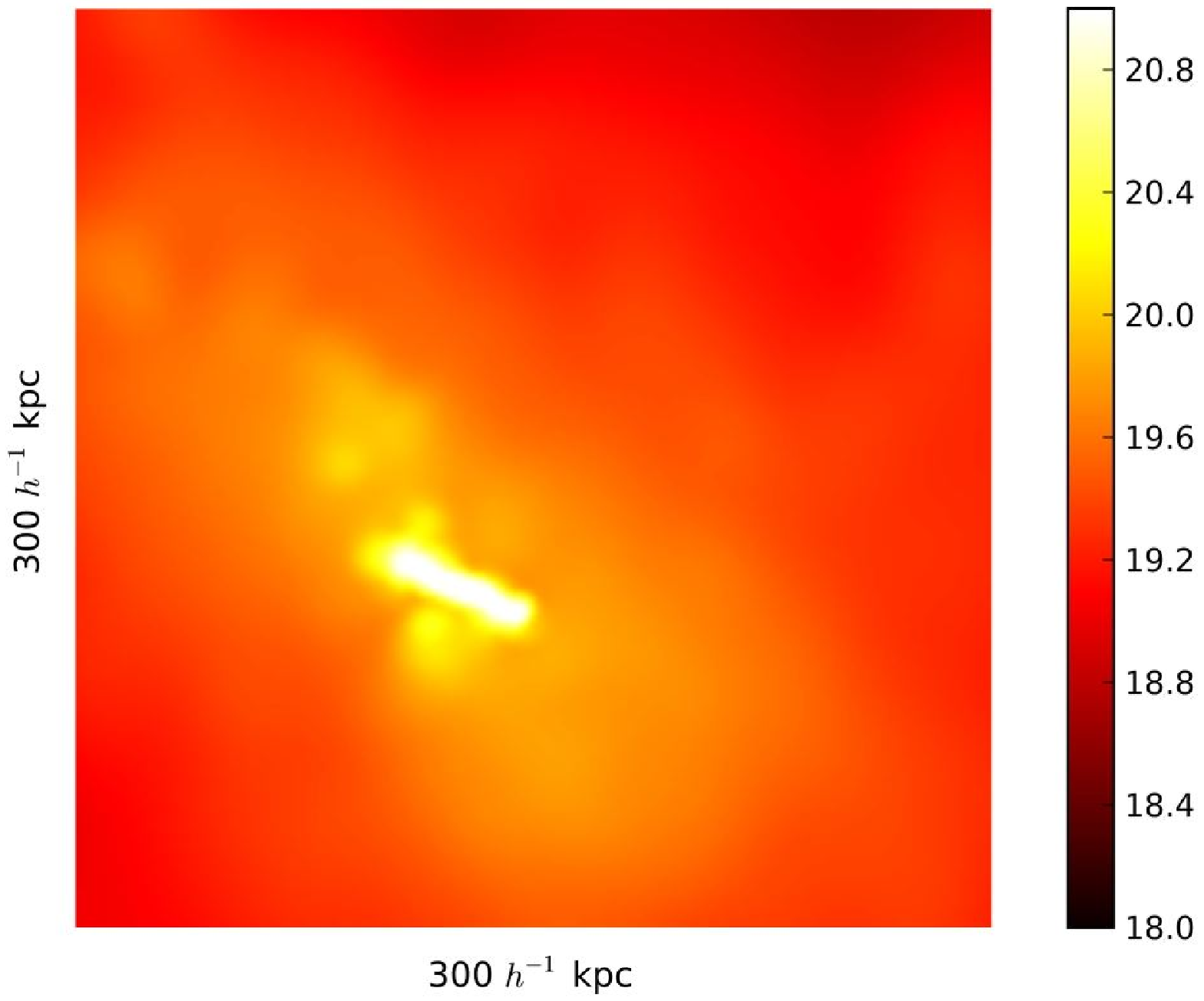}
  \includegraphics[width=0.35\textwidth]{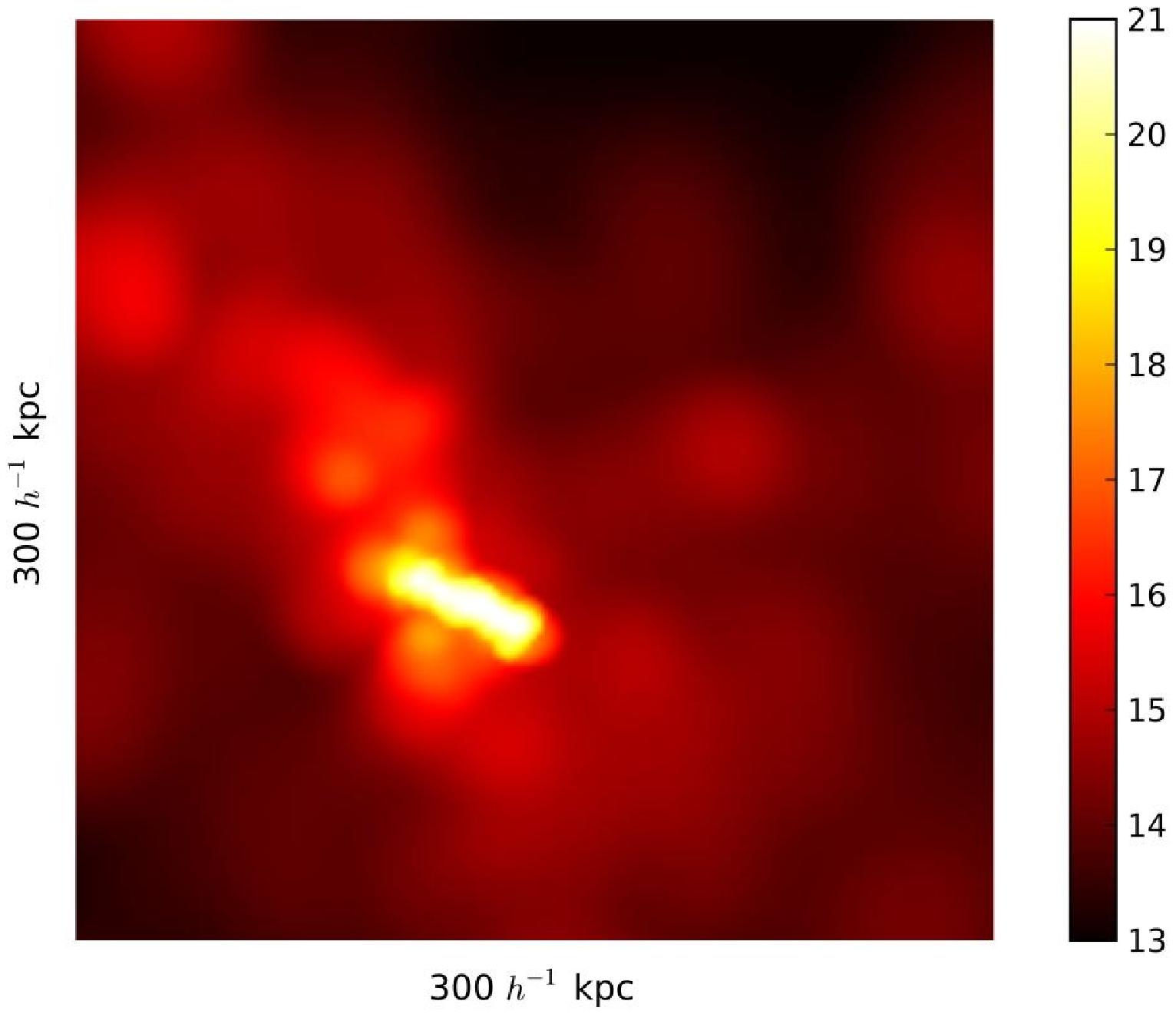}
  \includegraphics[width=0.29\textwidth]{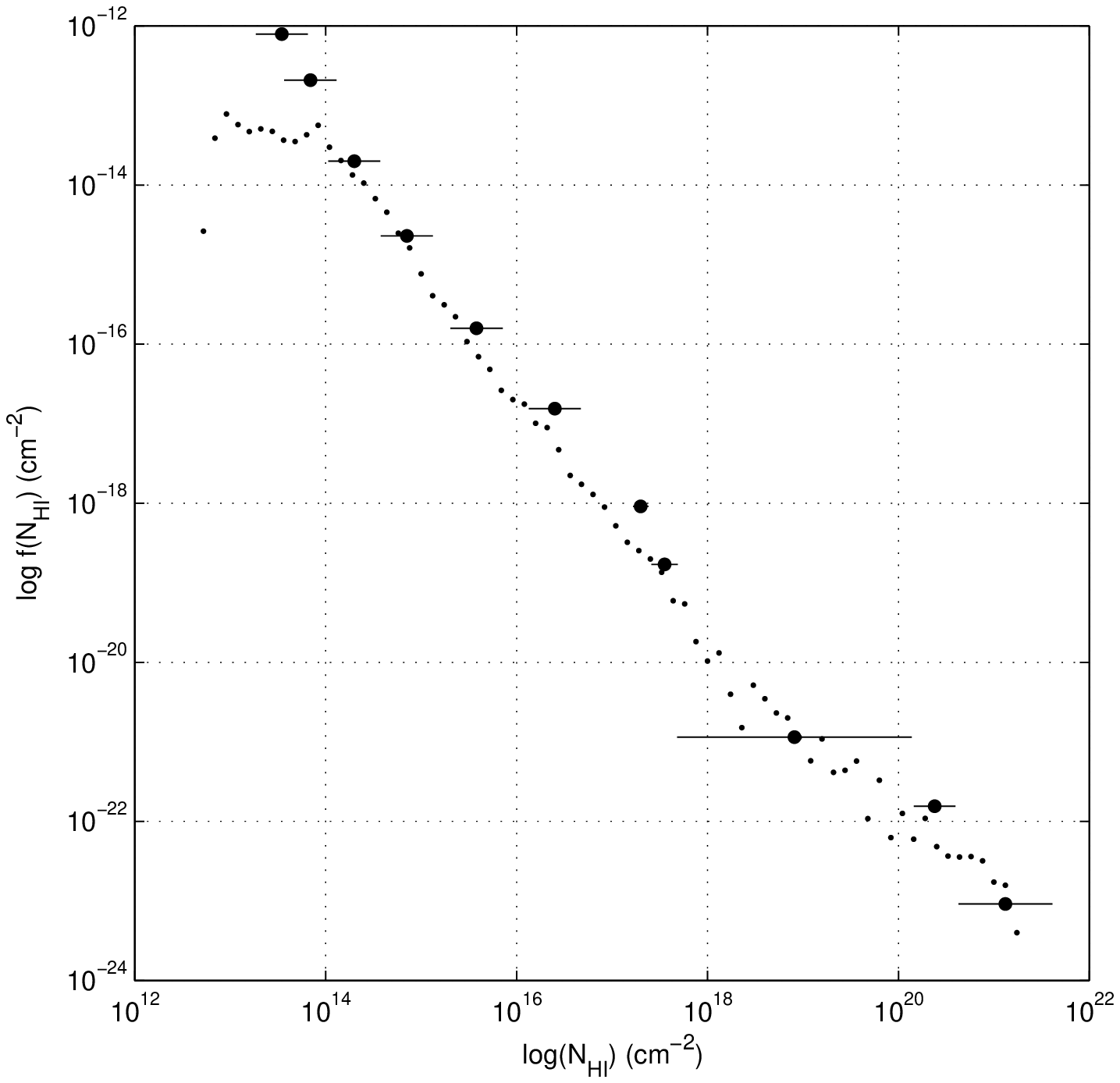}

  \includegraphics[width=0.35\textwidth]{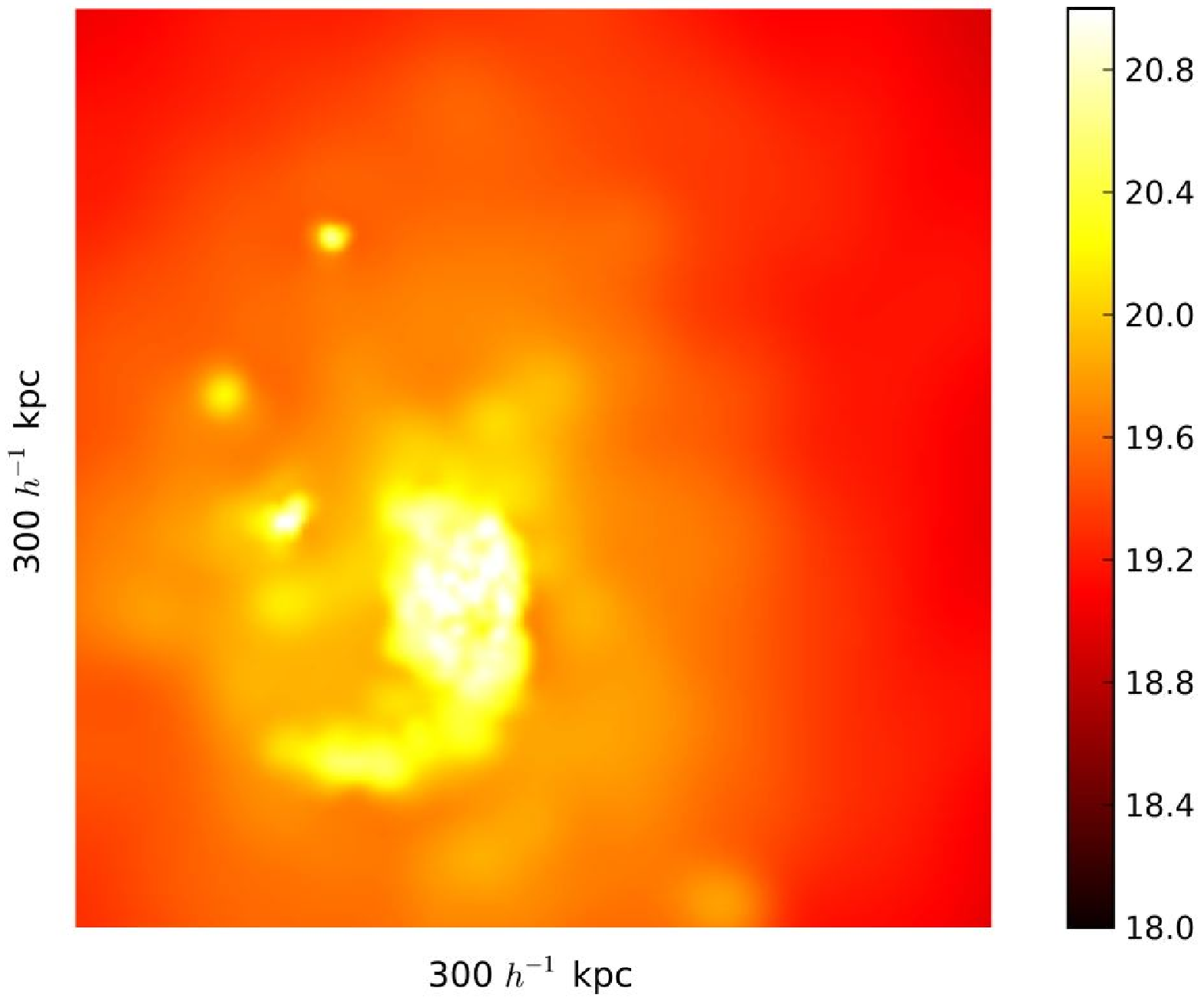}
  \includegraphics[width=0.35\textwidth]{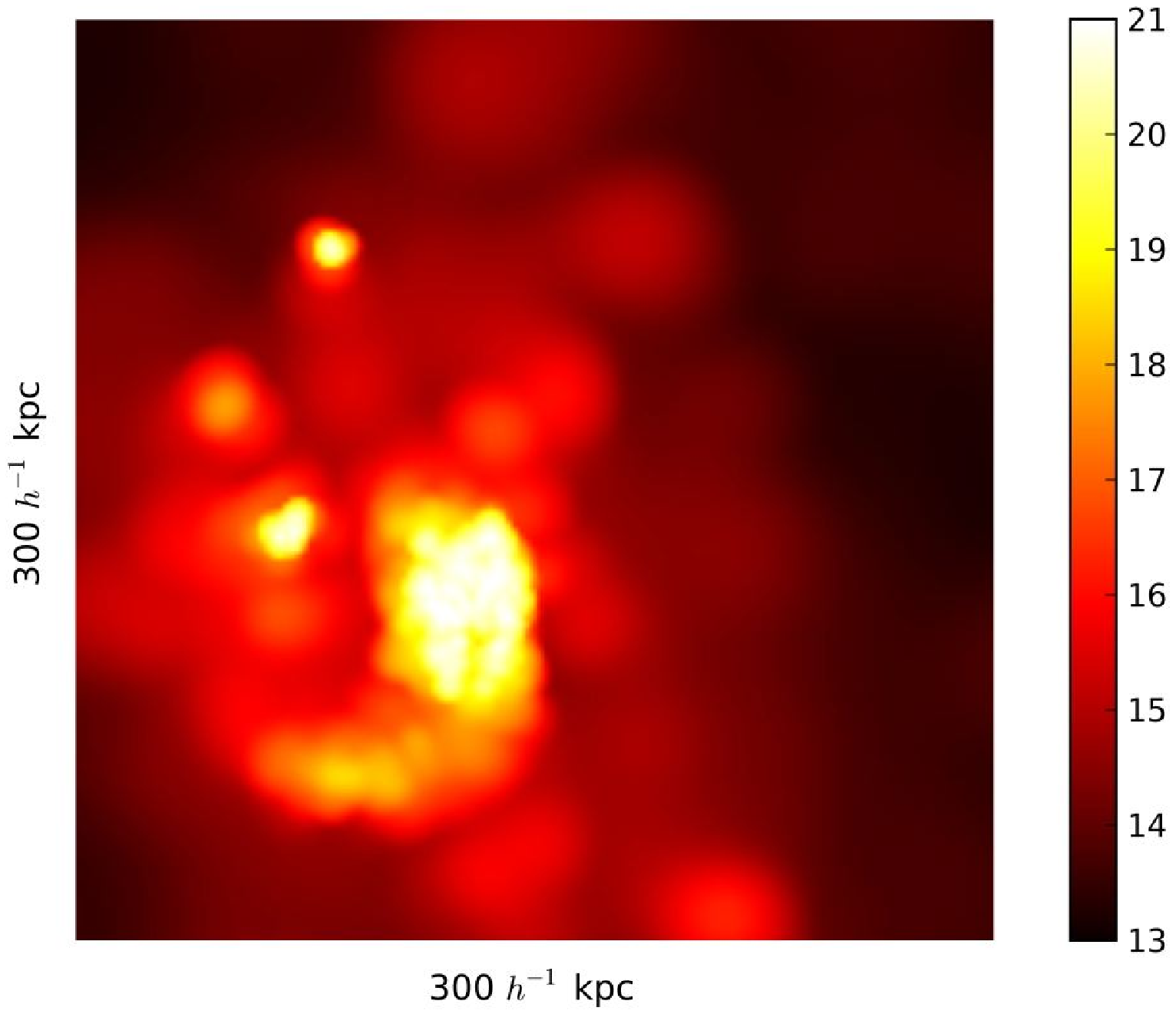}
  \includegraphics[width=0.29\textwidth]{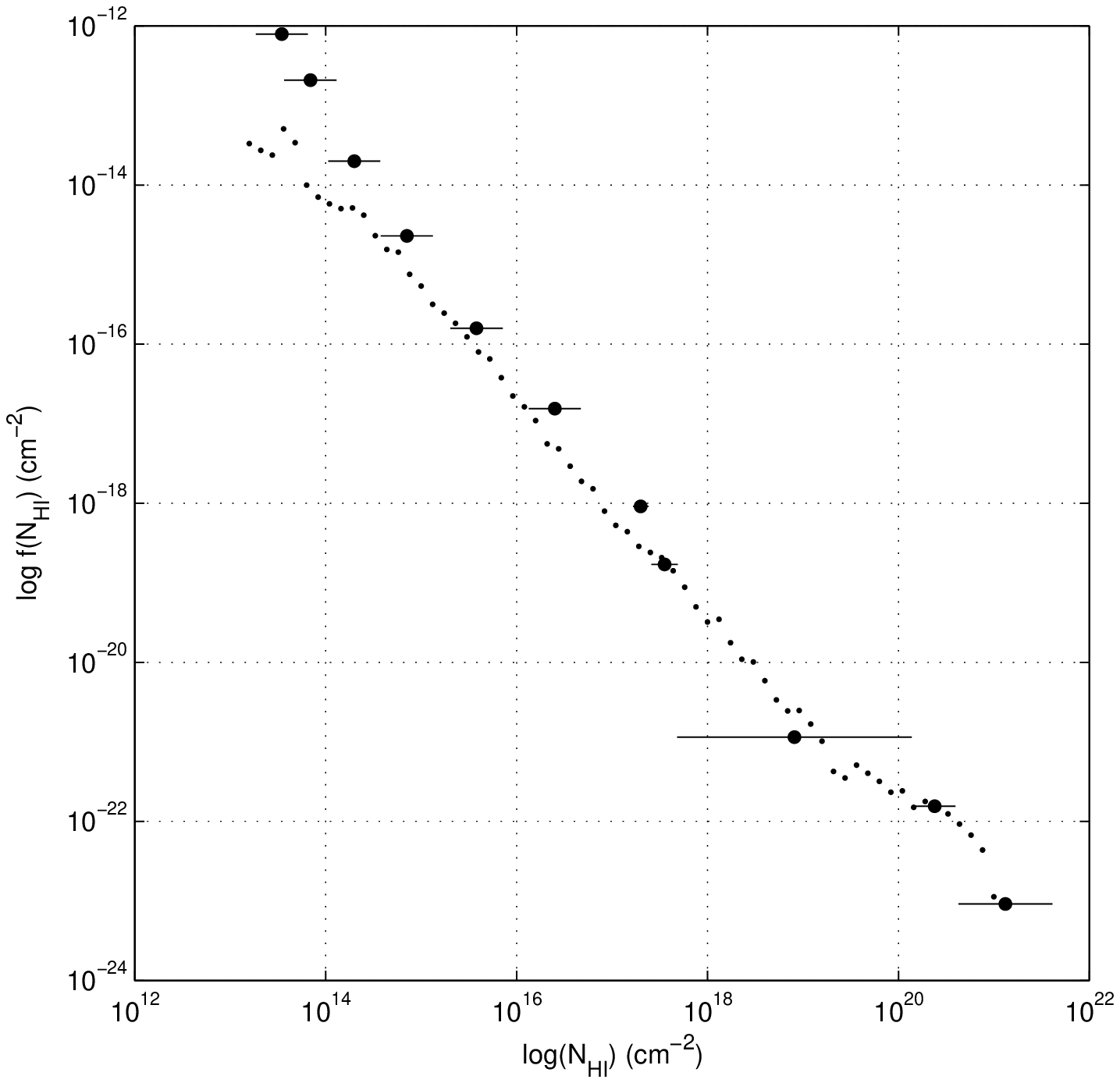}

  \caption{Four examples of high density regions in the reconstructed
    data, gridded to a cell size of 2 kpc. The left panels show the
    total hydrogen, while the middle panels show only the neutral
    component. Some objects have many satellites, as in the top
    panels, while others are much more isolated. All examples have
    extended \hbox{\rm H\,{\sc i}} at column densities around
    $\log(N_{HI}) = 16-17$ cm$^{-2}$. In the right panel, the individual
    \hbox{\rm H\,{\sc i}} column density distribution function is
    shown for each of the examples. Black dots correspond to the QSO
    absorption line data \citep{2002ApJ...567..712C}.}
  \label{2kpcres}
\end{figure*}

Figure~\ref{2kpcres} shows similar maps chosen from several
high-resolution regions, gridded to 2~kpc instead of 80~kpc.  The left
panels show a column density map of all the gas, while in the middle
panels the \hbox{\rm H\,{\sc i}} column densities are plotted. The right
panels show the \hbox{\rm H\,{\sc i}} column density distribution function
of the individual examples. The most complete distribution function is
obtained by summing the distribution functions of all the individual
objects, but even the individual distribution functions already display
the general trend of a flattening plateau around $N_{\rm{HI}} \sim
10^{19}$ cm$^{-2}$. Some objects have just a bright core with extended
emission, like the second example from the top. There are many objects
with small diffuse companions with maximum peak column densities of
$N_{\rm{HI}} \sim 10^{18}$ cm$^{-2}$. These companions are typically 20--40
kpc in size and are connected with filaments that have column densities
of $N_{\rm{HI}} \sim 10^{17}$ cm$^{-2}$ or even less. Comparing the plots
containing all the hydrogen and just the neutral hydrogen it can be seen
that the edge between low and high densities is much sharper for the
neutral hydrogen. The surface covered by column densities of $N_{\rm{HI}}
\sim 10^{17}$ cm$^{-2}$ is much larger than the surface
covered by column densities of $N_{\rm{HI}} \sim 10^{19}$ cm$^{-2}$.

\subsubsection{Neutral fraction}

\begin{figure}[t!]
  \includegraphics[width=0.5\textwidth]{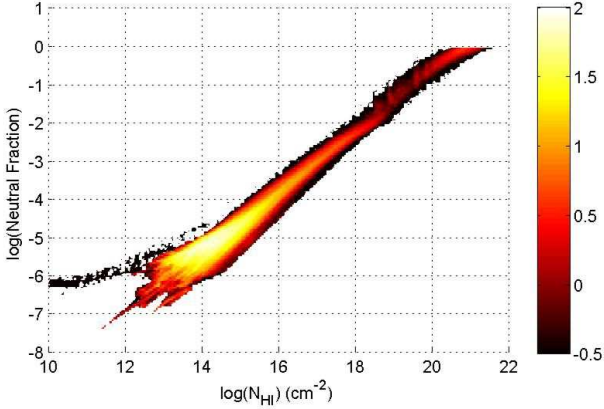}
   \caption{Neutral fraction plotted against \hbox{\rm H\,{\sc i}}
column density. The colorbar represents the probability of detecting a
certain combination of \hbox{\rm H\,{\sc i}} column density and
neutral fraction on logarithmic scale. At the highest densities
$N_{\rm{HI}} > 10^{20}$ cm$^{-2}$, the neutral fraction is almost
unity. Column densities around $N_{\rm{HI}} \sim 10^{19}$ cm$^{-2}$
have the smallest detection probability.}
  \label{HIvsFrac}
\end{figure}

The neutral fraction is plotted in a particularly instructive way in
Fig.~\ref{HIvsFrac}. Neutral fraction of the hydrogen gas is plotted
against \hbox{\rm H\,{\sc i}} column density, where the colorbar
represents the relative likelihood on a logarithmic scale of detecting
a given combination of neutral column and neutral fraction. The most
commonly occuring conditions are a neutral column density around
$N_{\rm{HI}} \sim 10^{14}$ cm$^{-2}$ with a neutral fraction of $\sim
10^{-5}$, representing Ly$\alpha$ forest gas.  The cutoff at low
column densities is artificial, owing to our gridding scheme.

At high densities, $N_{\rm{HI}} > 10^{20}$ cm$^{-2}$, the gas is
almost fully neutral and just below $N_{\rm{HI}} \sim 10^{19}$
cm$^{-2}$, the neutral fraction starts to drop very steeply below the
10 percent level. This is exactly the column density that is
considered to be the ``edge'' of \hbox{\rm H\,{\sc i}} galaxies, that
defines the border between optically thick and thin gas. This
transition from high to low neutral density happens on very small
scales of just a few kpc (\cite{1994ApJ...423..196D}). The surface area
with column densities in the range from $N_{\rm{HI}} \sim 10^{17}$ to
$10^{19}$ cm$^{-2}$ is relatively small. At lower column densities,
the probability of detecting \hbox{\rm H\,{\sc i}} in any given
direction increases. The well-defined correlation of neutral fraction
with neutral column for $N_{\rm{HI}} > 10^{16}$ cm$^{-2}$ defines a
straightforward correction for total gas mass accompanying an observed
neutral column density.

\begin{figure*}[t!]
  \includegraphics[width=0.5\textwidth]{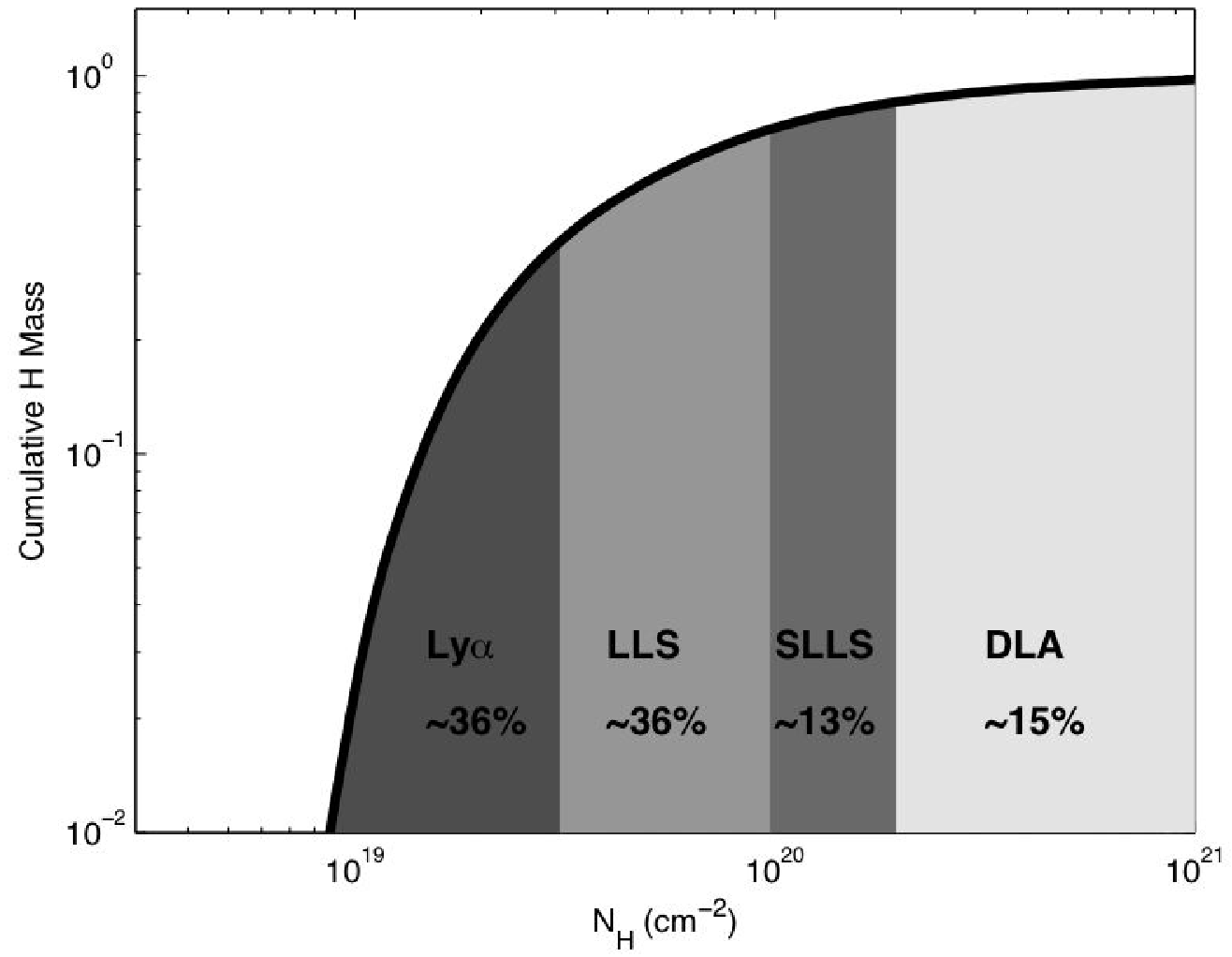}
  \includegraphics[width=0.5\textwidth]{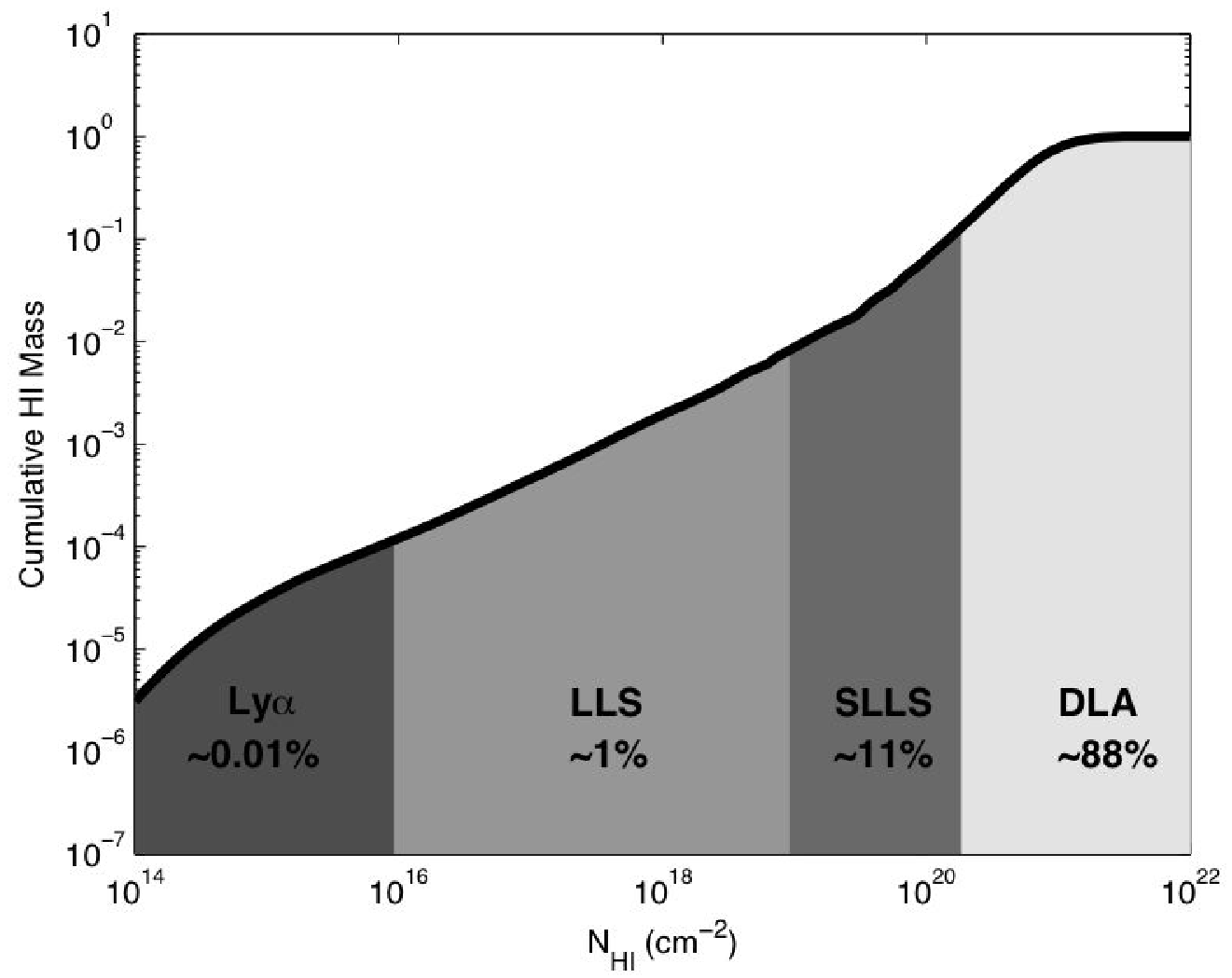}
  \caption{Left panel: Cumulative mass of total hydrogen as function
    of column density, the different phases are shown with different
    colours. The Ly$\alpha$ forest, the Lyman Limit System and
    galaxies have approximately the same mass component when
    considering all the gas. Right panel: Cumulative mass of neutral
    atomic hydrogen as function of column density. Although the surface
    covered by the LLSs is large, it contains only 1\% off all the
    neutral gas mass, while about 99\% resides in the galaxies}
  \label{CumMass}
\end{figure*}

In Fig.~\ref{CumMass} the cumulative mass is plotted as function of
total hydrogen column density (left panel) and the column density of
the \hbox{\rm H\,{\sc i}} gas (right panel). Note that the vertical
scale is different in the two panels. The plot is divided in different
regions, the galaxies or Damped Lyman-$\alpha$ Absorbers (DLA) are
coloured light gray. In neutral hydrogen, these are the column
densities above log$(N_{\rm{HI}})$ = 20.3. Lower column densities
belong to the Super Lyman Limit systems (SLLS), or sub-DLAs. In the
plot showing the neutral hydrogen an inflection point can be seen at a
column density of log$(N_{\rm{HI}})$ = 19. This is where the effect of
self shielding starts to decrease rapidly and the Lyman Limit regime
begins. At the lower end, below column densities of log$(N_{\rm{HI}})$
= 16 is the Lyman alpha forest, which is coloured dark gray. As can be
seen there is a huge difference in mass contribution for the different
phases, when comparing the neutral gas against the total gas
budget. In \hbox{\rm H\,{\sc i}}, about 99 percent of the mass is in
DLAs, Lyman Limit Systems account for about 1 percent of the mass and
the Lyman alpha forest contributes much less than a percent. When
looking at the total gas mass budget all three components (DLAs, LLSs
and the Ly-$\alpha$ forest) have approximately the same mass fraction.

\subsection{Two-Point Correlation Function}

\begin{figure}[t!]
  \includegraphics[width=0.5\textwidth]{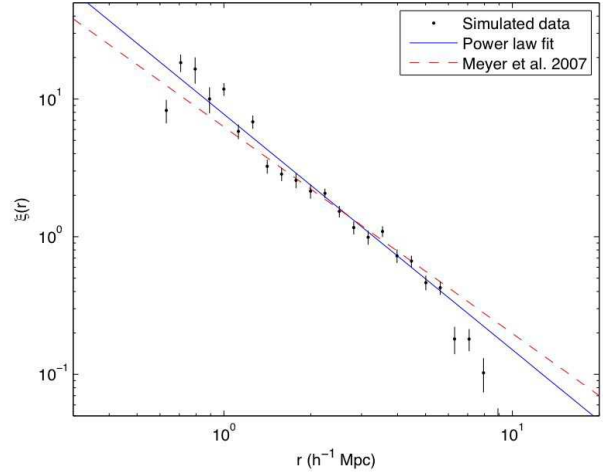}
   \caption{Two-point correlation function of \hbox{\rm H\,{\sc
         i}}-rich objects in our simulation, contrasted with a
     power-law fit to the
     observed relation of \cite{2007ApJ...654..702M}.}
  \label{corr_function}
\end{figure}

The two-point correlation function measures the degree of
clustering of galaxies in the spatial direction $(\xi (r))$, which
relates directly to the power spectrum through a Fourier transform
\citep[e.g.][]{1977ApJ...217..385G, 1982ApJ...254..437D}. The
spatial two-point correlation function is defined as the excess
probability, compared with that expected for a random distribution,
of finding a pair of galaxies at a separation $r_{1,2}$. For \hbox{\rm
H\,{\sc i}}, the clustering is weaker compared to optical galaxies
\citep{2007ApJ...654..702M}. On scales between $\sim 0.5$ kpc and 12 Mpc,
the correlation function for optical galaxies has been determined in SDSS
\citep{2005ApJ...621...22Z} and 2dFGS \citep{2002MNRAS.332..827N}. For
the \hbox{\rm H\,{\sc i}}-rich galaxies in the HIPASS catalogue, a scale
length is obtained of $r_0 = 3.45 \pm 0.25 h^{-1}$ Mpc and a slope of
$\gamma = 1.47 \pm 0.08$. 

In the past, several estimators have been given for the two-point
correlation function, we will use the Landy \& Szalay estimator as
described in \cite{1993ApJ...412...64L} as this estimator is used in
\citep{2007ApJ...654..702M} to determine the correlation for \hbox{\rm
H\,{\sc i}} galaxies. This estimator is given by:
\begin{equation}
\xi_{\rm{LS}} = \frac{1}{RR}[DD  - 2DR + RR ]
\end{equation}
where $DD$ are the galaxy-galaxy pairs, $RR$ the random-random pairs
and $DR$ the galaxy-random pairs. This estimator has to be normalised with the number of correlations in the simulated and random distributions:

\begin{equation}
\xi_{\rm{LS}} = \frac{1}{RR}\Big[\frac{DD}{(n_d (n_d-1))/2}  - \frac{2DR}{n_r n_d} + RR \Big]
\end{equation}
where $n_d$ is the number of detections or simulated objects and
$n_r$ is the number of galaxies in the random sample.

In Fig.~\ref{corr_function} the two-point correlation function is
plotted; the black dots represent the values obtained from the
simulation, while the dashed red line corresponds to the correlation
function that is fit to galaxies in the HIPASS catalogue by
\citep{2007ApJ...654..702M}. The solid line is our best fit, with
a scale length of $r_0 = 3.3 \pm 0.2 h^{-1}$ Mpc and a slope of
$\gamma = 1.7 \pm 0.2$, only data points where the radius is smaller
than 6 Mpc have been used for the fit.

There is very good correspondence between the simulated and observed
\hbox{\rm H\,{\sc i}}-correlation functions on scales between $\sim
0.5$ Mpc and $\sim 5$ Mpc. Accuracy at smaller scales is limited by
the finite resolution of the simulation. On the other hand, the
representation of large scales is limited by the physical size of the
box. In a 32 $h^{-1}$ Mpc box, the largest well-sampled structures are
about 5 Mpc in size. This difference is not surprising, because
  \cite{2007ApJ...654..702M} are able to sample structures up to 10
  Mpc, given their significantly larger survey
  volume. \cite{2007ApJ...654..702M} also looked at a limited sample
  of galaxies, applying the parameter cuts $M_{HI} > 10^{9.05}
  M_{\odot}$ and $D < 30$ Mpc. This limited sample is very similar to
  our sample of simulated objects and the power law parameters in this
  case are $r_0 = 3.2 \pm 1.4 h^{-1}$ Mpc and $\gamma = 1.5 \pm
  1.1$. Although the errors are larger, the results are very similar
  to the full sample and in excellent agreement with our simulations.

\subsection{\hbox{\rm H\,{\sc i}} Mass Function}

The \hbox{\rm H\,{\sc i}} Mass Function $\theta(M_{\rm{HI}})$ is
defined as the space density of objects in units of $h^3$
Mpc$^{-3}$. For fitting purposes a Schechter function
\citep{1976ApJ...203..297S} can be used of the form:
\begin{equation}
\theta(M_{\rm{HI}})dM_{\rm{HI}} = \theta^*\Big(\frac{M_{\rm{HI}}}{M_{\rm{HI}}^{*}}\Big)^\alpha
\exp \Big(-\frac{M_{\rm{HI}}}{M_{\rm{HI}}^{*}}\Big)\frac{dM_{\rm{HI}}}{M^*}
\end{equation}
characterised by the parameters $\alpha$, $M^{*}_{\rm{HI}}$ and
$\theta^*$, which define the slope of the power law, the \hbox{\rm
H\,{\sc i}} mass corresponding to the ``knee'' and the normalisation
respectively. In a logarithmic form the \hbox{\rm H\,{\sc i}} Mass
function can be written as:

\begin{equation}
\theta(M_{\rm{HI}}) = \ln(10)\theta^*\Big(\frac{M_{\rm{HI}}}{M^*_{\rm{HI}}}\Big)^{\alpha+1}\exp\Big(\frac{-M_{\rm{HI}}}{M^*_{\rm{HI}}}\Big)
\end{equation}

\begin{figure*}[t!]
  \includegraphics[width=0.5\textwidth]{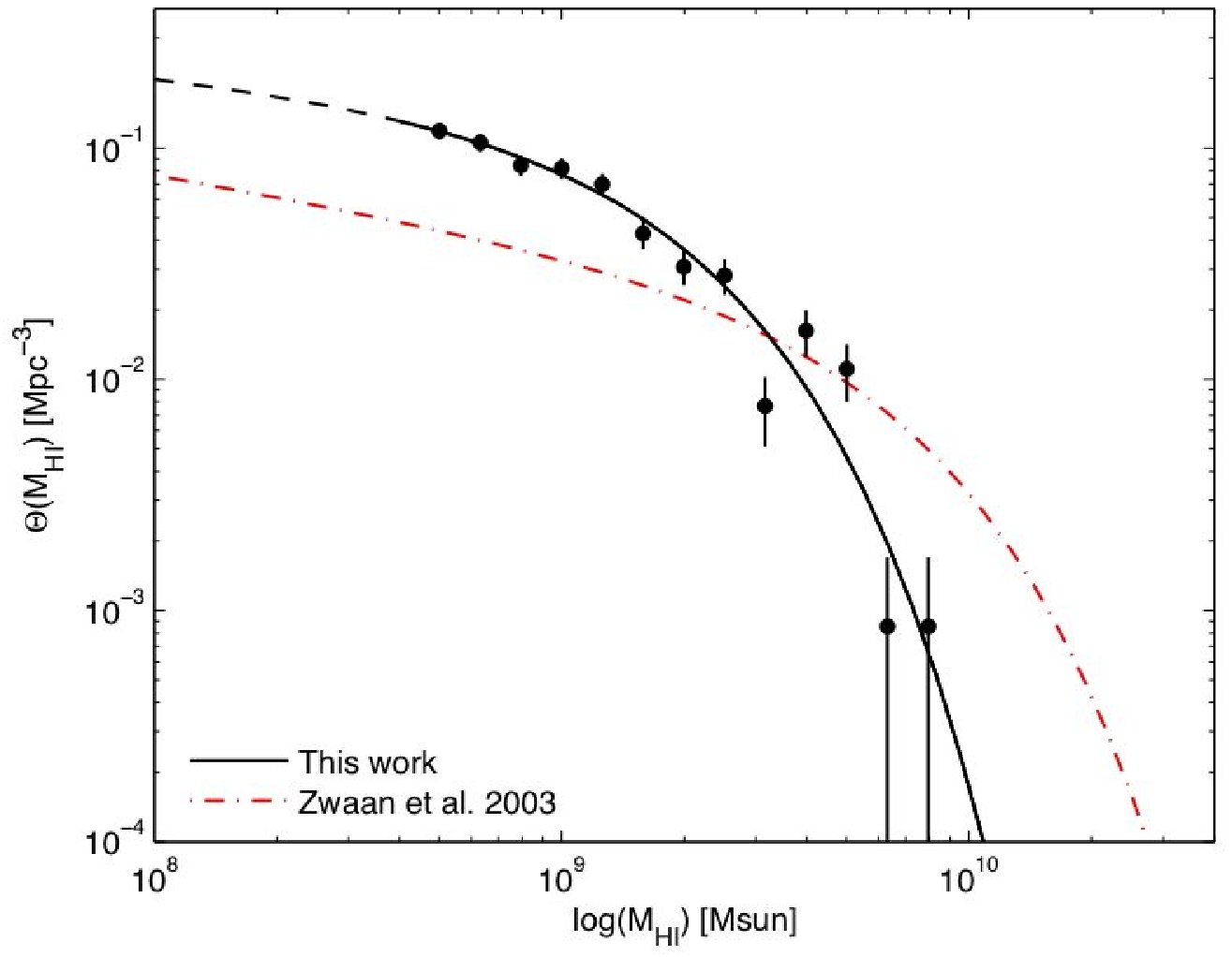}
  \includegraphics[width=0.5\textwidth]{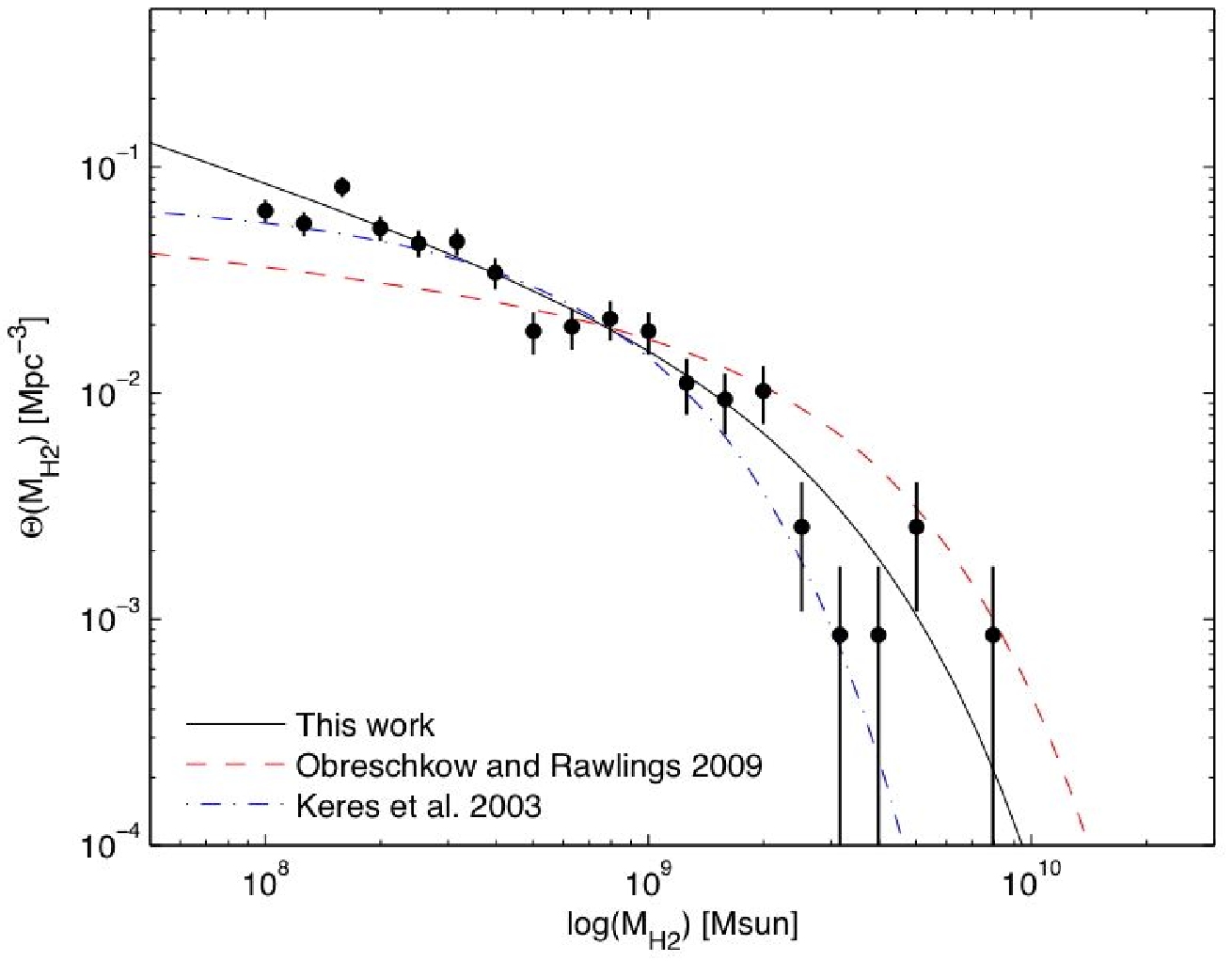}
   \caption{Left panel: \hbox{\rm H\,{\sc i}} Mass Function of the
     simulated data (black dots) with the best-fit Schechter function
     (solid black line), compared with the HIMF from
     \cite{2003AJ....125.2842Z} (dash-dotted red line). Our best fit
     line is dashed below $10^9 M_{\odot}$, because there are no data
     points there and the function is only an extrapolation. Right
     panel: H$_2$ Mass Function of the simulated data (black dots) and
     the best fit, compared with the H$_2$MF from \cite{2003ApJ...582..659K}
     (dash-dotted (blue) line) and \cite{2009MNRAS.tmp..289O} 
     (dashed (red) line). Both simulated Mass Functions show agreement with
     observations over about a decade in mass.}
  \label{HIMF}
\end{figure*}

The reconstructed structures in our high resolution grids can be used
to determine a simulated \hbox{\rm H\,{\sc i}} Mass Function for
structures above $\sim5 \times 10^8$ $M_\odot$. The result is plotted
in the left panel of Fig.\ref{HIMF}, where the \hbox{\rm H\,{\sc i}}
Mass function is shown with a bin size of 0.1 dex. Overlaid is the
best fit to the data, the fitting parameters that have been used are
$\theta^*$ = 0.059$\pm$0.047, $\log(M^*_{\rm{HI}}) = 9.2 \pm 0.3$ and
$\alpha = -1.16 \pm 0.45$. Note that in this case the value of
$\alpha$ is not very well-constrained, as this parameter defines the
slope of the lower end of the \hbox{\rm H\,{\sc i}} Mass Function, but
our simulation is unable to sample the mass function below a mass of
$M_{HI} \approx 5\times10^8 M_{\odot}$. The result is compared with
the \hbox{\rm H\,{\sc i}} mass function from
\cite{2003AJ....125.2842Z} (dash-dotted line). The reconstructed mass
function corresponds reasonably well with the mass functions obtained
from galaxies in HIPASS around $M^*$. At masses around $10^{10}$
$M_\odot$, the error bars are very large due to small number
statistics. A much larger simulation volume is required to sample this
regime properly. There is a hint of an excess in the simulation near
our resolution limit, this may simply reflect cosmic variance and will
be addressed in future studies. \cite{2003AJ....125.2842Z} compared
four different quadrants of the southern sky and found that at $M_{HI}
\sim 10^9 M_{\odot}$ the estimated space density varies by a factor of
about three, which is comparable to the factor $\sim2$ difference we
see between the simulation and observations.

\subsubsection{\hbox{\rm H\,}$_2$ Mass Function}
The H$_2$ mass Function can be determined in a similar way as the
\hbox{\rm H\,{\sc i}} Mass Function. The result is shown in the right
panel of Fig.~\ref{HIMF} where the simulated data points are fitted
with a Schechter function. Our best fit parameters are $\theta^*$ =
0.036$\pm$0.036, $\log(M^*_{\rm{H_2}}) = 8.7 \pm 0.3$ and $\alpha =
-1.01 \pm 0.57$. At the high end of the mass function the results are
affected by low number statistics. The simulated fit is compared with
the fits as determined by \cite{2009MNRAS.tmp..289O} (dashed line)
and \cite{2003ApJ...582..659K} (dash-dotted line). There is very
good agreement over the full mass range.

\begin{figure}[t!]
  \includegraphics[width=0.5\textwidth]{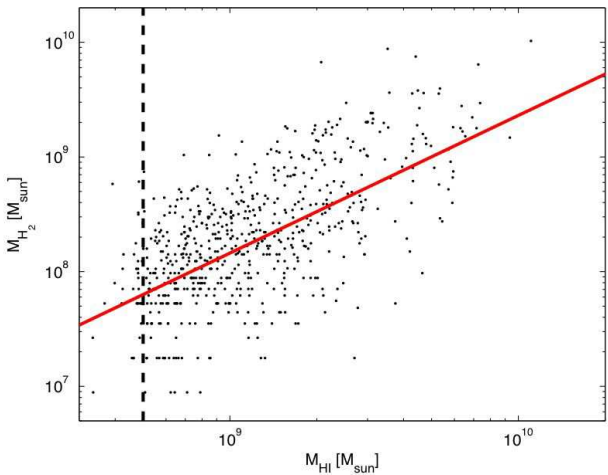}
     \caption{Derived H$_2$ masses are plotted against \hbox{\rm
           H\,{\sc i}} masses for individual simulated objects. The
         distribution can be fit using a simple power law (solid (red)
         line), although the scatter is very large. The dashed
         vertical line represents the sample cut-off for \hbox{\rm
           H\,{\sc i}} masses in the simulated data.}
  \label{H2vsHI}
\end{figure}

In Fig.~\ref{H2vsHI} the derived H$_2$ masses are plotted as function of
\hbox{\rm H\,{\sc i}} mass. The dashed vertical line represents the
completeness limit of \hbox{\rm H\,{\sc i}} masses. The data can be
fitted using a power law (solid line) which looks like $M_{H_2} =
(M_{HI}/m_0)^\beta$ with a scaling parameter of $m_0=158\pm43$ and a
slope of $\beta = 1.2\pm 0.1$.

\subsection{Stars, Dark Matter and Molecular Hydrogen}

In addition to the SPH-particles, the simulations also contain dark
matter and stars. The distribution of these components can be
reconstructed and compared with the distribution of neutral
hydrogen. For reconstructing the dark matter and stars a very simple
adaptive gridding scheme has been used. This gridding
scheme was adopted, because the dark matter and star particles do not
have a variable smoothing kernel like the gas particles. They do have
a smoothing kernel defined by the softening length of 2.5 kpc
$h^{-1}$, however this is a fixed value. The gridding method as
described in section 3.3 using a spline kernel can not be used for
this reason. Exactly the same regions have been gridded as those
previously determined \hbox{\rm H\,{\sc i}} objects. First the objects
have been gridded with a cell size of 10 kpc using a nearest
grid-point algorithm. The resulting moment maps have been used to
determine which high density regions contain many particles. In a
second step the particles have been gridded to five independent cubes
using a 2 kpc cell size. Based on the density of simulation particles
in the 10 kpc resolution cube, an individual particle is assigned to
one of five different cubes for gridding. The density threshold is
determined by the number of particles in the 10 kpc resolution cube
integrated along the line of sight. The threshold numbers are 2, 6,
18, 54 and everything above 54 particles respectively. The five cubes
are integrated along the line of sight and smoothed using a Gaussian
kernel with a standard deviation of 7, 5, 3.5, 2.5 and 1.5 pixels of
2~kpc respectively. Finally the five smoothed maps are added
together. These smoothing kernels where chosen to insure that each
individual cube covering a different density regime has a smooth
density distribution, but preserves as much resolution as practical.

\begin{figure*}[t!]
 
  \includegraphics[width=0.33\textwidth]{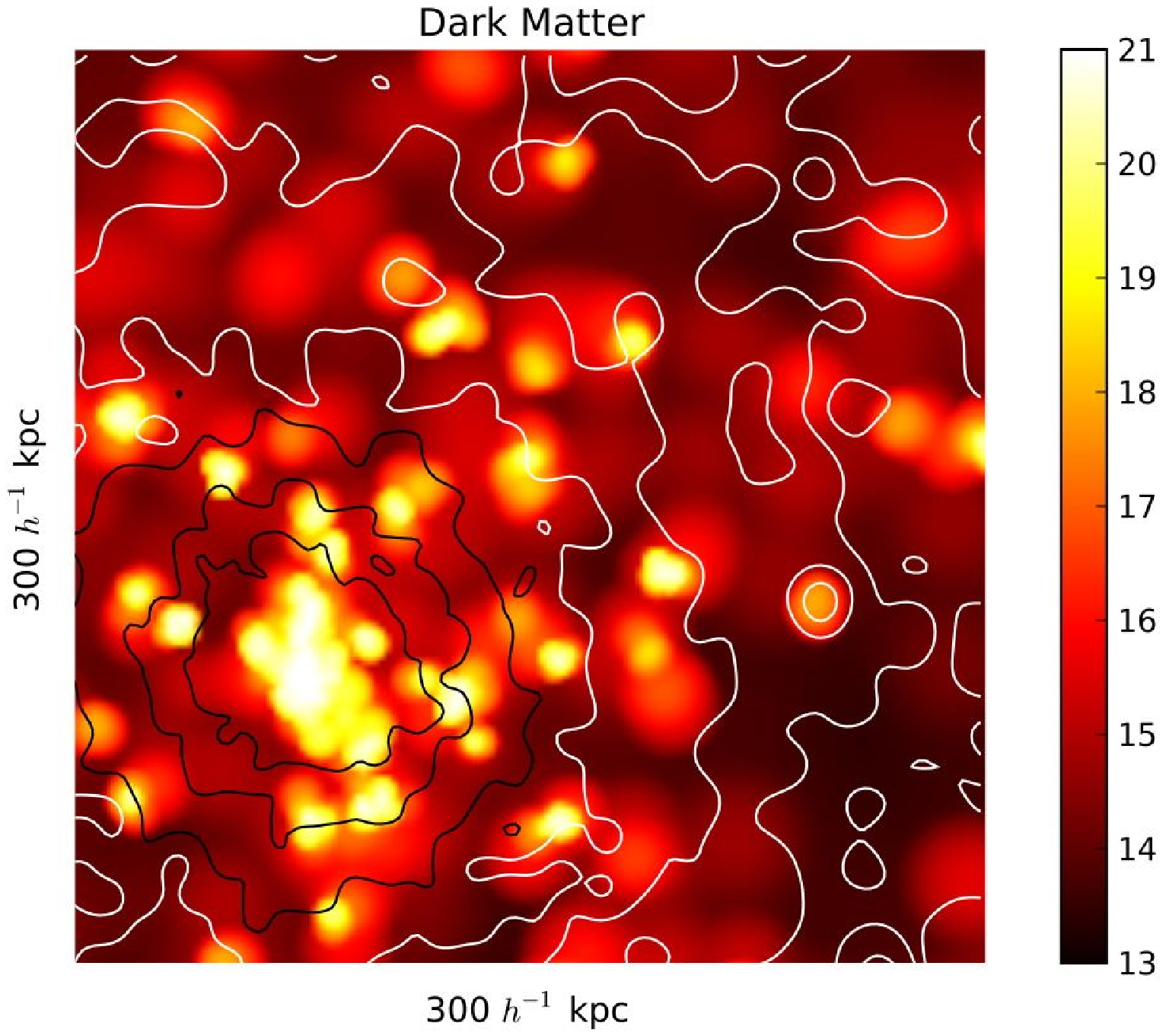}
  \includegraphics[width=0.33\textwidth]{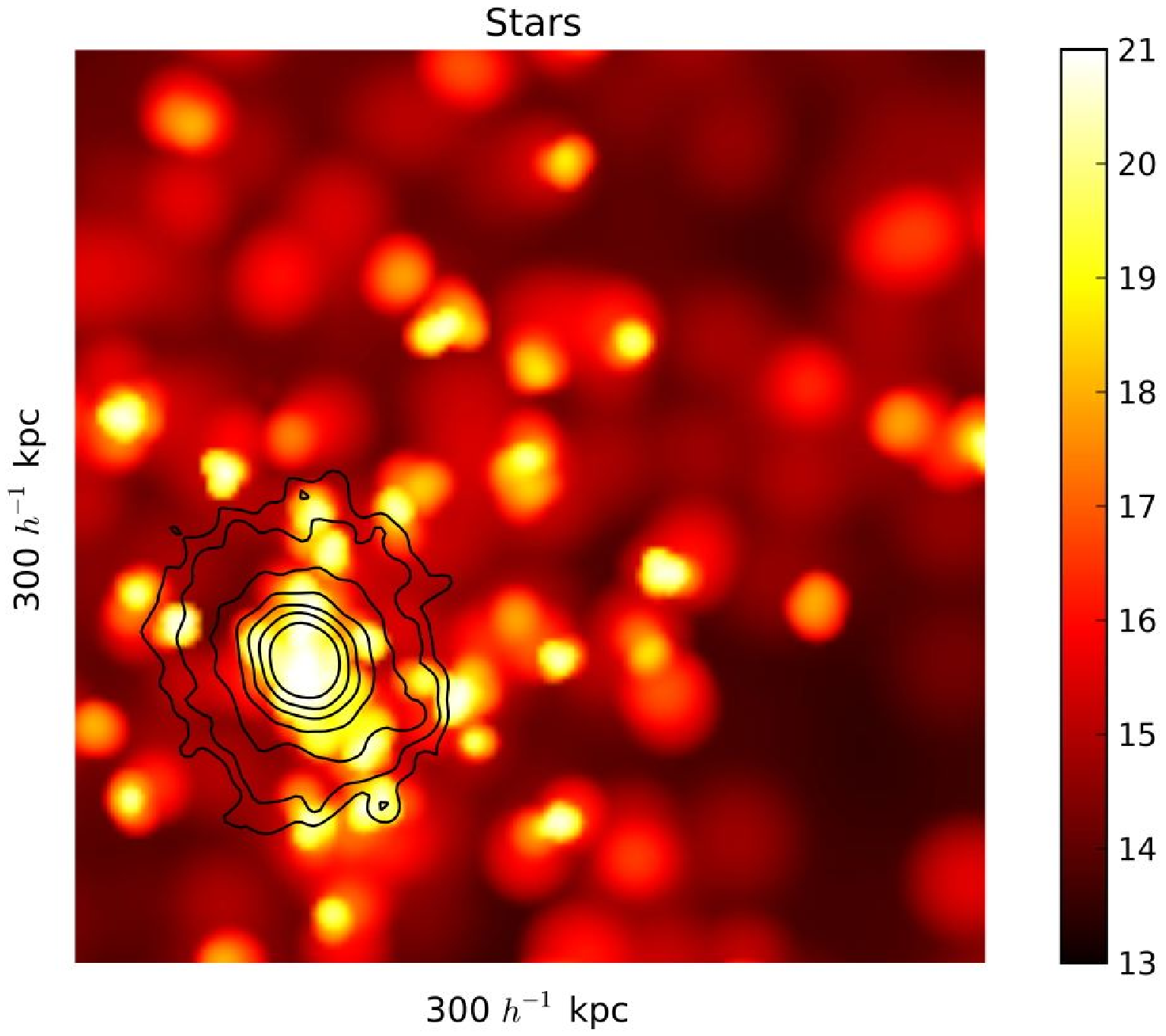}
  \includegraphics[width=0.33\textwidth]{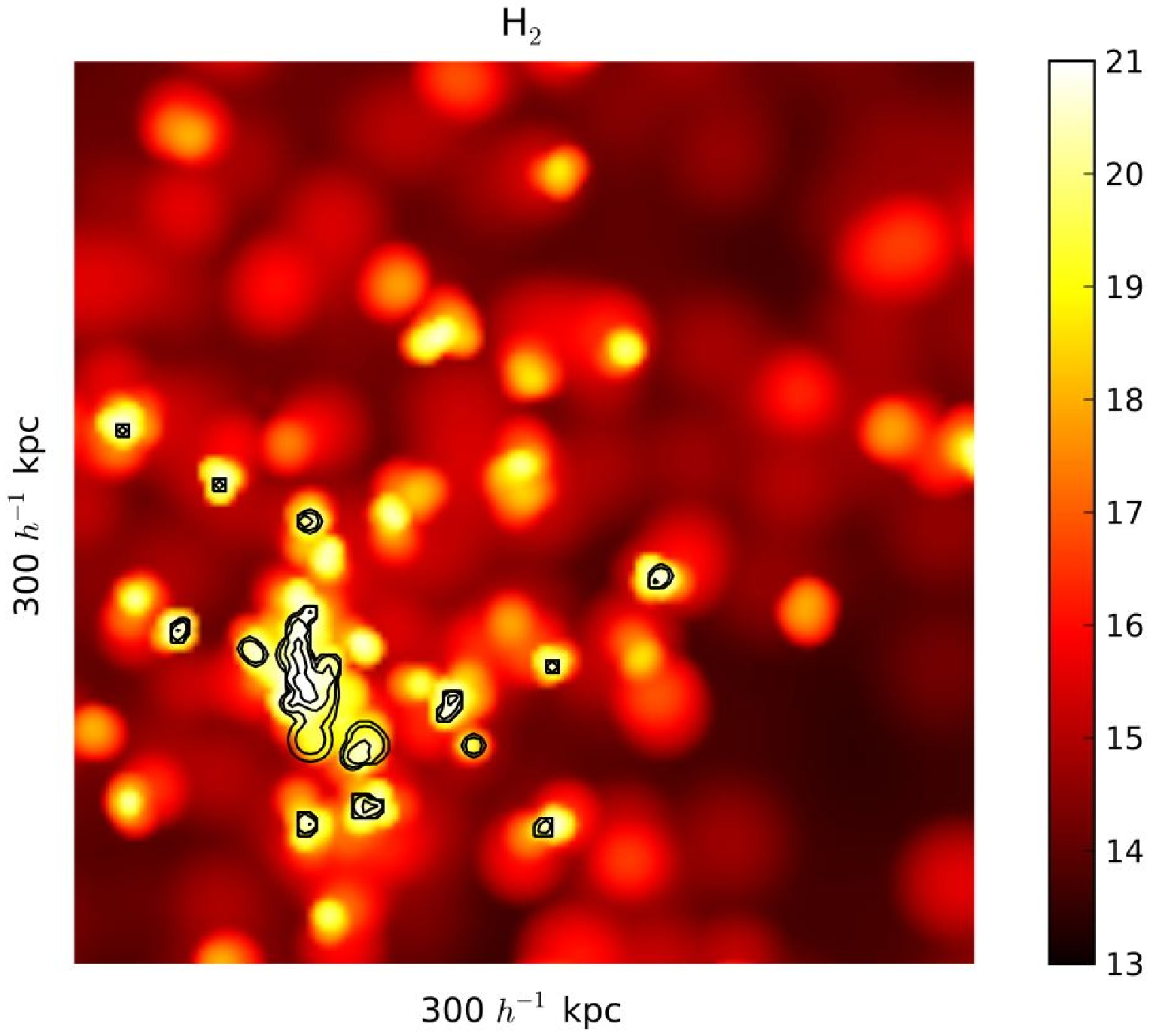}
 
  \includegraphics[width=0.33\textwidth]{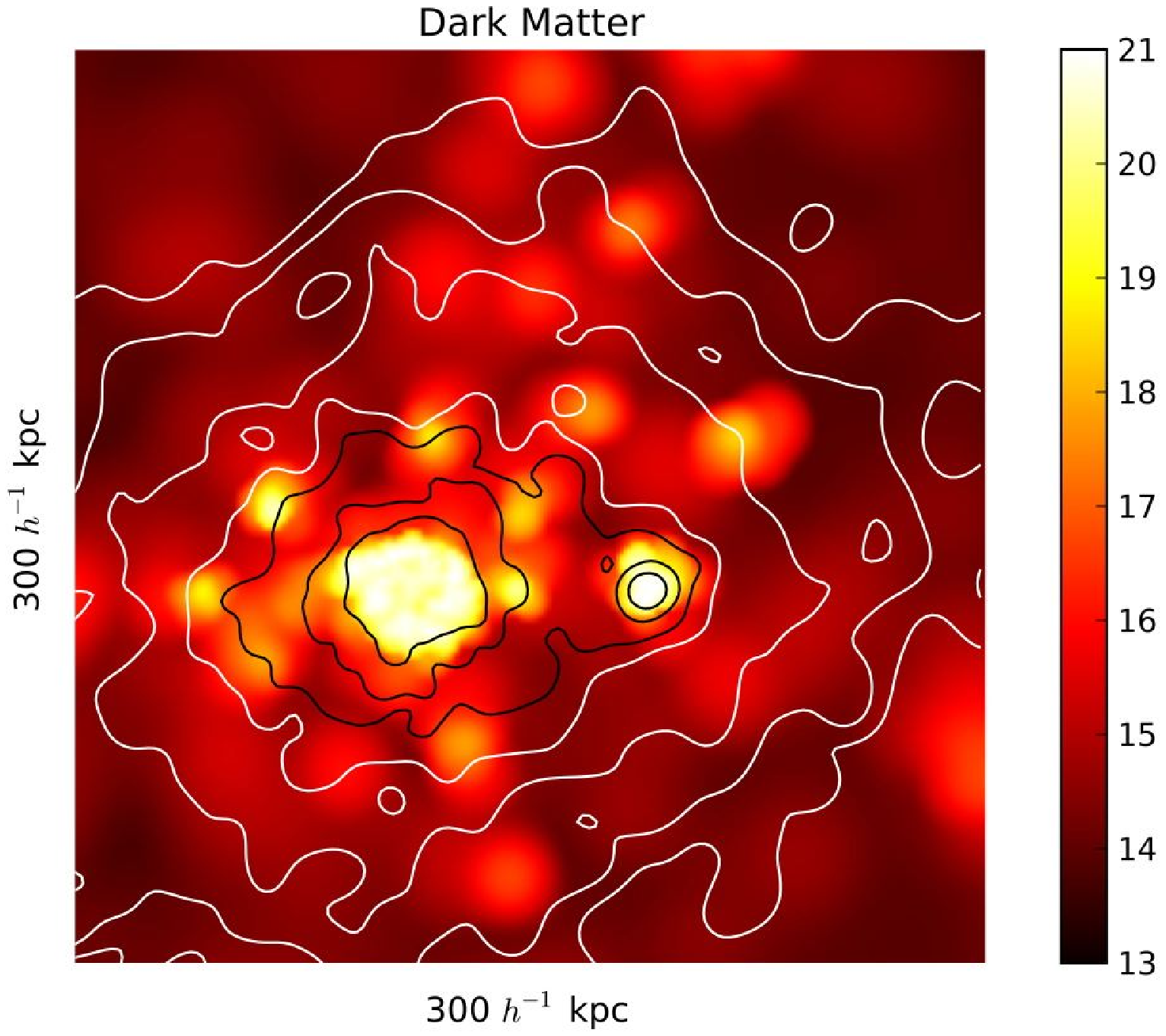}
  \includegraphics[width=0.33\textwidth]{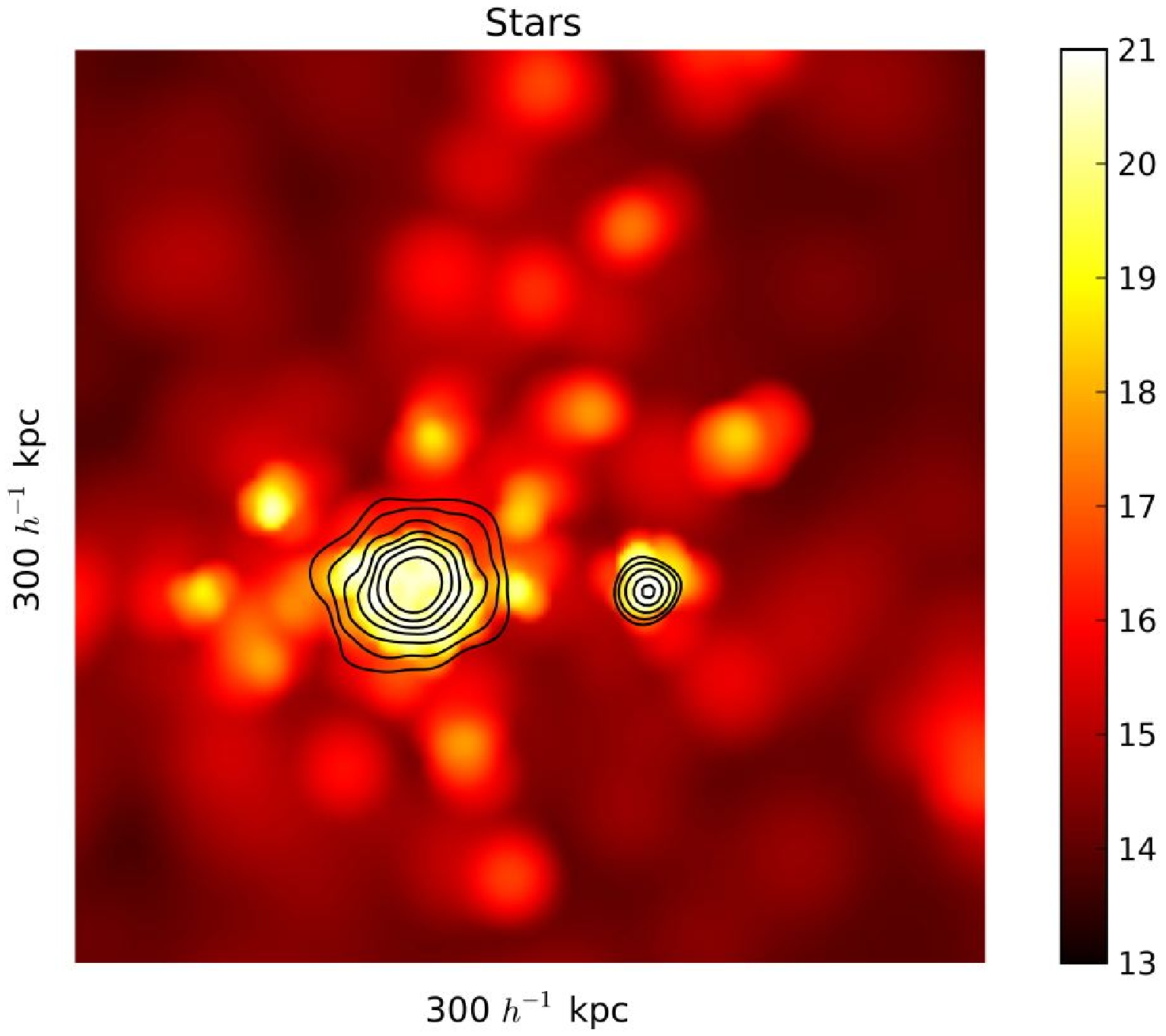}
  \includegraphics[width=0.33\textwidth]{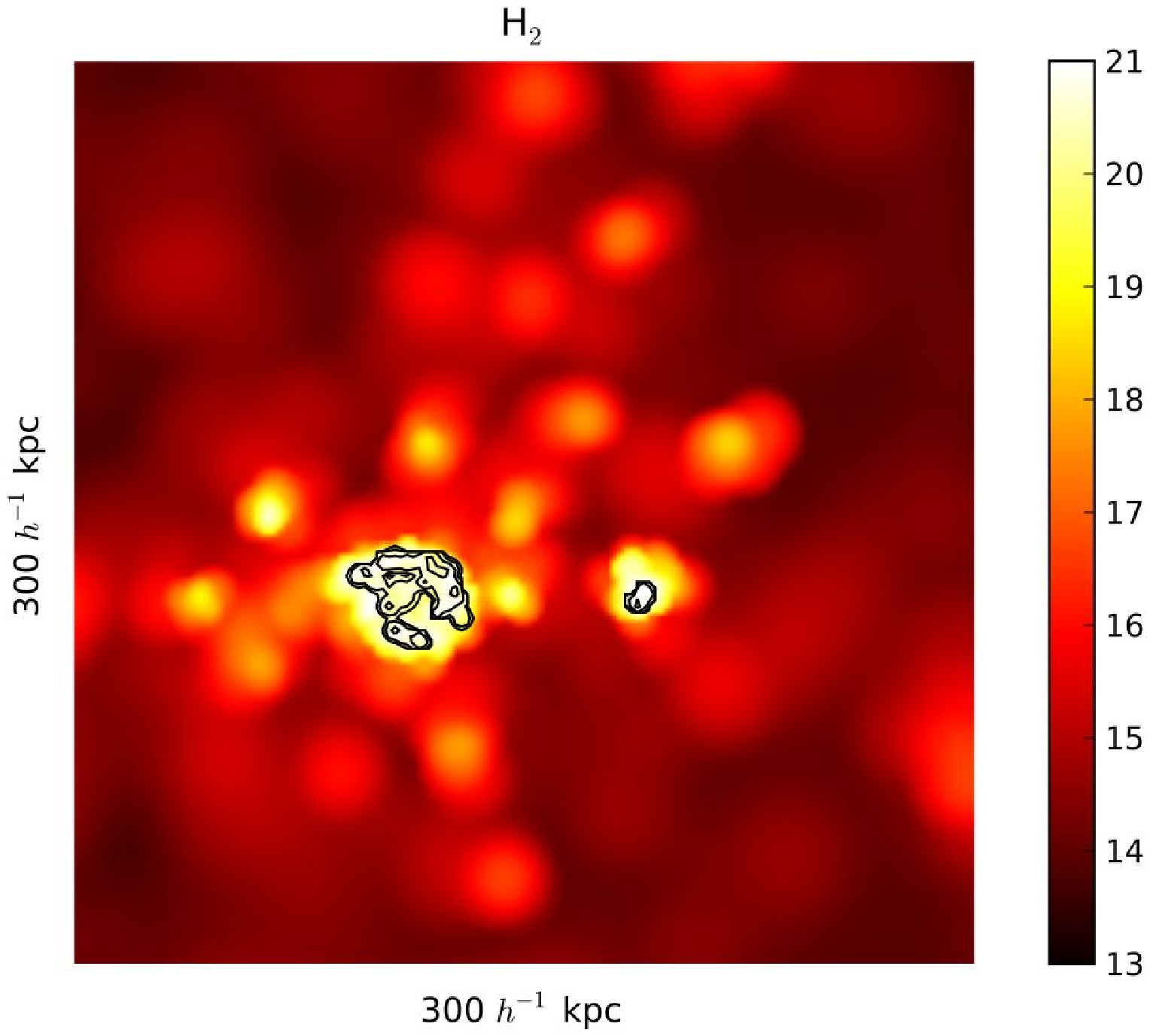}
 
  \includegraphics[width=0.33\textwidth]{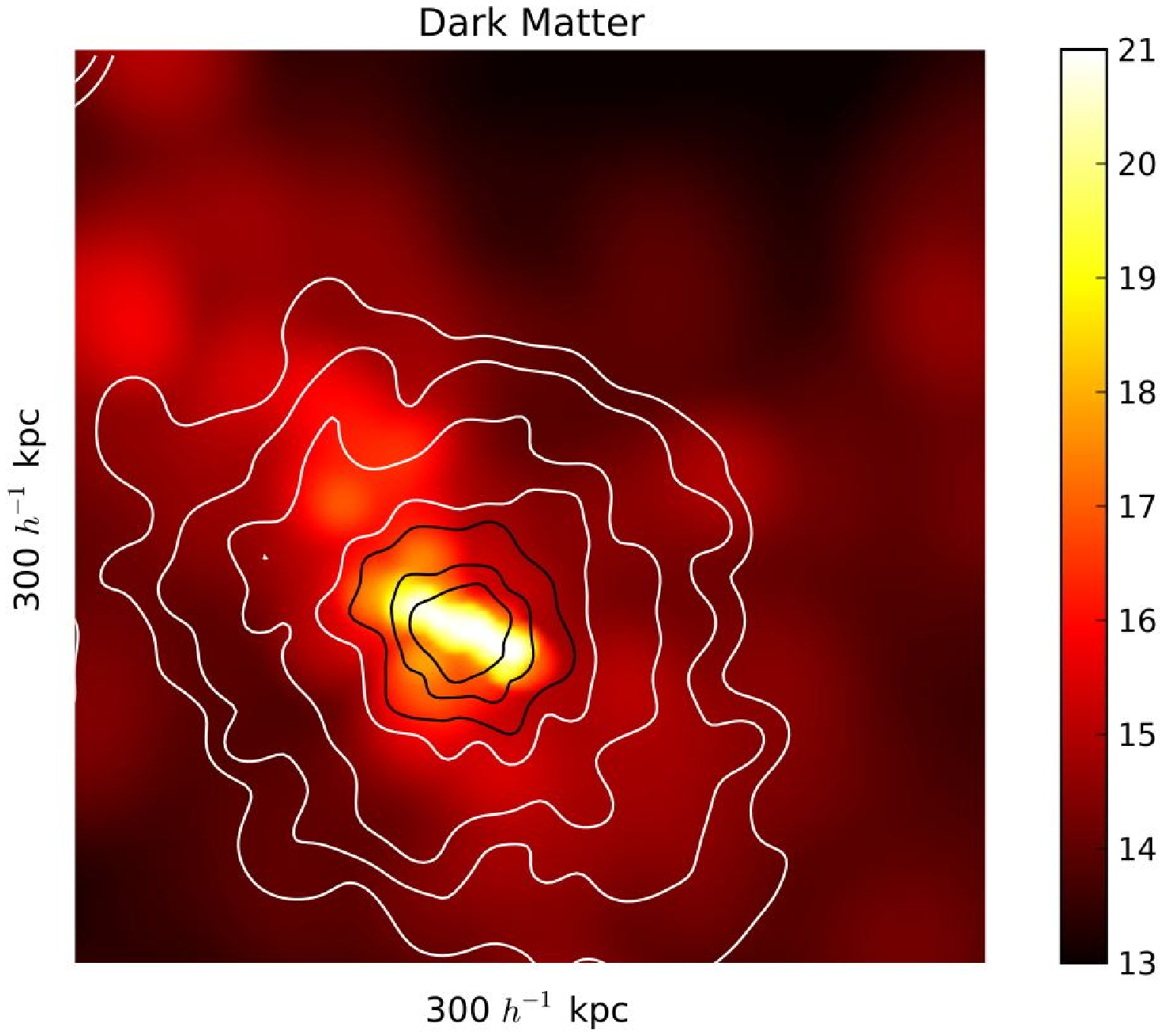}
  \includegraphics[width=0.33\textwidth]{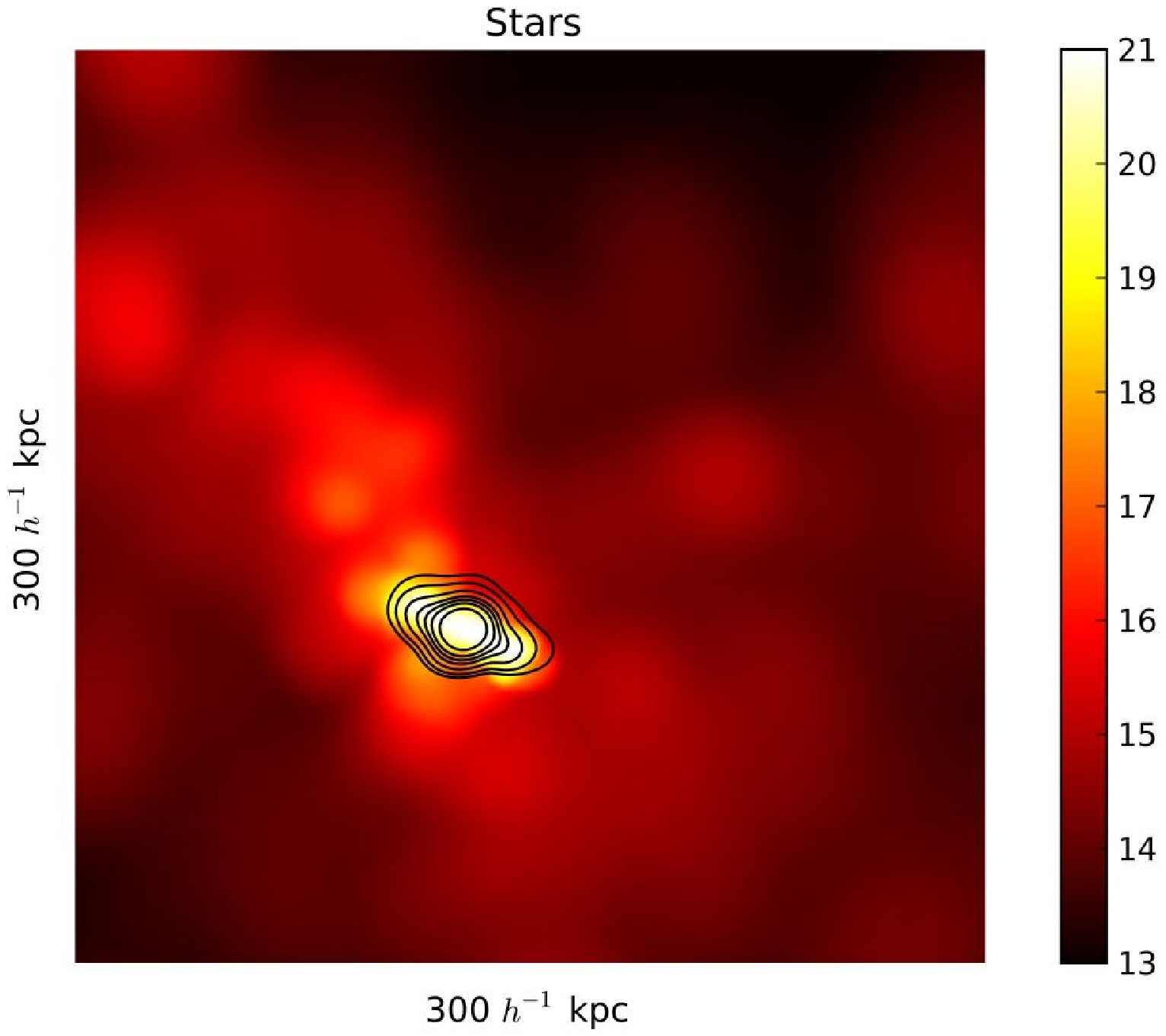}
  \includegraphics[width=0.33\textwidth]{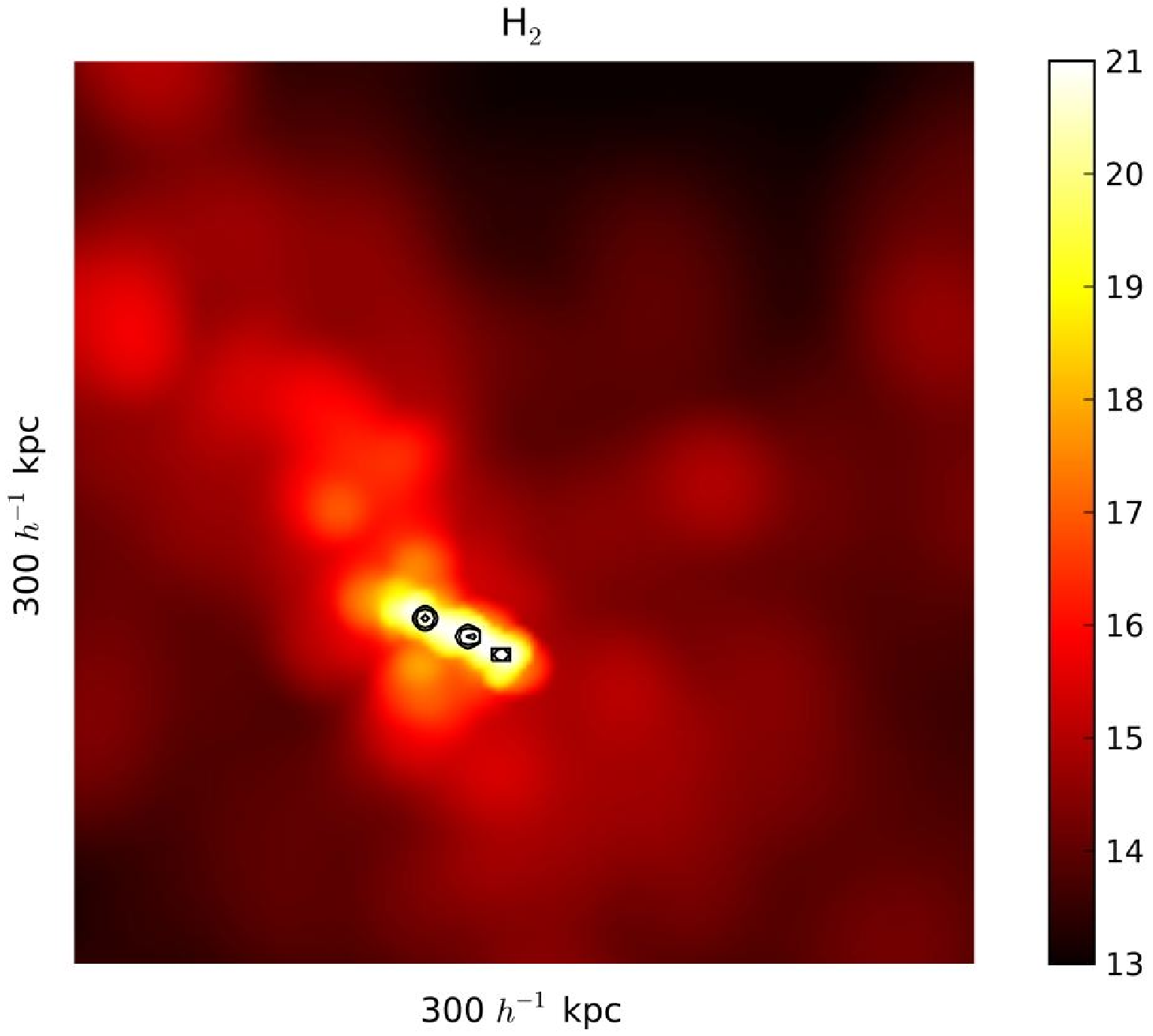}

  \includegraphics[width=0.33\textwidth]{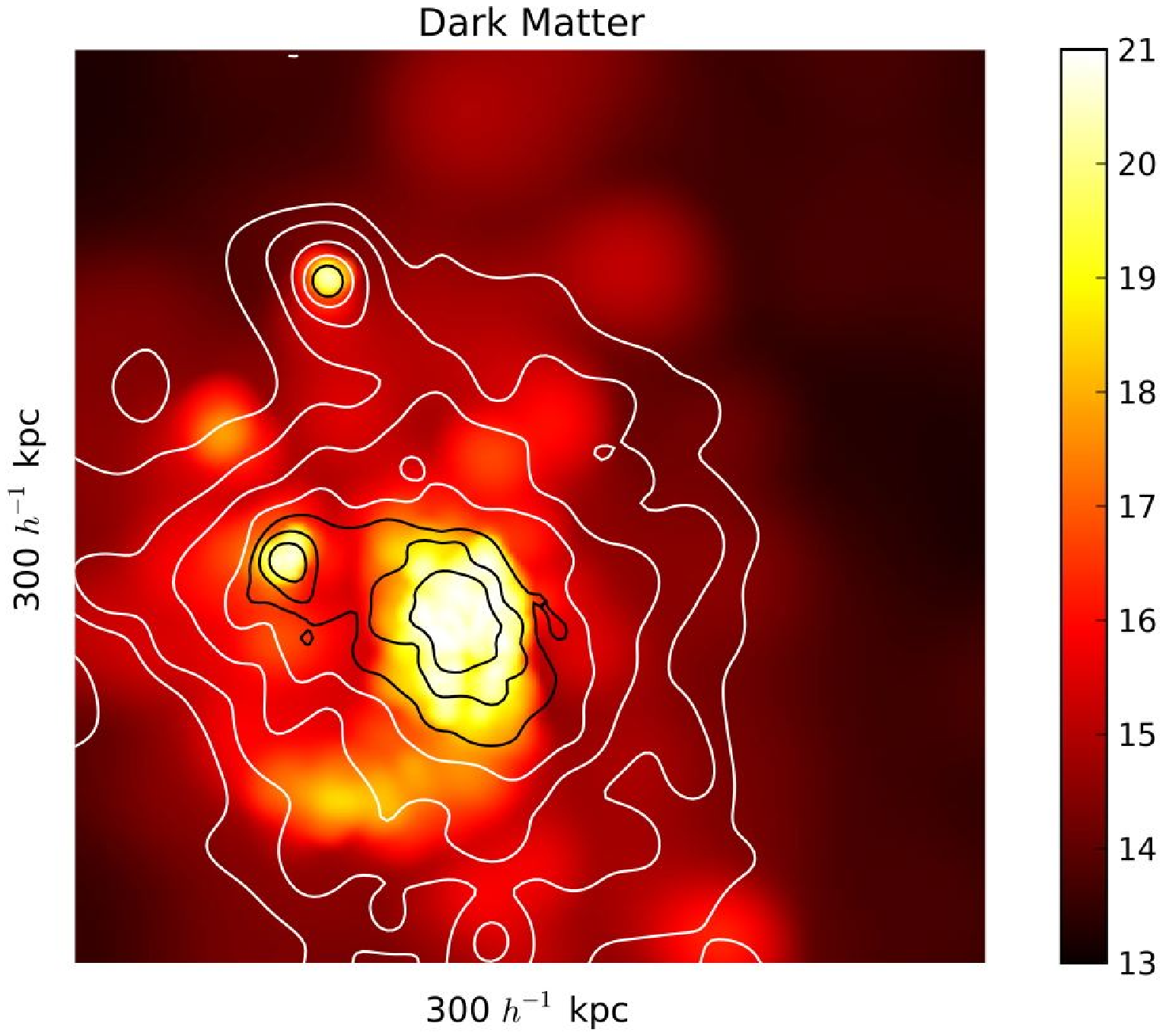}
  \includegraphics[width=0.33\textwidth]{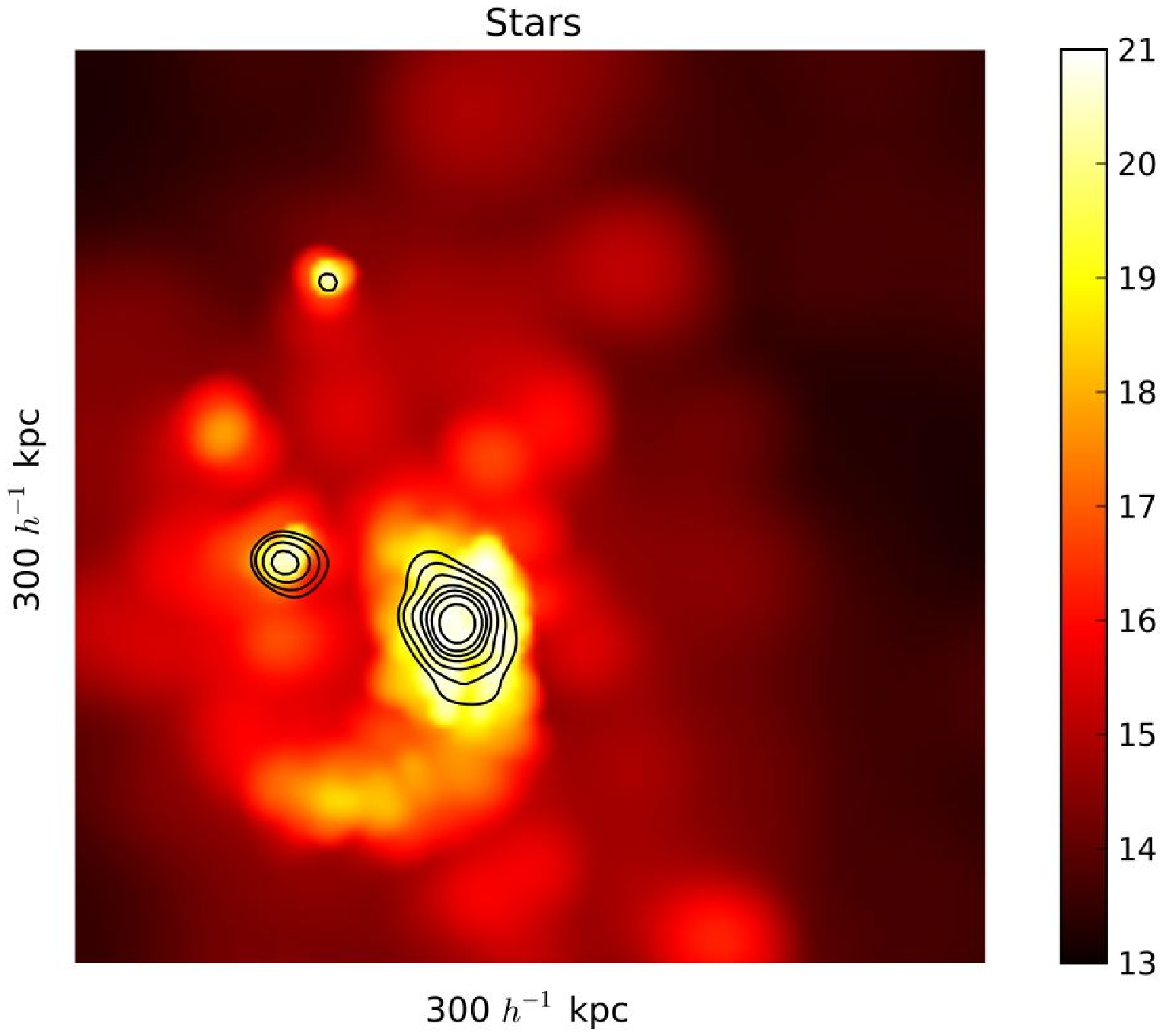}
  \includegraphics[width=0.33\textwidth]{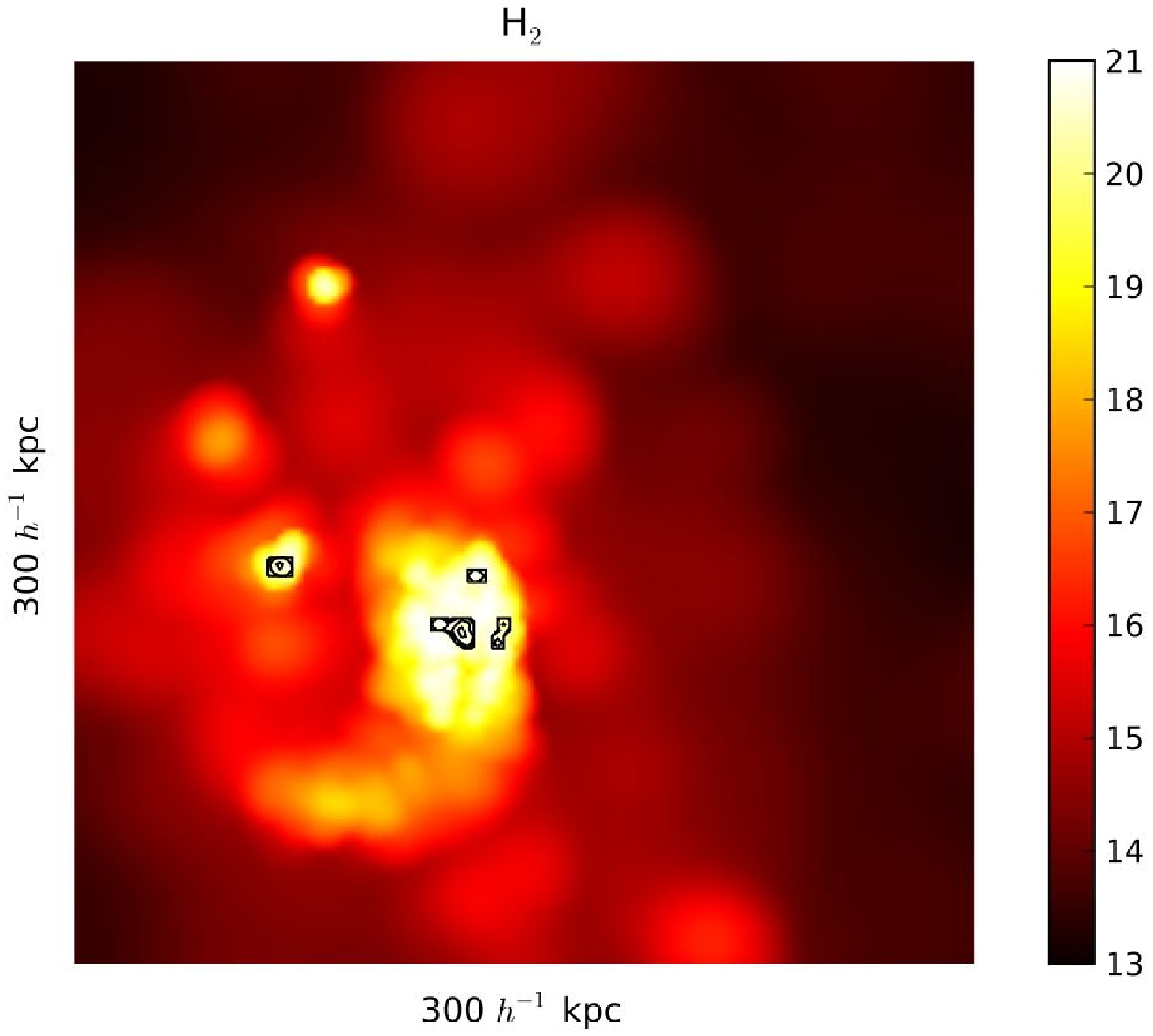}

  \caption{Column density maps of four reconstructed objects as seen
    in neutral hydrogen with contours of Dark Matter (left panels),
    Stars (middle panels) and molecular hydrogen (right
      panels). For both the Dark Matter and the stars contour levels
    are at $N$ = 3, 5, 10, 20, 30, 50 and 100 $\times 10^6$
    $M_{\odot}$ kpc$^{-2}$. For the molecular hydrogen contours
      are drawn at $N_{H_2}= 10^{18}$, $10^{19}$, $10^{20}$ and
      $10^{21}$ cm$^{-2}$. Stars are concentrated in the very dense
    parts of the \hbox{\rm H\,{\sc i}} objects, dark matter is more
    extended, however the extended \hbox{\rm H\,{\sc i}} does not
    always trace the dark matter. The \hbox{\rm H\,{\sc i}}
      satellites or companions are within the same Dark Matter Halo,
      but do not always contain stars.}
  \label{dm_stars}
\end{figure*}

In Figure~\ref{dm_stars} four examples are shown of the dark matter
and stellar distributions overlaid on the of \hbox{\rm H\,{\sc i}}
column density maps. The right panels in this image show the
  contours of molecular hydrogen. The stars are concentrated in the
bright and dense parts of the \hbox{\rm H\,{\sc i}} objects
corresponding to the bulges and disks of galaxies. The third example
shows an edge-on extended gas disk (its thickness is likely an
artifact of our numerical resolution).  The neutral hydrogen is much
more extended than the stars, and the smaller, more diffuse \hbox{\rm
  H\,{\sc i}} clouds do not have a stellar counterpart in
general. Many objects have \hbox{\rm H\,{\sc i}} satellites or
  companions, as in the first two examples. These companions or
  smaller components do not always have a stellar or molecular
  counterpart, although the \hbox{\rm H\,{\sc i}} column densities can
  reach high values up to $N_{HI} \sim 10^{20}$ cm$^{-2}$ as
  seen in Galactic high velocity clouds.

Interestingly, these \hbox{\rm H\,{\sc i}} clouds only occasionally
trace dark matter substructures, hence in many cases they are not
obvious large-scale accretion events (since the most massive accreting
clumps should be accompanied by dark matter).  The origin of these
diffuse clouds, perhaps analogous to high velocity clouds, may be from
a ``halo fountain" of gas cycling in and out of galaxies owing to
galactic outflows, as speculated in \cite{2008MNRAS.387..577O}, or
represent more diffuse accretion from the extended environment.
Studying the kinematics and metallicities of these clouds may reveal
signatures of their origin.  In the future we plan to investigate such
signatures in the simulations to assess how observations of diffuse
\hbox{\rm H\,{\sc i}} clouds around galaxies can inform our
understanding of the processes of galaxy assembly.

\section{Discussion}

To make observational predictions based on numerical simulations, the
first essential step is to establish that the simulation can
adequately reproduce all observational constraints. We have carried
out a critical comparison of our simulated \hbox{\rm H\,{\sc i}} data
with observations using a wide range of statistical
measures. Essential in creating simulated data is the minimization of
the number of free parameters over and above the many that are
already inherent in the simulation;
\citep{2006MNRAS.373.1265O,2008MNRAS.387..577O}. In our analysis, the
only additional assumptions we make are that the transitions from
ionized to atomic and from atomic to molecular hydrogen can be
described by a simple threshold effect at two different values of
the thermal pressure, while demanding that the recombination
time be less than the sound-crossing time. The threshold
values are determined by fixing the average \hbox{\rm H\,{\sc i}}
density and the density ratio $\Omega_{H_2}/\Omega_{HI}$ at
z~=~0 to those determined observationally. In choosing this simple
prescription we are strongly limited by current numerical
capabilities. Although we did not solve the complete radiative
transfer problem, we do get quite complex behaviour emerging. The
range of threshold values we explored are consistent with
expected values in literature.

The statistics of the reconstructed and observed \hbox{\rm H\,{\sc
i}} distributions agree quite well, making it plausible that the
associated \hbox{\rm H\,{\sc i}} structures in the simulation may be
similar to those ocurring in nature. The simulation cannot reproduce
structures that resemble actual galaxies in detail. Besides the finite
mass resolution of the SPH-particles of $\sim 10^7$ $M_\odot$, there are
the inevitable limitations on the included physical processes and their
practical implementation. Nonetheless, we may begin to explore the fate of
partially neutral gas in at least the diffuse outskirts of major galaxies.

Despite the limitations, the simulations can reproduce many observed
statistical aspects of \hbox{\rm H\,{\sc i}} in galaxies, which
is very encouraging for further exploration of this approach. The
adopted self-shielding threshold provides good results for this one
simulation. Future work will test the variations that are encountered
with different feedback mechanisms. The method can also be applied to
look at the evolution of neutral hydrogen with redshift. Furthermore,
mock observations can be created to make predictions for what future
telescopes should detect. This is particularly relevant for assessing
performance requirements for the various facilities now under development
with a strong \hbox{\rm H\,{\sc i}} focus, such as the Square Kilometre Array
(SKA) and many of the SKA pathfinders.

\subsection{Resolution Effects}
Simulations are limited by their volume and the mass resolution
  of the particles (over and above the limitations that result from
  incomplete physics). Although we are able to reconstruct structures
  of several Mpc in scale, the simulated volume is relatively
  small. To be able to reconstruct the largest structures encountered
  in the universe, and effectively overcome cosmic variance, a
  simulated volume is needed that is about about 300 (rather than 32)
  Mpc on a side. In the current volume, the most massive structures
  are suffering from low number statistics. A drawback of using a
  larger volume is that the mass resolution of individual particles
  decreases rapidly, making it impossible to resolve structures on
  even multi-kiloparsec scales. The approach we have employed to
  approximate the effects of self-shielding is extremely simple. The
  actual processes acting on sub-kiloparsec scales are undoubtedly
  much more complex. Clumping will occur on the scales of molecular
  clouds, which will dramatically increase the local densities. The
  threshold thermal-pressure we determine to approximate the atomic to
  molecular hydrogen transition only has possible relevance on the
  kiloparsec scales of our modelled gas particles. At the smaller
  physical scales where the transition actually occurs, the physical
  pressures will likely be substantially different. 

  We note that realistic simulations of cosmological volumes are
  extremely challenging, and that even the current state-of-the-art is
  not particularly successful at reproducing objects that resemble
  observed galaxies in great detail. The simulation we employ
  represents a very good compromise between simulations focused on
  larger and smaller scales. We are mainly interested in the diffuse
  intergalactic structures on multi-kiloparsec scales, that would be
  observable when doing 21cm \hbox{\rm H\,{\sc i}} observations with
  sufficient sensitivity. We have enough resolution to map and resolve
  these structures and to reconstruct the extended environments of
  galaxies and galaxy groups. Our simulation is not suitable for
  making predictions about the inner cores of galaxies or resolving
  the star formation process in molecular clouds.

Future work will test our analysis method on both larger and
smaller scales. Simulations in a larger volume can provide a more
sensitive test of the reconstructed \hbox{\rm H\,{\sc i}} Mass
function and the two-point correlation function. Simulations in a
smaller volume will probably require a more advanced method of
modelling the atomic to molecular hydrogen transition. However,
substantial insights into the more diffuse phenomena, such as accretion
and feedback processes around individual galaxies, are very likely
within reach.

\subsection{Future Observations}

This simulation can not only be used for getting a better understanding
of the \hbox{\rm H\,{\sc i}} distribution, especially at low column
densities, but is also very suitable for making predictions for future
observations. Currently many radio telescopes are being built as
pathfinders toward the Square Kilometre Array (SKA). A few examples are
the Australia SKA Pathfinder (ASKAP), the Allen Telescope Array (ATA),
Karoo Array Telescope (MeerKAT) and the Low frequency array (LOFAR),
and many more. Although the SKA will be the final goal, each of these
telescopes is a good detector on its own and is planned to be operational
in the relatively near future. Of course each telescope will have different
characteristics, but in general it will be possible to do surveys deeper,
faster and over a broader bandwidth.

We will discuss the simulated maps of two current single dish telescopes,
Parkes and Arecibo, and compare these with the capabilities of two future
telescopes, the ASKAP and the SKA. Parkes and Arecibo are both single
dish telescopes with a multi-beam receiver that have recently been used
to do large area surveys, the \hbox{\rm H\,{\sc i}} Parkes All Sky Survey
(HIPASS, \cite{2001MNRAS.322..486B}) and the Arecibo Legacy Fast ALFA
Survey (ALFALFA \cite{2005AJ....130.2598G} ).

To make a fair comparison, all four telescope will get 500 hours of
observing time to map 30 square degrees of the sky. These numbers are
chosen as 500 hours is a reasonable timescale to make a deep image and
a 30 square degree field is needed to map the extended environment of
a galaxy. Furthermore, 30 square degrees is the planned instantaneous
field-of-view of ASKAP.

We focus on two cases that illustrate the capabilities of present
and upcoming telescopes.  In the first example, observations will be
simulated at a distance where the beam of the telescopes has a physical
size of 25 kpc. This approximate beam size is needed to resolve diffuse
filaments and extended companions from the simulation.  In the second
example observations will be simulated at a fixed distance of 6 Mpc,
which is the limiting distance for the Parkes telescope according to
the above argument.

Right ascension and declination coordinates are added according to the
distance, centered on an RA of 12 hours and a Dec of 0 degrees. For
each telescope the sensitivity is determined that can be achieved
using the given conditions. The technical properties are listed in
Table~\ref{mock_data}. For the Parkes telescope we use the sensitivity
that is achieved when re-reducing HIPASS data (Popping 2009 in prep.),
which corresponds to $\sim 8$mJy/Beam over 26 km s$^{-1}$ for a
typical HIPASS field after integrating over $\sim 560$s per square
degree. For the Arecibo telescope we used the sensitivity of the
Arecibo Galaxy Environment Survey (AGES) \citep{2006MNRAS.371.1617A}
assuming 0.95 mJy/beam over 10 km s$^{-1}$ after integrating for 10
hours per square degree. The expected sensitivities for ASKAP are
described in the initial Array Configuration paper which can be found
on {\it
http://www.atnf.csiro.au/projects/askap/newdocs/configs-3.pdf}. ASKAP
is expected to achieve a sensitivity of 7.3 mJy/beam over 21 km
s$^{-1}$ after one hour of integration time. For the SKA we assume a
$A_{eff}/T_{sys} = 2000$ m$^2$ K$^{-1}$ at 2.5 arcmin resolution,
which is 20 times higher than ASKAP. Furthermore we assume a field of
view similar to ASKAP, 30 square degrees instantaneous.

All the flux densities can be converted into brightness temperature using the equation:
\begin{equation}
T_b = \frac{\lambda^2S}{2k\Omega}
\end{equation}
where $\lambda$ is the observed wavelength, $S$ is the flux density,
$k$ the Boltzmann constant and $\Omega$ is the beam solid angle of the
telescope. When using the 21 cm line of \hbox{\rm H\,{\sc i}}, this equation can be written as:
\begin{equation}
T_b = \frac{606}{b_{min}b_{maj}}S
\end{equation}
where $b_{min}$ and $b_{maj}$ are respectively the beam minor and
major axis in arcsec and $S$ is the flux in units of mJy/Beam. The
integrated 21cm line intensity can directly be converted into an \hbox{\rm
H\,{\sc i}} column density using:
\begin{equation}
N_{HI} = 1.823 \cdot 10^{18} \int T_b dv
\end{equation}
with $[N_{HI}]$ = cm$^{-2}$, $[T_b]$ = K and $[dv]$ = km s$^{-1}$.\\

\begin{figure*}[t!]
 
  \includegraphics[angle=270,width=0.33\textwidth]{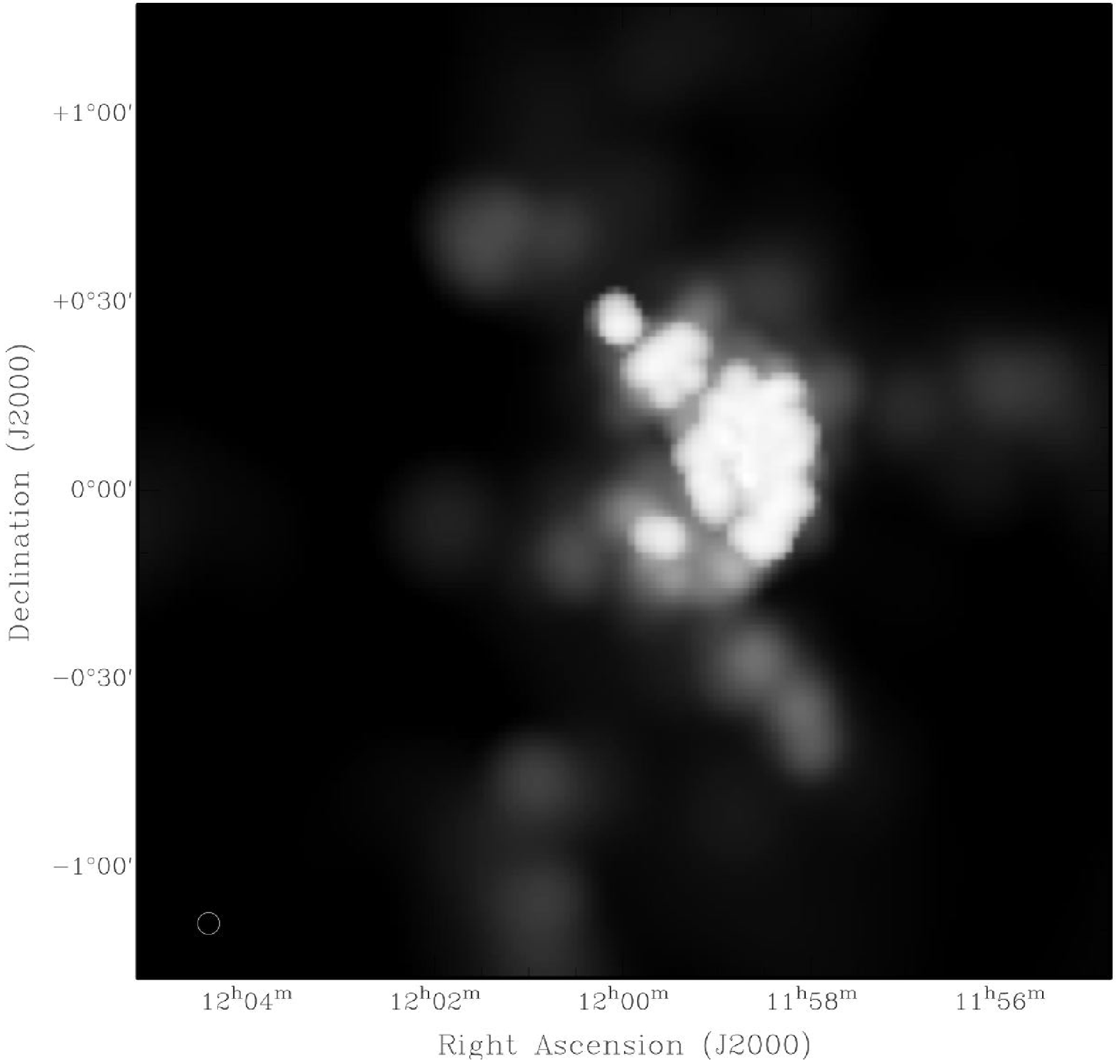}
  \includegraphics[angle=270,width=0.33\textwidth]{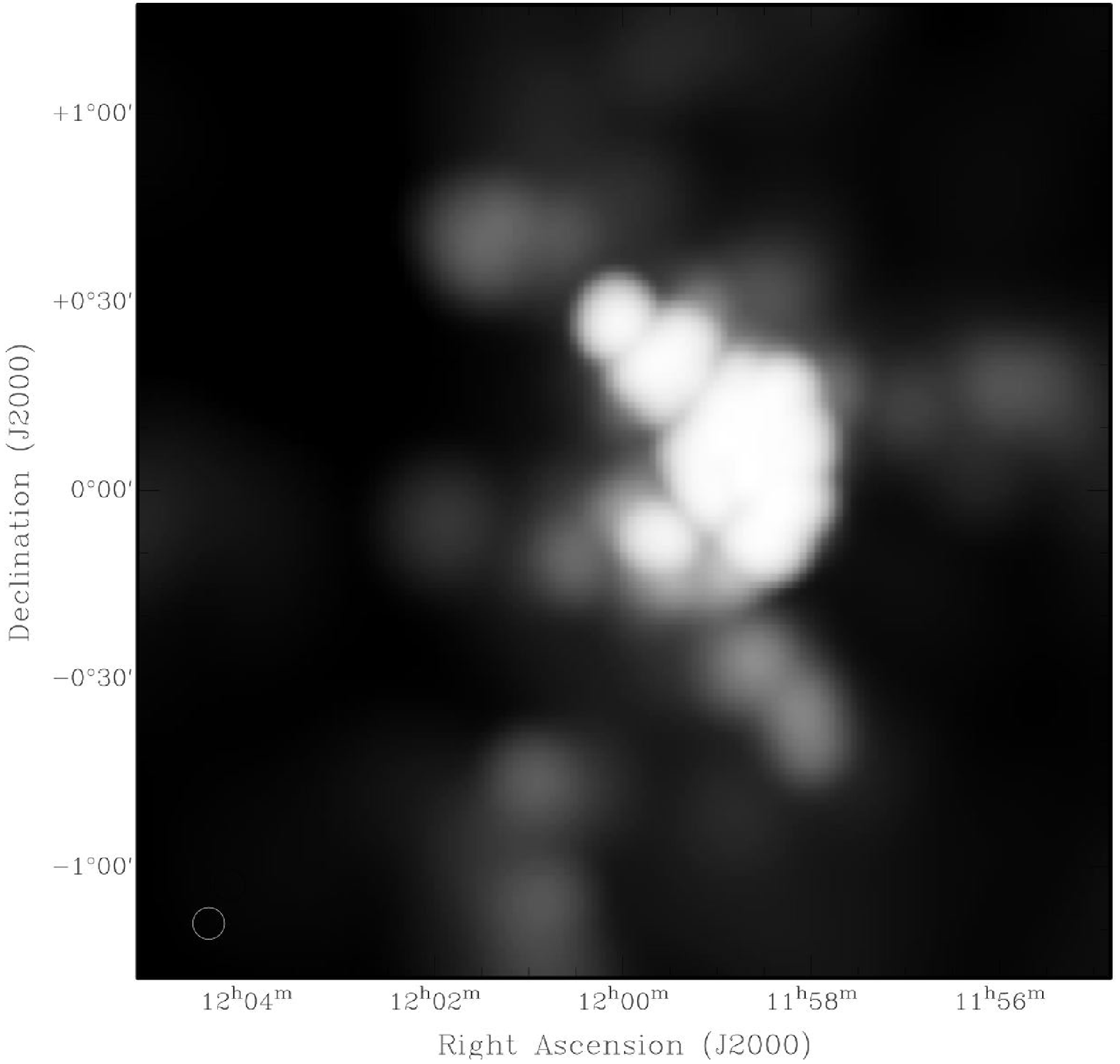}
  \includegraphics[angle=270,width=0.33\textwidth]{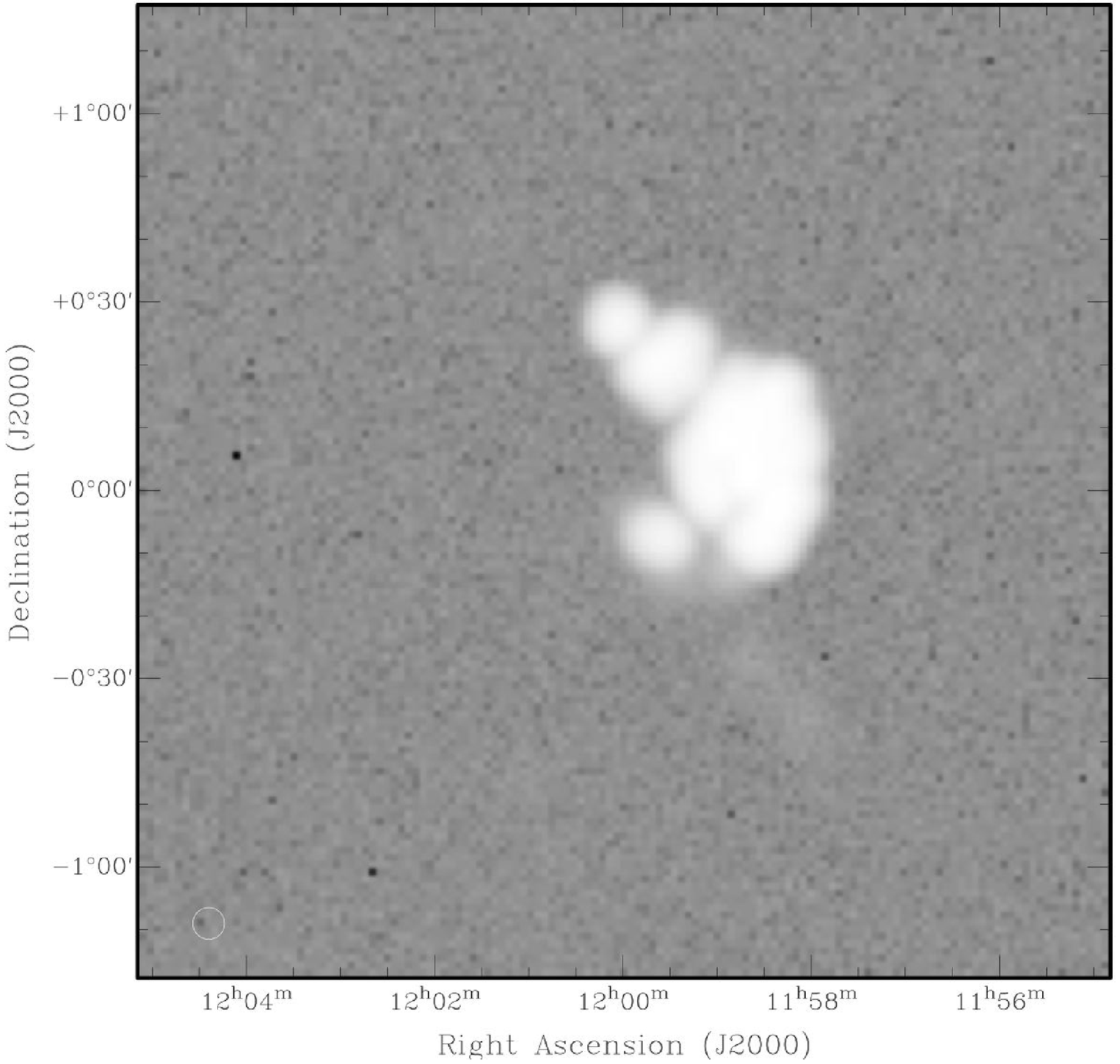}

  \caption{Left panel: simulated object at a distance of 6 Mpc, with a
    6 kpc intrinsic beam size. Middle panel: simulated object smoothed
    to a 8.7 kpc beam size, corresponding to the ASKAP beam at 6
    Mpc. Right panel: noise is added corresponding to the ASKAP
    sensitivity limit after a 500 hour observation of 30 square degrees.}
  \label{mock}
\end{figure*}

The second column in Table~\ref{mock_data} gives the beam size of each
telescope, the third column gives the distance at which the beam has
a physical size of 25 kpc, and the fourth column gives the physical
beam size at a distance of 6 Mpc. The sensitivities in column five are
converted to a column density limit when sampling a line of approximately
25 km s$^{-1}$ width in the last column. We assume that in any analysis
only the channels will be selected containing diffuse emission and
that the line width of diffuse regions will be of the order of 25 km
s$^{-1}$. Galaxies can have a much larger linewidth, however detecting
these is not an observational challenge. The reconstructed maps have an
intrinsic resolution of 2 kpc with a minimum smoothing kernel size of
three pixels. This yields an initial beam size of 6 kpc, which will be
smoothed to the appropriate beam size of each telescope, corresponding to
the simulated distance. Using the calculated sensitivity limits random
gaussian noise is generated and added to the maps. The steps are shown
in Figure~\ref{mock}, the left panel shows the original reconstructed
column density map that is smoothed to the beam of ASKAP in the middle
panel. At this stage most of the diffuse and extended emission can still
be recognized. Noise is added in the right panel, most of the diffuse
emission disappears in the noise.

\begin{table*}[t]
\begin{center}
\begin{tabular}{lccccc}
\hline
\hline
Telescope & Beam (arcmin) & $D$ (beam 25 kpc) & Beam at $D=6$ Mpc & RMS (25 km s$^{-1}$) & $N_{HI}$ (25 km s$^{-1}$) \\
\hline

Parkes  & 14.4'  & 6  Mpc & 25 kpc & 0.9   mJy/Beam  &  $3.3\cdot10^{16}$ cm$^{-2}$  \\
Arecibo & 3.5' & 25 Mpc & 6.1 kpc & 0.45  mJy/Beam  &  $2.9\cdot10^{17}$ cm$^{-2}$\\
ASKAP   & 3' & 28 Mpc & 5.2 kpc & 0.3  mJy/Beam  &  $2.6\cdot10^{17}$ cm$^{-2}$\\
SKA     & 2.5'  & 33 Mpc & 4.4 kpc & 0.015 mJy/Beam  &  $1.8\cdot10^{16}$ cm$^{-2}$\\

\hline
\hline 
\end{tabular}
\end{center}
\caption{Sensitivity limits for four different telescopes 
  for an assumed linewidth of 25 km s$^{-1}$ after a total integration time of
  500 hours to image an area of 30
  square degrees.}
\label{mock_data}
\end{table*}

\begin{figure*}[t!]
 
  \includegraphics[angle=270,width=0.5\textwidth]{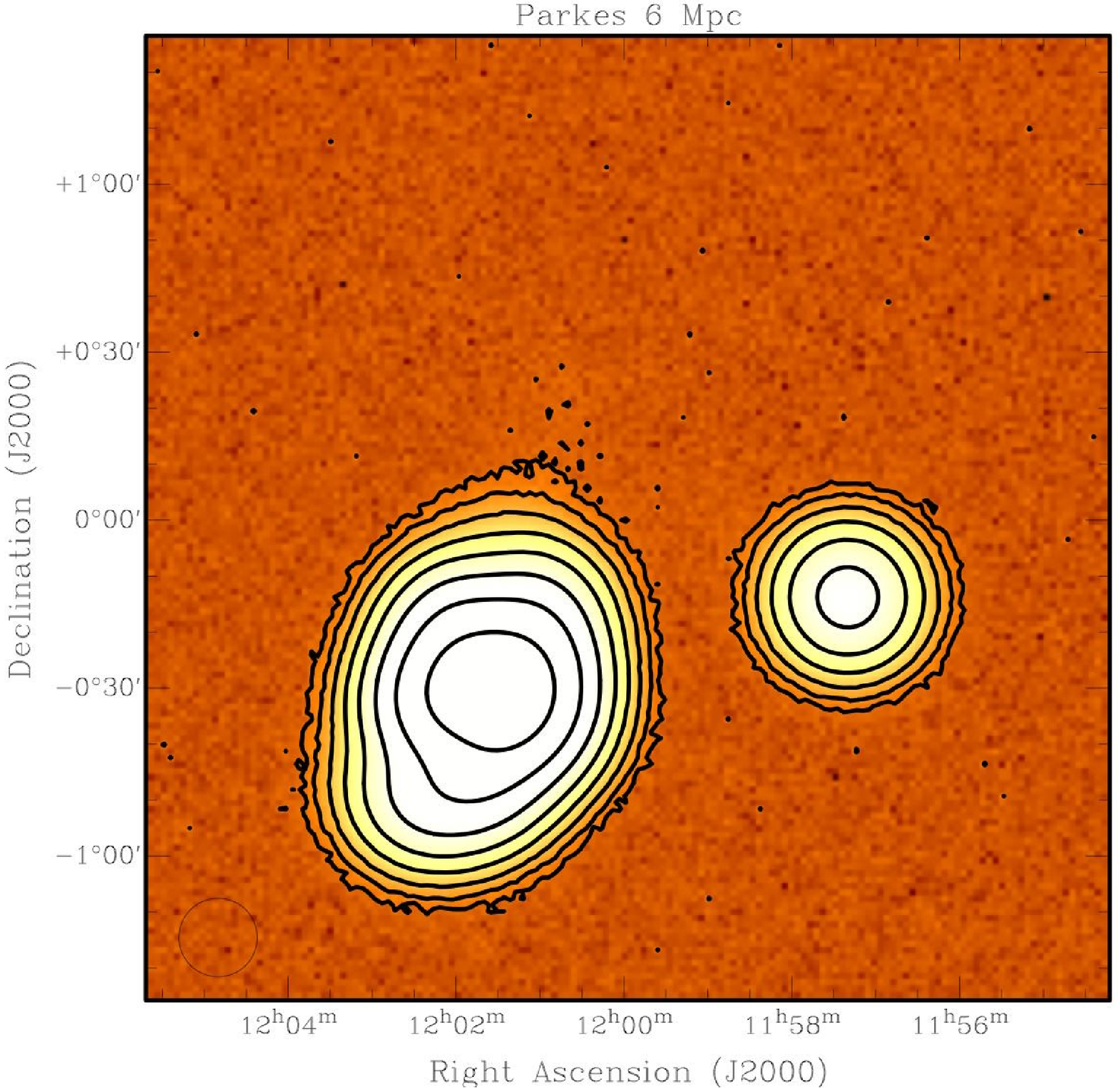}
  \includegraphics[angle=270,width=0.5\textwidth]{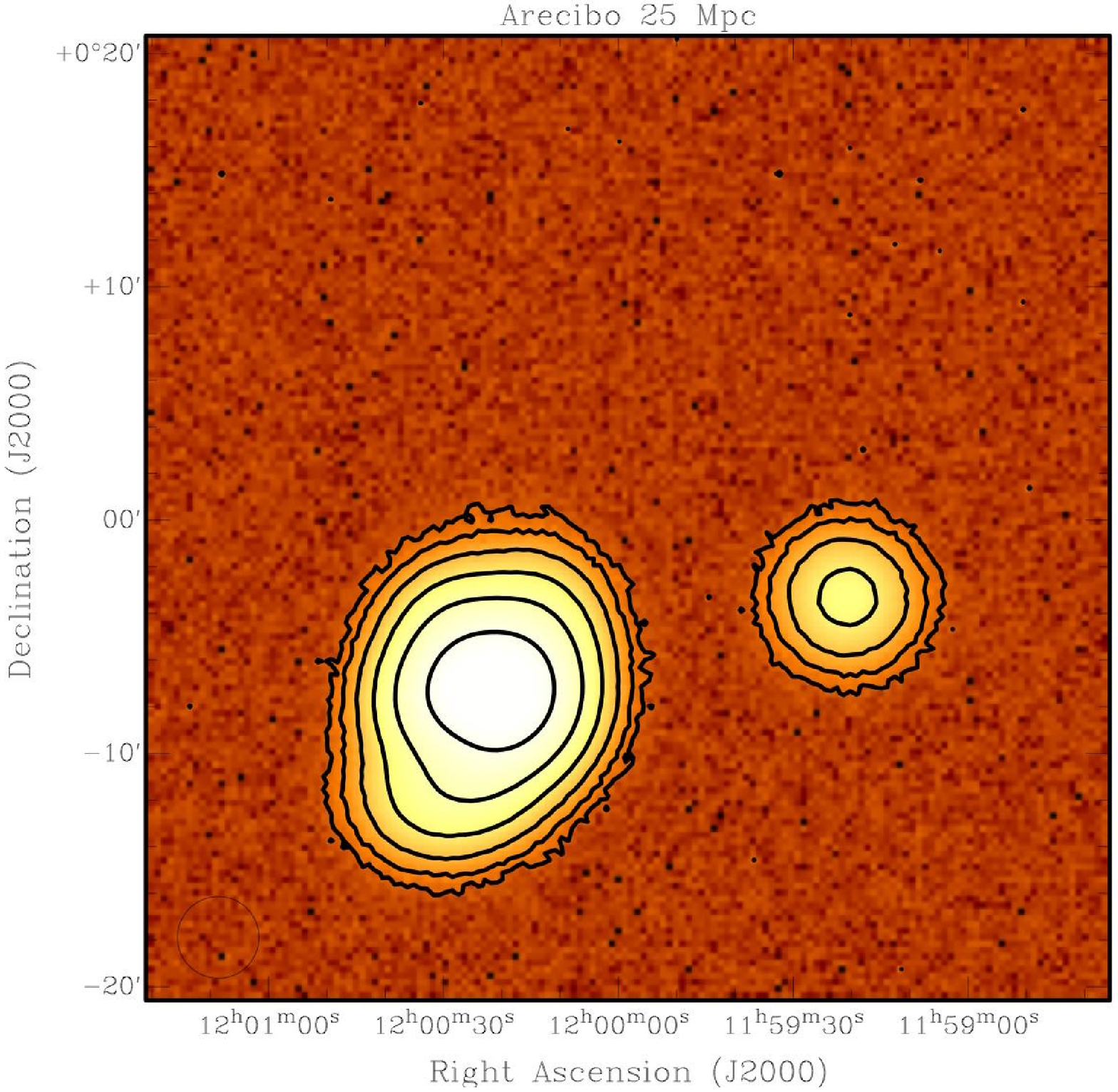}

  \includegraphics[angle=270,width=0.5\textwidth]{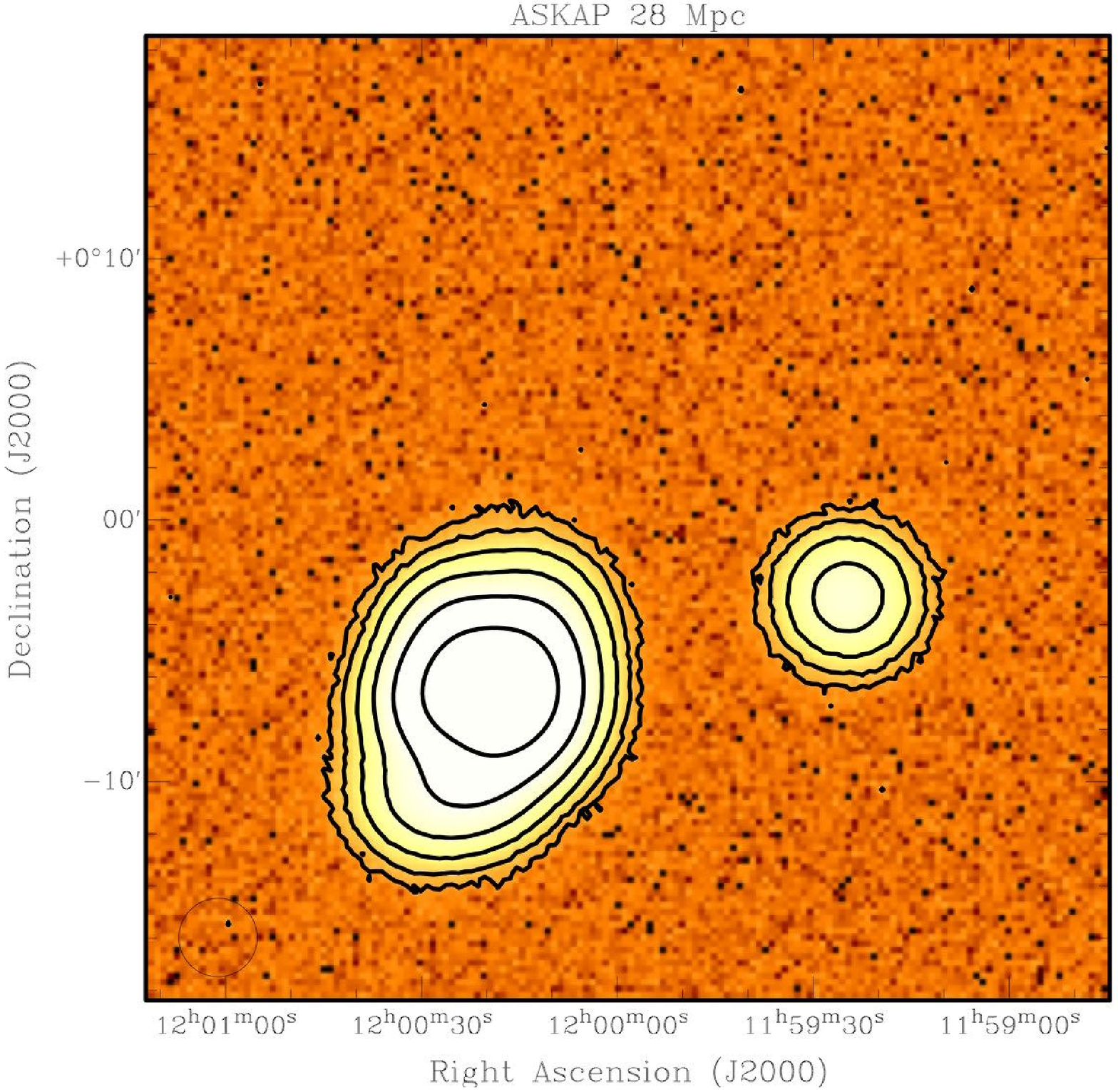}
  \includegraphics[angle=270,width=0.5\textwidth]{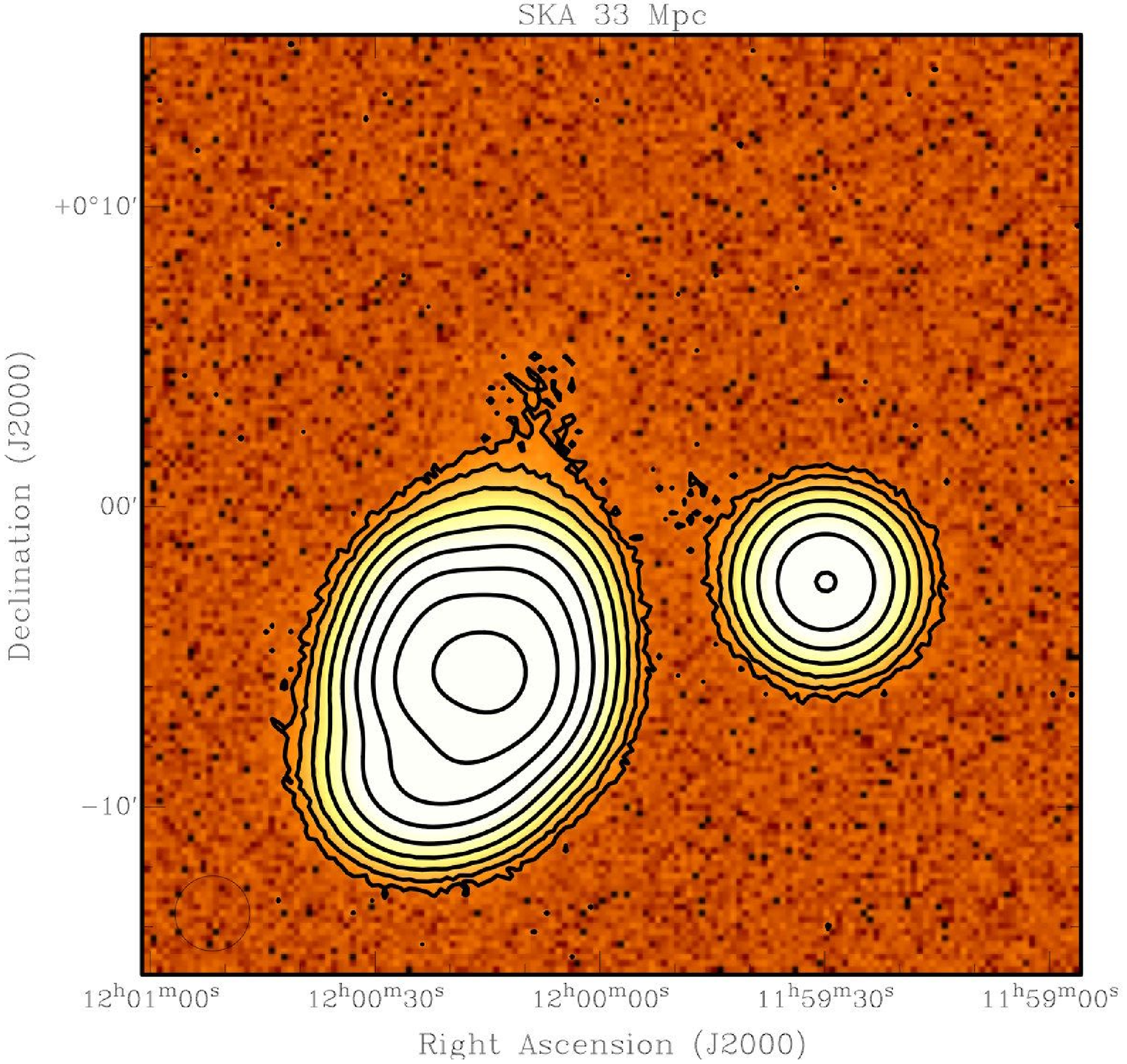}

  \caption{Simulated observations of the same object at a distance at
    which the beam size corresponds to 25 kpc for Parkes (top left),
    Arecibo (top right), ASKAP (bottom left) and the SKA (bottom
    right). All contours begin at a 3 $\sigma$ level, after
    a 500 hour observation of a 30 square degree field with each
    telescope. Every subsequent contour level is a factor 3 higher than the
    previous one. Note that the angular scale is different in each
    panel.}
  \label{mock25}
\end{figure*}

\begin{figure*}[t!]
 
  \includegraphics[angle=270,width=0.5\textwidth]{figures/parkes_6.eps}
  \includegraphics[angle=270,width=0.5\textwidth]{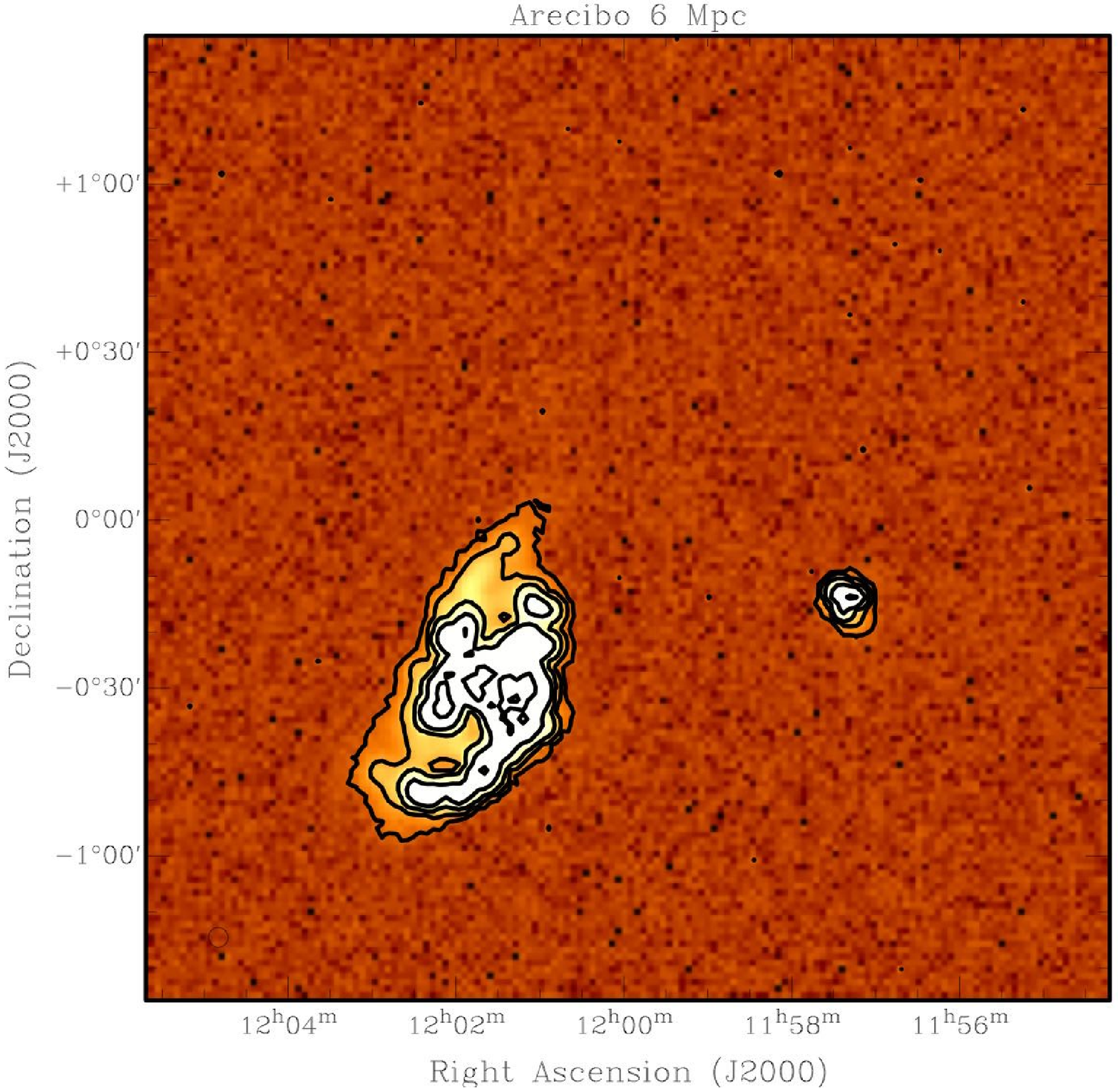}

  \includegraphics[angle=270,width=0.5\textwidth]{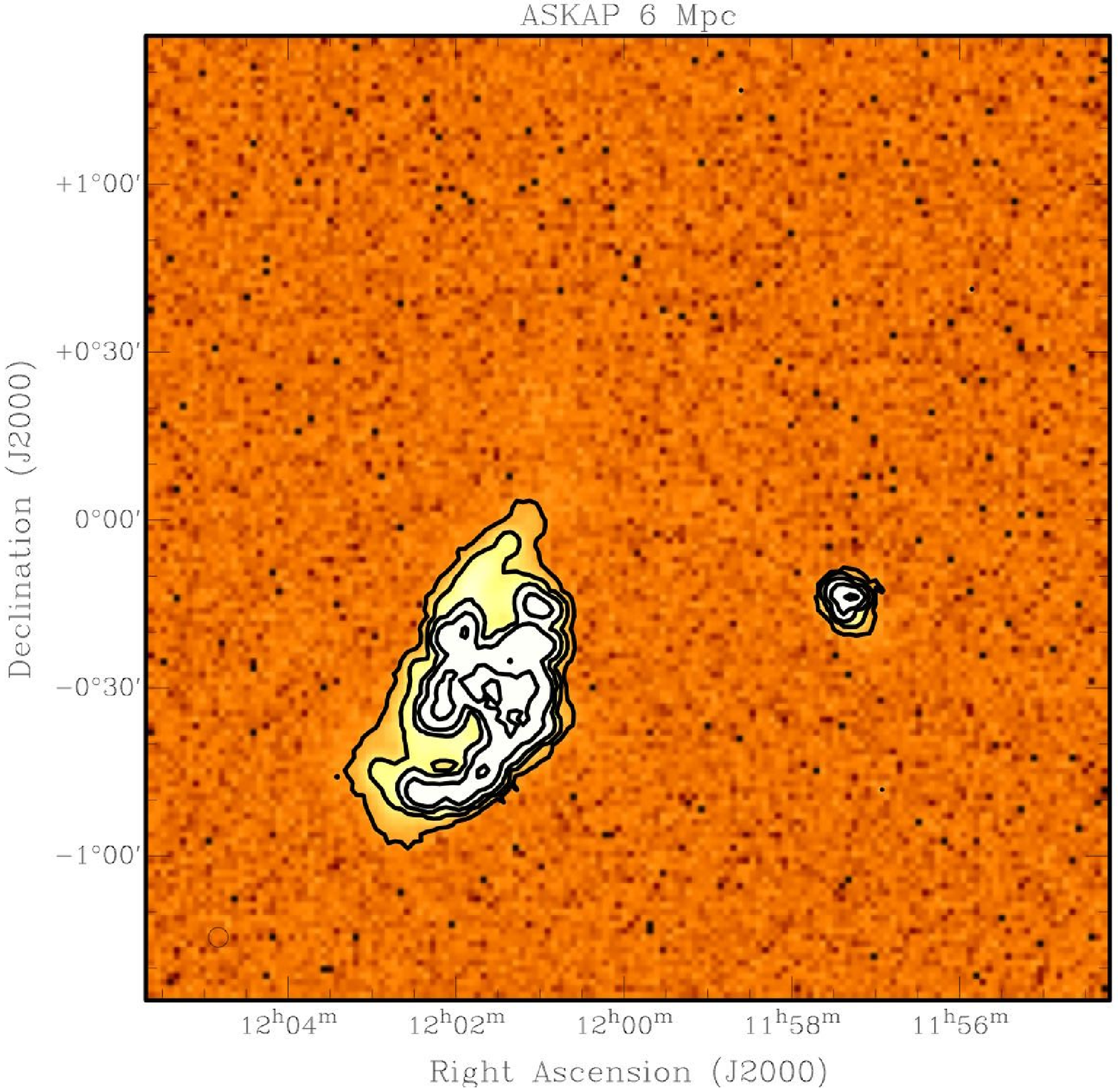}
  \includegraphics[angle=270,width=0.5\textwidth]{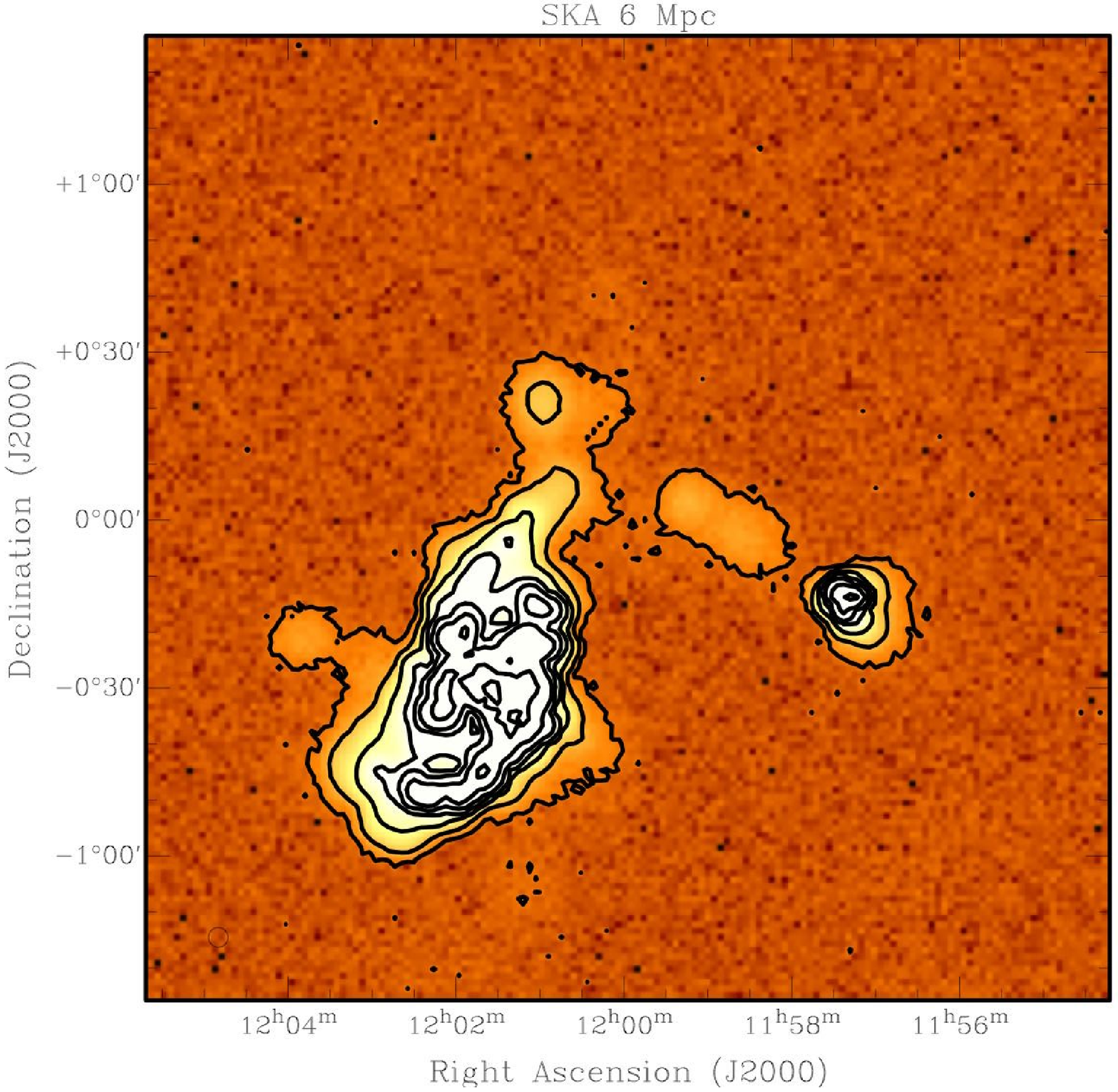}

  \caption{Simulated observations of the same object at a fixed
    distance of 6 Mpc for Parkes (top left), Arecibo (top right),
    ASKAP (bottom left) and the SKA (bottom right). Contour
    levels start at a 3 $\sigma$ level after 500 hour
    observation of a 30 square degree field. For Parkes every subsequent
    contour is a factor 3 higher than the previous one. For ASKAP and
    Arecibo, the contours interval is a factor 7. For the SKA the
    contours start a $4\cdot 10^{16}$ cm$^{-2}$ and increase by a factor
    6. Parkes is not really competitive in detecting substructures at
    this distance. The other telescopes all have sufficient
    resolution, however only the SKA is sensitive enough to detect the
    faint diffuse sub-structures.}
  \label{mock6}
\end{figure*}

In Figure~\ref{mock25} the same field is shown as in Figure~\ref{mock},
placed at a distance where the beam has a size of 25 kpc. This means
that the object looks similar for all four telescopes as it fills the
same number of beams, although there is a big difference in angular scale. 

For each panel contour levels are drawn starting at 3 $\sigma$
and increasing as noted in the figure caption, so the actual
values are different in each panel, but can be determined using the
sensitivity limits given in table.~\ref{mock_data}. All panels look
very similar, as most of the structure and substructure of the
original map is smoothed into two large blobs and essentially all the
diffuse structures are lost in the noise. Only with the SKA can some
extended contours still be recognized at the top of the left
object. However, it is very difficult to distinguish extended emission
from companions, unless the companions are clearly separated by at
least a beam width from the primary object.

An example of this can be seen in Figure~\ref{mock6} where the same
object is shown simulated with four telescopes but now at a similar
distance of 6 Mpc. A smaller beam size now yields higher resolution
and more detected structure. The differences between the four panels
are now obvious. Observed with Parkes in the top left panel the object
has essentially no resolved structure. Clearly the beam size is too
large and only suitable for very nearby galaxies, closer than 6
Mpc. Arecibo, with a much smaller beam, can resolve the inner core of
the object and can just detect the brightest companion. Again the
contour levels have values starting at 3 $\sigma$, so
the outer contour in the top right panel corresponds to
$9\cdot10^{17}$ cm$^{-2}$. ASKAP has a very similar sensitivity
limit as Arecibo with a comparable beam size. However it is notable
that it reaches Arecibo sensitivities with a much smaller collecting
area. Furthermore, these deep integrations have not really been
explored, so the given sensitivities are theoretical limits. It is
very likely that correcting for systematic effects, like the shape of
the spectral bandpass, will be more achievable with an interferometer
like ASKAP rather than a large single dish telescope. In the SKA image
essentially all companions can be clearly distinguished down to a
contour level of $\sim4\cdot10^{16}$ cm$^{-2}$. Note that this value
is lower than the 3$\sigma$ value in Table~\ref{mock_data}, this is
because the beam size of the SKA is smaller than the beam size of the
simulation at a distance of 6 Mpc. To adjust for this we adopt the
beam size of the reconstructed data and smooth the noise to this
larger beam size. In this case the noise value will decrease,
resulting in a higher sensitivity of $1.3\cdot10^{16}$ cm$^{-2}$.

\section{Conclusion}

We have used a hydrodynamic simulation to predict the neutral
hydrogen distribution in the universe. The simulation employs a random
cube of 32 $h^{-1}$ Mpc on a side at redshift zero, with an SPH mass
resolution of $\sim 10^7$ $M_\odot$. The physics in the simulation
includes subgrid treatments of star formation and feedback
mechanisms. 

We have developed a method to extract the neutral hydrogen component
from the total gas budget. At low volume densities the balance is
calculated between photo-ionization and radiative recombination. For
high densities a correction has to be applied for self-shielding, as
the gas becomes optically thick for ionizing photons. In the densest
regions, the atomic hydrogen is turned into molocular hydrogen that
will subsequently form stars. The molecular hydrogen and the
self-shielding transition are both modelled by a critical thermal
pressure. Above the first threshold limit ($P/k = 155$ cm$^{-3}$ K),
the gas is assumed to recombine and the neutral fraction is set to
unity. At even higher pressures ($P/k = 810$ cm$^{-3}$ K), the atomic
hydrogen is assumed to become molecular hydrogen, so the atomic
fraction becomes zero. These processes only apply to simulated
particles for which the recombination time is shorter than the
sound-crossing time on kpc scales. The two threshold pressures are
tuned to reproduce the observed average \hbox{\rm H\,{\sc i}} density
of $\bar{\rho}_{HI} = 6.1 \times 10^7 h$ $M_{\odot}$ Mpc$^{-3}$ and an
assumed molecular density of $\bar{\rho}_{H_2} = 1.8 \times 10^7 h$
$M_{\odot}$ Mpc$^{-3}$, corresponding to a molecular to atomic density
ratio of $\eta_{z=0}=0.3$.

 A wide range of statistical comparisons have been made between
the reconstructed \hbox{\rm H\,{\sc i}} distribution and existing
observational constraints including: the two-point correlation function,
the \hbox{\rm H\,{\sc i}} mass function and the \hbox{\rm H\,{\sc i}}
column density distribution. There is agreement between all these
statistical measures of the observations and the simulations, which
is a very encouraging result. Based on this agreement, the simulated
\hbox{\rm H\,{\sc i}} distribution may be a plausible description of
the \hbox{\rm H\,{\sc i}} universe, at least on the intermediate spatial
scales that are both well-resolved and well-sampled. 

We also compare the distribution of neutral and molecular
hydrogen with the distribution of dark matter and stars in the
simulation. Massive \hbox{\rm H\,{\sc i}} structures generally have
associated stars, but the more diffuse clouds do not contain large
stellar components or in many cases even concentrations of dark
matter. The method to extract neutral hydrogen from an SPH output cube
can be applied to other simulations, to allow comparison of different
models of galaxy formation. Furthermore, the results can be used to
create mock observations and make predictions for future observations.
This preliminary study shows that as \hbox{\rm H\,{\sc i}}
observations of diffuse gas outside of galactic disks continue to
improve, simulations will play a vital role in guiding and
interpreting such data to help us better understand the role that
\hbox{\rm H\,{\sc i}} plays in galaxy formation.

\begin{acknowledgements}
We would like to thank Thijs van der Hulst for useful discussions and
comments on the original manuscript. We appreciate the constructive
  comments of the anonymous referee.
\end{acknowledgements}

\bibliographystyle{aa}
\bibliography{names,bibliography}

\begin{thebibliography}{56}
\expandafter\ifx\csname natexlab\endcsname\relax\def\natexlab#1{#1}\fi

\bibitem[{{Auld} {et~al.}(2006){Auld}, {Minchin}, {Davies}, {Catinella}, {van
  Driel}, {Henning}, {Linder}, {Momjian}, {Muller}, {O'Neil}, {Sabatini},
  {Schneider}, {Bothun}, {Cortese}, {Disney}, {Hoffman}, {Putman}, {Rosenberg},
  {Baes}, {de Blok}, {Boselli}, {Brinks}, {Brosch}, {Irwin}, {Karachentsev},
  {Kilborn}, {Koribalski}, \& {Spekkens}}]{2006MNRAS.371.1617A}
{Auld}, R., {Minchin}, R.~F., {Davies}, J.~I., {et~al.} 2006, \mnras, 371, 1617

\bibitem[{{Barnes} {et~al.}(2001){Barnes}, {Staveley-Smith}, {de Blok},
  {Oosterloo}, {Stewart}, {Wright}, {Banks}, {Bhathal}, {Boyce}, {Calabretta},
  {Disney}, {Drinkwater}, {Ekers}, {Freeman}, {Gibson}, {Green}, {Haynes}, {te
  Lintel Hekkert}, {Henning}, {Jerjen}, {Juraszek}, {Kesteven}, {Kilborn},
  {Knezek}, {Koribalski}, {Kraan-Korteweg}, {Malin}, {Marquarding}, {Minchin},
  {Mould}, {Price}, {Putman}, {Ryder}, {Sadler}, {Schr{\"o}der}, {Stootman},
  {Webster}, {Wilson}, \& {Ye}}]{2001MNRAS.322..486B}
{Barnes}, D.~G., {Staveley-Smith}, L., {de Blok}, W.~J.~G., {et~al.} 2001,
  \mnras, 322, 486

\bibitem[{{Blitz} \& {Rosolowsky}(2004)}]{2004ApJ...612L..29B}
{Blitz}, L. \& {Rosolowsky}, E. 2004, \apjl, 612, L29

\bibitem[{{Blitz} \& {Rosolowsky}(2006)}]{2006ApJ...650..933B}
{Blitz}, L. \& {Rosolowsky}, E. 2006, \apj, 650, 933

\bibitem[{{Boulares} \& {Cox}(1990)}]{1990ApJ...365..544B}
{Boulares}, A. \& {Cox}, D.~P. 1990, \apj, 365, 544

\bibitem[{{Braun} \& {Thilker}(2004)}]{2004A&A...417..421B}
{Braun}, R. \& {Thilker}, D.~A. 2004, \aap, 417, 421

\bibitem[{{Cen} \& {Ostriker}(1999)}]{1999ApJ...514....1C}
{Cen}, R. \& {Ostriker}, J.~P. 1999, \apj, 514, 1

\bibitem[{{Corbelli} \& {Bandiera}(2002)}]{2002ApJ...567..712C}
{Corbelli}, E. \& {Bandiera}, R. 2002, \apj, 567, 712

\bibitem[{{Dav{\'e}} {et~al.}(2001){Dav{\'e}}, {Cen}, {Ostriker}, {Bryan},
  {Hernquist}, {Katz}, {Weinberg}, {Norman}, \& {O'Shea}}]{2001ApJ...552..473D}
{Dav{\'e}}, R., {Cen}, R., {Ostriker}, J.~P., {et~al.} 2001, \apj, 552, 473

\bibitem[{{Dav{\'e}} {et~al.}(2006){Dav{\'e}}, {Finlator}, \&
  {Oppenheimer}}]{2006MNRAS.370..273D}
{Dav{\'e}}, R., {Finlator}, K., \& {Oppenheimer}, B.~D. 2006, \mnras, 370, 273

\bibitem[{{Dav{\'e}} {et~al.}(1999){Dav{\'e}}, {Hernquist}, {Katz}, \&
  {Weinberg}}]{1999ApJ...511..521D}
{Dav{\'e}}, R., {Hernquist}, L., {Katz}, N., \& {Weinberg}, D.~H. 1999, \apj,
  511, 521

\bibitem[{{Dav{\'e}} {et~al.}(2008){Dav{\'e}}, {Oppenheimer}, \&
  {Sivanandam}}]{2008MNRAS.391..110D}
{Dav{\'e}}, R., {Oppenheimer}, B.~D., \& {Sivanandam}, S. 2008, \mnras, 391,
  110

\bibitem[{{Dav{\'e}} \& {Tripp}(2001)}]{2001ApJ...553..528D}
{Dav{\'e}}, R. \& {Tripp}, T.~M. 2001, \apj, 553, 528

\bibitem[{{Davis} \& {Huchra}(1982)}]{1982ApJ...254..437D}
{Davis}, M. \& {Huchra}, J. 1982, \apj, 254, 437

\bibitem[{{Dekel} {et~al.}(2008){Dekel}, {Birnboim}, {Engel}, {Freundlich},
  {Goerdt}, {Mumcuoglu}, {Neistein}, {Pichon}, {Teyssier}, \&
  {Zinger}}]{2008arXiv0808.0553D}
{Dekel}, A., {Birnboim}, Y., {Engel}, G., {et~al.} 2008, ArXiv e-prints

\bibitem[{{Dove} \& {Shull}(1994)}]{1994ApJ...423..196D}
{Dove}, J.~B. \& {Shull}, J.~M. 1994, \apj, 423, 196

\bibitem[{{Eisenstein} \& {Hu}(1999)}]{1999ApJ...511....5E}
{Eisenstein}, D.~J. \& {Hu}, W. 1999, \apj, 511, 5

\bibitem[{{Finlator} \& {Dav{\'e}}(2008)}]{2008MNRAS.385.2181F}
{Finlator}, K. \& {Dav{\'e}}, R. 2008, \mnras, 385, 2181

\bibitem[{{Fukugita} {et~al.}(1998){Fukugita}, {Hogan}, \&
  {Peebles}}]{1998ApJ...503..518F}
{Fukugita}, M., {Hogan}, C.~J., \& {Peebles}, P.~J.~E. 1998, \apj, 503, 518

\bibitem[{{Giovanelli} {et~al.}(2005){Giovanelli}, {Haynes}, {Kent},
  {Perillat}, {Saintonge}, {Brosch}, {Catinella}, {Hoffman}, {Stierwalt},
  {Spekkens}, {Lerner}, {Masters}, {Momjian}, {Rosenberg}, {Springob},
  {Boselli}, {Charmandaris}, {Darling}, {Davies}, {Lambas}, {Gavazzi},
  {Giovanardi}, {Hardy}, {Hunt}, {Iovino}, {Karachentsev}, {Karachentseva},
  {Koopmann}, {Marinoni}, {Minchin}, {Muller}, {Putman}, {Pantoja}, {Salzer},
  {Scodeggio}, {Skillman}, {Solanes}, {Valotto}, {van Driel}, \& {van
  Zee}}]{2005AJ....130.2598G}
{Giovanelli}, R., {Haynes}, M.~P., {Kent}, B.~R., {et~al.} 2005, \aj, 130, 2598

\bibitem[{{Groth} \& {Peebles}(1977)}]{1977ApJ...217..385G}
{Groth}, E.~J. \& {Peebles}, P.~J.~E. 1977, \apj, 217, 385

\bibitem[{{Haardt} \& {Madau}(2001)}]{2001cghr.confE..64H}
{Haardt}, F. \& {Madau}, P. 2001, in Clusters of Galaxies and the High Redshift
  Universe Observed in X-rays

\bibitem[{{Keres} {et~al.}(2003){Keres}, {Yun}, \&
  {Young}}]{2003ApJ...582..659K}
{Keres}, D., {Yun}, M.~S., \& {Young}, J.~S. 2003, \apj, 582, 659

\bibitem[{{Kere\v s} {et~al.}(2008){Kere\v s}, {Katz}, {Fardal}, {Dave}, \&
  {Weinberg}}]{2008arXiv0809.1430K}
{Kere\v s}, D., {Katz}, N., {Fardal}, M., {Dave}, R., \& {Weinberg}, D.~H.
  2008, ArXiv e-prints

\bibitem[{{Kere{\v s}} {et~al.}(2005){Kere{\v s}}, {Katz}, {Weinberg}, \&
  {Dav{\'e}}}]{2005MNRAS.363....2K}
{Kere{\v s}}, D., {Katz}, N., {Weinberg}, D.~H., \& {Dav{\'e}}, R. 2005,
  \mnras, 363, 2

\bibitem[{{Landy} \& {Szalay}(1993)}]{1993ApJ...412...64L}
{Landy}, S.~D. \& {Szalay}, A.~S. 1993, \apj, 412, 64

\bibitem[{{Mellema} {et~al.}(2006){Mellema}, {Iliev}, {Pen}, \&
  {Shapiro}}]{2006MNRAS.372..679M}
{Mellema}, G., {Iliev}, I.~T., {Pen}, U.-L., \& {Shapiro}, P.~R. 2006, \mnras,
  372, 679

\bibitem[{{Meyer} {et~al.}(2007){Meyer}, {Zwaan}, {Webster}, {Brown}, \&
  {Staveley-Smith}}]{2007ApJ...654..702M}
{Meyer}, M.~J., {Zwaan}, M.~A., {Webster}, R.~L., {Brown}, M.~J.~I., \&
  {Staveley-Smith}, L. 2007, \apj, 654, 702

\bibitem[{{Monaghan} \& {Lattanzio}(1985)}]{1985A&A...149..135M}
{Monaghan}, J.~J. \& {Lattanzio}, J.~C. 1985, \aap, 149, 135

\bibitem[{{Murray} {et~al.}(2005){Murray}, {Quataert}, \&
  {Thompson}}]{2005ApJ...618..569M}
{Murray}, N., {Quataert}, E., \& {Thompson}, T.~A. 2005, \apj, 618, 569

\bibitem[{{Noordermeer} {et~al.}(2005){Noordermeer}, {van der Hulst},
  {Sancisi}, {Swaters}, \& {van Albada}}]{2005A&A...442..137N}
{Noordermeer}, E., {van der Hulst}, J.~M., {Sancisi}, R., {Swaters}, R.~A., \&
  {van Albada}, T.~S. 2005, \aap, 442, 137

\bibitem[{{Norberg} {et~al.}(2002){Norberg}, {Baugh}, {Hawkins}, {Maddox},
  {Madgwick}, {Lahav}, {Cole}, {Frenk}, {Baldry}, {Bland-Hawthorn}, {Bridges},
  {Cannon}, {Colless}, {Collins}, {Couch}, {Dalton}, {De Propris}, {Driver},
  {Efstathiou}, {Ellis}, {Glazebrook}, {Jackson}, {Lewis}, {Lumsden},
  {Peacock}, {Peterson}, {Sutherland}, \& {Taylor}}]{2002MNRAS.332..827N}
{Norberg}, P., {Baugh}, C.~M., {Hawkins}, E., {et~al.} 2002, \mnras, 332, 827

\bibitem[{{Obreschkow} \& {Rawlings}(2009)}]{2009MNRAS.tmp..289O}
{Obreschkow}, D. \& {Rawlings}, S. 2009, \mnras, 289

\bibitem[{{Oppenheimer} \& {Dav{\'e}}(2006)}]{2006MNRAS.373.1265O}
{Oppenheimer}, B.~D. \& {Dav{\'e}}, R. 2006, \mnras, 373, 1265

\bibitem[{{Oppenheimer} \& {Dav{\'e}}(2008)}]{2008MNRAS.387..577O}
{Oppenheimer}, B.~D. \& {Dav{\'e}}, R. 2008, \mnras, 387, 577

\bibitem[{{Oppenheimer} \& {Dav{\'e}}(2009)}]{2009MNRAS.395.1875O}
{Oppenheimer}, B.~D. \& {Dav{\'e}}, R. 2009, \mnras, 395, 1875

\bibitem[{{Osterbrock}(1989)}]{1989agna.book.....O}
{Osterbrock}, D.~E. 1989, {Astrophysics of gaseous nebulae and active galactic
  nuclei} (Research supported by the University of California, John Simon
  Guggenheim Memorial Foundation, University of Minnesota, et al.~Mill Valley,
  CA, University Science Books, 1989, 422 p.)

\bibitem[{{Pelupessy}(2005)}]{2005PhDT........17P}
{Pelupessy}, F.~I. 2005, {PhD thesis} (Leiden University)

\bibitem[{{Pontzen} {et~al.}(2008){Pontzen}, {Governato}, {Pettini}, {Booth},
  {Stinson}, {Wadsley}, {Brooks}, {Quinn}, \& {Haehnelt}}]{2008MNRAS.390.1349P}
{Pontzen}, A., {Governato}, F., {Pettini}, M., {et~al.} 2008, \mnras, 390, 1349

\bibitem[{{Rosenberg} \& {Schneider}(2002)}]{2002ApJ...567..247R}
{Rosenberg}, J.~L. \& {Schneider}, S.~E. 2002, \apj, 567, 247

\bibitem[{{Schechter}(1976)}]{1976ApJ...203..297S}
{Schechter}, P. 1976, \apj, 203, 297

\bibitem[{{Spergel} {et~al.}(2007){Spergel}, {Bean}, {Dor{\'e}}, {Nolta},
  {Bennett}, {Dunkley}, {Hinshaw}, {Jarosik}, {Komatsu}, {Page}, {Peiris},
  {Verde}, {Halpern}, {Hill}, {Kogut}, {Limon}, {Meyer}, {Odegard}, {Tucker},
  {Weiland}, {Wollack}, \& {Wright}}]{2007ApJS..170..377S}
{Spergel}, D.~N., {Bean}, R., {Dor{\'e}}, O., {et~al.} 2007, \apjs, 170, 377

\bibitem[{{Springel} \& {Hernquist}(2002)}]{2002MNRAS.333..649S}
{Springel}, V. \& {Hernquist}, L. 2002, \mnras, 333, 649

\bibitem[{{Springel} \& {Hernquist}(2003{\natexlab{a}})}]{2003MNRAS.339..289S}
{Springel}, V. \& {Hernquist}, L. 2003{\natexlab{a}}, \mnras, 339, 289

\bibitem[{{Springel} \& {Hernquist}(2003{\natexlab{b}})}]{2003MNRAS.339..312S}
{Springel}, V. \& {Hernquist}, L. 2003{\natexlab{b}}, \mnras, 339, 312

\bibitem[{{Swaters} {et~al.}(2002){Swaters}, {van Albada}, {van der Hulst}, \&
  {Sancisi}}]{2002A&A...390..829S}
{Swaters}, R.~A., {van Albada}, T.~S., {van der Hulst}, J.~M., \& {Sancisi}, R.
  2002, \aap, 390, 829

\bibitem[{{Tielens}(2005)}]{2005pcim.book.....T}
{Tielens}, A.~G.~G.~M. 2005, {The Physics and Chemistry of the Interstellar
  Medium}

\bibitem[{{Verner} \& {Ferland}(1996)}]{1996ApJS..103..467V}
{Verner}, D.~A. \& {Ferland}, G.~J. 1996, \apjs, 103, 467

\bibitem[{{Walterbos} \& {Braun}(1996)}]{1996ASPC..106....1W}
{Walterbos}, R.~A.~M. \& {Braun}, R. 1996, in The Minnesota Lectures on
  Extragalactic Neutral Hydrogen, Vol. 106, 1--

\bibitem[{{Weinberg} {et~al.}(1997){Weinberg}, {Miralda-Escude}, {Hernquist},
  \& {Katz}}]{1997ApJ...490..564W}
{Weinberg}, D.~H., {Miralda-Escude}, J., {Hernquist}, L., \& {Katz}, N. 1997,
  \apj, 490, 564

\bibitem[{{Wolfire} {et~al.}(2003){Wolfire}, {McKee}, {Hollenbach}, \&
  {Tielens}}]{2003ApJ...587..278W}
{Wolfire}, M.~G., {McKee}, C.~F., {Hollenbach}, D., \& {Tielens}, A.~G.~G.~M.
  2003, \apj, 587, 278

\bibitem[{{Wong} \& {Blitz}(2002)}]{2002ApJ...569..157W}
{Wong}, T. \& {Blitz}, L. 2002, \apj, 569, 157

\bibitem[{{Zehavi} {et~al.}(2005){Zehavi}, {Eisenstein}, {Nichol}, {Blanton},
  {Hogg}, {Brinkmann}, {Loveday}, {Meiksin}, {Schneider}, \&
  {Tegmark}}]{2005ApJ...621...22Z}
{Zehavi}, I., {Eisenstein}, D.~J., {Nichol}, R.~C., {et~al.} 2005, \apj, 621,
  22

\bibitem[{{Zheng} \& {Miralda-Escud{\'e}}(2002)}]{2002ApJ...568L..71Z}
{Zheng}, Z. \& {Miralda-Escud{\'e}}, J. 2002, \apjl, 568, L71

\bibitem[{{Zwaan} {et~al.}(2003){Zwaan}, {Staveley-Smith}, {Koribalski},
  {Henning}, {Kilborn}, {Ryder}, {Barnes}, {Bhathal}, {Boyce}, {de Blok},
  {Disney}, {Drinkwater}, {Ekers}, {Freeman}, {Gibson}, {Green}, {Haynes},
  {Jerjen}, {Juraszek}, {Kesteven}, {Knezek}, {Kraan-Korteweg}, {Mader},
  {Marquarding}, {Meyer}, {Minchin}, {Mould}, {O'Brien}, {Oosterloo}, {Price},
  {Putman}, {Ryan-Weber}, {Sadler}, {Schr{\"o}der}, {Stewart}, {Stootman},
  {Warren}, {Waugh}, {Webster}, \& {Wright}}]{2003AJ....125.2842Z}
{Zwaan}, M.~A., {Staveley-Smith}, L., {Koribalski}, B.~S., {et~al.} 2003, \aj,
  125, 2842

\bibitem[{{Zwaan} {et~al.}(2005){Zwaan}, {van der Hulst}, {Briggs},
  {Verheijen}, \& {Ryan-Weber}}]{2005MNRAS.364.1467Z}
{Zwaan}, M.~A., {van der Hulst}, J.~M., {Briggs}, F.~H., {Verheijen}, M.~A.~W.,
  \& {Ryan-Weber}, E.~V. 2005, \mnras, 364, 1467

\end{thebibliography}
\end{document}